\documentclass[a4paper,11pt]{article}
\pdfoutput=1 

\usepackage{jheppub} 

\usepackage[T1]{fontenc} 

\title{Comprehensive analysis of beta decays  \\ 
within and beyond the Standard Model}

\usepackage{graphicx}
 \usepackage{dcolumn}
\usepackage{bm}
\usepackage{color}
\usepackage[normalem]{ulem}
\usepackage{hyperref}
\usepackage{braket} 
\usepackage{amsmath}
\usepackage{cancel}

\begin{document}
\newcommand{\fref}[1]{Fig.~\ref{fig:#1}} 
\newcommand{\eref}[1]{Eq.~\eqref{eq:#1}} 
\newcommand{\erefn}[1]{ (\ref{eq:#1})}
\newcommand{\erefs}[2]{Eqs.~(\ref{eq:#1}) - (\ref{eq:#2}) } 
\newcommand{\aref}[1]{Appendix~\ref{app:#1}}
\newcommand{\sref}[1]{Section~\ref{sec:#1}}
\newcommand{\cref}[1]{Chapter~\ref{ch:.#1}}
\newcommand{\tref}[1]{Table~\ref{tab:#1}}

\newcommand{\nn}{\nonumber \\}  
\newcommand{\nnl}{\nonumber \\}  
\newcommand{\nl}{& \nonumber \\ &}
\newcommand{\bnl}{\right .  \nonumber \\  \left .}
\newcommand{\dbnl}{\right .\right . & \nonumber \\ & \left .\left .}

\newcommand{\beq}{\begin{equation}} 
\newcommand{\eeq}{\end{equation}} 
\newcommand{\ba}{\begin{array}}  
\newcommand{\ea}{\end{array}} 
\newcommand{\bea}{\begin{eqnarray}}  
\newcommand{\eea}{\end{eqnarray} }  
\newcommand{\be}{\begin{eqnarray}}  
\newcommand{\ee}{\end{eqnarray} }  
\newcommand{\bal}{\begin{align}}
\newcommand{\eal}{\end{align}}   
\newcommand{\bi}{\begin{itemize}}  
\newcommand{\ei}{\end{itemize}}  
\newcommand{\ben}{\begin{enumerate}}  
\newcommand{\een}{\end{enumerate}}  
\newcommand{\bc}{\begin{center}}
\newcommand{\ec}{\end{center}} 
\newcommand{\bt}{\begin{table}}
\newcommand{\et}{\end{table}}  
\newcommand{\btb}{\begin{tabular}}
\newcommand{\etb}{\end{tabular}}  
\newcommand{\bvec}{\left ( \ba{c}}
\newcommand{\evec}{\ea \right )}

\newcommand{\cO}{{\mathcal O}} 
\newcommand{\co}{{\mathcal O}} 
\newcommand{\cL}{{\mathcal L}} 
\newcommand{\cl}{{\mathcal L}} 
\newcommand{\cM}{{\mathcal M}}

\newcommand{\const}{\mathrm{const}}

\newcommand{\ev}{ \mathrm{eV}}
\newcommand{\kev}{\mathrm{keV}}
\newcommand{\mev}{\mathrm{MeV}}
\newcommand{\gev}{\mathrm{GeV}}
\newcommand{\tev}{\mathrm{TeV}}

\newcommand{\mpl}{M_{\mathrm Pl}}

\def\mgut{\, M_{\rm GUT}}
\def\tgut{\, t_{\rm GUT}}
\def\mpl{\, M_{\rm Pl}}
\def\mkk{\, M_{\rm KK}}
\newcommand{\msusy}{M_{\rm soft}}

\newcommand{\dslash}[1]{#1 \! \! \! {\bf /}}
\newcommand{\ddslash}[1]{#1 \! \! \! \!  {\bf /}}

\def\ads{AdS$_5$\,}
\def\adse{AdS$_5$}
\def\intdk{\int {d^4 k \over (2 \pi)^4}} 

\def\ra{\rangle}
\def\la{\langle}  

\def\sgn{{\rm sgn}}
\def\pa{\partial}  
\newcommand{\dlr}{\overleftrightarrow{\partial}}
\newcommand{\Dlr}{\overleftrightarrow{D}}
\newcommand{\re}{{\mathrm{Re}} \,}
\newcommand{\im}{{\mathrm{Im}} \,}
\newcommand{\tr}{\mathrm T \mathrm r}  

\newcommand{\Ra}{\Rightarrow}
\newcommand{\lra}{\leftrightarrow}
\newcommand{\llra}{\longleftrightarrow}

\newcommand\simlt{\stackrel{<}{{}_\sim}}
\newcommand\simgt{\stackrel{>}{{}_\sim}}   
\newcommand{\zt}{$\mathbb Z_2$ }

\newcommand{\ha}{{\hat a}}
\newcommand{\hab}{{\hat b}}
\newcommand{\hac}{{\hat c}} 

\newcommand{\ti}{\tilde}  
\def\hc{{\rm h.c.}} 
\def\ov{\overline}  
  

\newcommand{\eps}{\epsilon}
\newcommand{\eS}{\epsilon_S}
\newcommand{\eT}{\epsilon_T}
\newcommand{\eP}{\epsilon_P}
\newcommand{\eL}{\epsilon_L}
\newcommand{\eR}{\epsilon_R}
\newcommand{\teps}{{\tilde{\epsilon}}}
\newcommand{\teS}{{\tilde{\epsilon}_S}}
\newcommand{\teT}{{\tilde{\epsilon}_T}}
\newcommand{\teP}{{\tilde{\epsilon}_P}}
\newcommand{\teL}{{\tilde{\epsilon}_L}}
\newcommand{\teR}{{\tilde{\epsilon}_R}}
\newcommand{\eLc}{{\epsilon_L^{(c)}}}
\newcommand{\eLv}{{\epsilon_L^{(v)}}}
\newcommand{\eSP}{\epsilon_{S,P}}
\newcommand{\teSP}{{\tilde{\epsilon}_{S,P}}}

\newcommand{\lz}{\lambda_z}
\newcommand{\dgz}{\delta g_{1,z}}
\newcommand{\dkg}{\delta \kappa_\gamma}

\def\cog{\color{OliveGreen}}
\def\cor{\color{Red}}
\def\copu{\color{purple}}
\def\coro{\color{RedOrange}}
\def\coma{\color{Maroon}}
\def\cob{\color{Blue}}
\def\cobr{\color{Brown}}
\def\cobl{\color{Black}}
\def\cost{\color{WildStrawberry}}

\newcommand{\tl}{{\tilde{\lambda}}}
\newcommand{\dll}{{\frac{\delta\lambda}{\lambda}}}

\preprint{IFIC/20-49, FTUV/20-1027}

\emailAdd{adam.falkowski@ijclab.in2p3.fr}
\emailAdd{martin.gonzalez@ific.uv.es}
\emailAdd{naviliat@lpccaen.in2p3.fr}

\author[a]{Adam Falkowski,}
\author[b]{Mart\'{i}n Gonz\'{a}lez-Alonso,}
\author[c,d]{and Oscar Naviliat-Cuncic}


\affiliation[a]{Universit\'{e} Paris-Saclay, CNRS/IN2P3, IJCLab, 91405 Orsay, France}
\affiliation[b]{Departament de F\'isica Te\`orica, IFIC, Universitat de Val\`encia - CSIC, Apt.  Correus 22085, E-46071 Val\`encia, Spain}
\affiliation[c]{Laboratoire de Physique Corpusculaire de Caen, CNRS/IN2P3, ENSICAEN, Universit\'e de Caen Normandie, 14050 Caen, France}
\affiliation[d]{National Superconducting Cyclotron Laboratory and Department of Physics and Astronomy, Michigan State University, East Lansing, 48824 MI, USA}

\abstract{
Precision measurements in allowed nuclear beta decays and neutron decay are reviewed and analyzed both within the Standard Model and looking for new physics. The analysis incorporates the most recent experimental and theoretical developments. The results are interpreted in terms of Wilson coefficients describing the effective interactions between leptons and nucleons (or quarks)   that are responsible for beta decay. 
New global fits are performed incorporating a comprehensive list of precision  measurements in  neutron decay, superallowed $0^+\rightarrow 0^+$ transitions, and other nuclear decays that include, for the first time, data from mirror beta transitions.
The results confirm the $V$-$A$ character of the interaction and translate into updated values for $V_{ud}$ and $g_A$ at the $10^{-4}$ level. 
We also place new stringent limits on exotic couplings involving left-handed and right-handed neutrinos, which benefit significantly from the inclusion of mirror decays in the analysis.
}

\maketitle

\section{Introduction} 

Precision measurements are essential ingredients in  particle physics research. 
Their role is twofold. 
On the one hand, they allow us to extract numerical values of the free parameters of the Standard Model (SM), 
such as the gauge couplings or the Cabibbo-Kobayashi-Maskawa (CKM) matrix elements.
On the other hand, they provide us with information about non-SM particles and their interactions, even if these particles are beyond reach of existing particle colliders.

Studies of {\em beta decay} processes are an important and active field  within the precision  program~\cite{Cirigliano:2013xha, Gonzalez-Alonso:2013uqa, Vos:2015eba,Gonzalez-Alonso:2018omy}. 
In the SM, beta decays are mediated by the exchange of a $W$ boson between light quark and lepton currents. 
Since the interaction strength between the $W$ and quarks in the SM is controlled by the CKM matrix, beta decays provide an opportunity to  extract the $V_{ud}$ element of that matrix.
In fact, this is currently the most precise (by far!) method to determine $V_{ud}$~\cite{Zyla:2020zbs}.  

Hypothetical particles such as $W'$ bosons or leptoquarks may alter the rates and angular correlations of beta decays, as compared to the SM predictions.  
Rather than studying each such model successively, 
it is convenient to resort to effective field theory (EFT) techniques. In this approach, information about the underlying physics at high-energies that is relevant for beta decays is condensed into a small number of {\em Wilson coefficients}, which parametrize the strength of  effective interactions between nucleons, electrons, and neutrinos at low energies.   
The general EFT Lagrangian describing these interactions at the leading order was written more than 60 years ago by Lee and Yang~\cite{Lee:1956qn}:
\bea
\label{eq:INT_Lleeyang}
\cL_{\rm Lee-Yang} &=& 
-  \bar{p}\gamma^\mu n 
\left( C_V \bar{e} \gamma_\mu  \nu
- C_V' \bar{e} \gamma_\mu \gamma_5 \nu  \right) 
+  \bar{p}\gamma^\mu \gamma_5 n 
\left( C_A \bar{e} \gamma_\mu \gamma_5 \nu
-  C_A' \bar{e} \gamma_\mu  \nu  \right)  
\nonumber\\
 &- &  
  \bar{p}n \left( 
C_S \bar{e}  \nu -  C_S' \bar{e} \gamma_5 \nu  \right)
- \frac{1}{2}\bar{p}\sigma^{\mu\nu} n 
 \left( C_T \bar{e} \sigma_{\mu\nu} \nu 
-  C_T' \bar{e} \sigma_{\mu\nu} \gamma_5 \nu  \right) 
\nonumber\\
& - &  \bar{p} \gamma_5 n  \left( C_P \bar{e}\gamma_5 \nu 
-  C_P' \bar{e}\nu  \right)
+ \hc   
\eea
Soon after that work, observables in beta decays were calculated in  terms of the Wilson coefficients $C_X^{(')}$~\cite{Jackson:1957auh,EBEL1957213}. This program led to the determination of the $V$-$A$ structure of the weak interaction, and it is currently used in searches for additional non-standard interactions.
Along with refined experimental techniques and with careful calculations of small SM contributions, this program is strengthened by model-independent analyses of the data. 
The latter is the focus of this work, which is achieved through the construction of a likelihood  function for the Wilson coefficients $C_X^{(')}$. 
Given the likelihood, it is straightforward to extract information about any underlying high-energy theory  - be it the SM or one of its extensions - once the map between $C_X^{(')}$ and the high-energy parameters is established. 
The map can be calculated via the usual EFT techniques of matching at consecutive heavy particle thresholds, and renormalization group running in between  thresholds (see e.g. Ref.~\cite{Pich:1998xt}).  

Although a number of global fits of beta decay data have been performed in the past \cite{Paul:1970gfz,Boothroyd:1984fz,Severijns:2006dr,Gonzalez-Alonso:2018omy}, the complete likelihood function for the $C_X^{(')}$, including all correlations, has never been constructed.
Instead, only partial results are available, with selected Wilson coefficients simultaneously included in the fits. 
Such results have a limited value, since many theories beyond the SM generate an intricate pattern of Wilson coefficients, especially when the effect of mixing under renormalization group is taken into account.     
One of the goals of the present work is to implement a complete and  model-independent EFT approach in the field of beta decays that includes the state-of-the art measurements and theoretical developments. 
We obtain a likelihood function for the Wilson coefficients in \eref{INT_Lleeyang} while allowing {\em all} $C_X^{(')}$ to be present at the same time, and taking into account all correlations.

In this analysis we use a comprehensive list of sensitive observables for {\em allowed} beta transitions, that is, the ones controlled by the Fermi and/or Gamow-Teller (GT) nuclear matrix elements. 
A key novelty of the present work is that we analyze, for the first time in a systematic and consistent fashion, the role of {\em mirror beta decays} in new physics searches.  
The name {\em mirror} refers to mixed Fermi-GT $J^+\!\!\!\to J^+$ transitions with isospin $T=1/2$ nuclei in both the initial and final states.\footnote{%
Formally speaking, neutron decay is also a mirror transition, 
but we keep it in a separate category, and reserve the name ``mirror'' for  transitions involving nuclei with $A>1$.}  
Because of the high degree of theoretical control over the nuclear matrix elements, the importance of mirror decays for the precision program has long been recognized~\cite{Severijns:2008ep,NaviliatCuncic:2008xt}. 
In particular, they offer an alternative path to determining the $V_{ud}$ parameter~\cite{NaviliatCuncic:2008xt}, 
which is subject to a vibrant experimental program for the determination of the relevant spectroscopic quantities and correlations. 
While the results are currently inferior in accuracy to $V_{ud}$ determinations from superallowed and neutron decays, they are affected by completely different systematic uncertainties, and thus they offer an important cross-check.

Another goal of this work is to analyze whether mirror transitions play a (numerically) important role in new physics searches. 
This is a reasonable possibility since the multi-dimensional parameter space of the Wilson coefficients in \eref{INT_Lleeyang} is much larger than in the SM case. 
Indeed, we find that including mirror decays in the global fit leads to improvement of model-independent constraints on $C_X^{(')}$ by approximately a factor of two.
The results translate into stringent constraints on non-standard currents involving left- and right-handed neutrinos. 
We find that current data are well described by the SM. 
At the same time, the most general beyond-the-SM (BSM) fit shows a $3.2\sigma$ preference for non-standard tensor current interactions involving the right-handed neutrino. 
This tension is driven by a single recent measurement 
($a_n$ by the aSPECT collaboration~\cite{Beck:2019xye}) and thus it should be taken with caution. 

The structure of this article is the following. In~\sref{theory} we review the theoretical formalism used for the quantitative description of mirror and other beta transitions.  
In~\sref{experiment} we describe the experimental results relevant for this analysis. 
\sref{fits} contains the main results: the confidence intervals  for the Wilson coefficients of the effective Lagrangian in \eref{INT_Lleeyang}. 
We conclude in \sref{conclusions}, and briefly discuss the prospect of improving the existing constraints by more precise measurements in mirror beta transitions.

\section{Theoretical formalism}
\label{sec:theory} 

\subsection{Lagrangian}
\label{sec:Lagrangian}
At a fundamental level, nuclear beta decays probe charged-current interactions between the first generation of quarks and leptons.
In this paper we adopt the EFT approach to parametrize these interactions. 
The central assumption is that, at the energy scale corresponding to beta decays, there is no other light degrees of freedom except for those of the SM and (eventually) the right-handed electron neutrino. 
Given this field content, the leading order effective Lagrangian at the scale $\mu \simeq 2$~GeV contains the following interactions relevant for  beta decays
\bea
\label{eq:TH_Lrweft}
\cL  \supset 
- \frac{V_{ud}}{v^2} \Big[ \
&&\!\!\!\!\left( 1 +  \eL \right) \
\bar{e}  \gamma_\mu  \nu_L  \cdot \bar{u}   \gamma^\mu  (1 - \gamma_5)  d \Big.
~+~
\teL  \,\bar{e}  \gamma_\mu \nu_R  \cdot \bar{u}   \gamma^\mu  (1 - \gamma_5)  d
\nnl
&& +\eR   \,  \bar{e}  \gamma_\mu \nu_L
\cdot \bar{u}   \gamma^\mu  (1 + \gamma_5)  d
~+~
\teR   \,  \bar{e}  \gamma_\mu    \nu_R
\cdot \bar{u}   \gamma^\mu  (1 + \gamma_5)  d
\nnl
&& +{1 \over 4} \eT    \   \bar{e}   \sigma_{\mu \nu} \nu_L    \cdot  \bar{u}   \sigma^{\mu \nu} (1 - \gamma_5) d
~+~
\Big.
{1 \over 4} \teT      \   \bar{e}   \sigma_{\mu \nu}  \nu_R    \cdot  \bar{u}
 \sigma^{\mu \nu} (1 + \gamma_5) d  
 \nnl
&& + \eS  \, \bar{e}  \nu_L  \cdot  \bar{u} d
~+~
\teS  \, \bar{e}  \nu_R  \cdot  \bar{u} d
- \eP  \,  \bar{e}  \nu_L  \cdot  \bar{u} \gamma_5 d
~-~
\teP  \,  \bar{e}  \nu_R  \cdot  \bar{u} \gamma_5 d
\Big] 
+ {\rm h.c.}~ \qquad 
\label{eq:leff-lowE}
\eea
where $u$, $d$, $e$ are the up quark, down quark, and electron fields,
$\nu_{L,R} \equiv (1\pm\gamma_5)\nu/2$ are the left-handed and right-handed electron neutrino fields, 
$V_{ud}$ is the $[\cdot]_{11}$ real entry of the unitary CKM matrix, 
and $v \approx 246.22$~GeV is related to the Fermi constant by 
$G_F = (\sqrt 2 v^2)^{-1}$. 
We treat the neutrinos as massless.\footnote{
For neutrinos with any appreciable coupling to matter, the cosmological constraints require $\sum m_\nu < 0.12$~eV~\cite{Aghanim:2018eyx}, making their masses completely negligible for the beta processes we include in this analysis.
The EFT framework adopted in this work assumes the absence of any other non-SM light degrees of freedom at the energy scale relevant for beta decays. 
} 
The Wilson coefficients $\epsilon_X$ and $\tilde \epsilon_X$, 
$X = L,R,S,P,T$,  parametrize possible effects of non-SM particles, which have been integrated out.\footnote{%
Note that the normalization of $\epsilon_T$ and $\tilde \epsilon_T$   differs by $1/4$ from that used in previous works, e.g.,  Ref.~\cite{Gonzalez-Alonso:2018omy}. 
The ``new" normalization is more natural in the sense that typical new physics models generating tensor interactions give similar contribution to $\epsilon_T$ and $\epsilon_{S,P}$, see for instance Ref.~\cite{deBlas:2017xtg}.}
We assume, for concreteness, that all $\epsilon_X$ and $\tilde \epsilon_X$ are real, since the observables included in the analysis are not sensitive to their imaginary parts at the linear order in new physics. 
In the SM limit we have $\epsilon_X = \tilde \epsilon_X = 0$ for all $X$.
The situation where the right-handed neutrino is absent from the low-energy EFT (e.g. because it has a large Majorana mass) can be described by setting  $\tilde \epsilon_X = 0$ for all $X$.

The quark-level Lagrangian in \eref{TH_Lrweft} is what matters in particle physics. 
Its parameters can be readily related to masses and couplings of specific BSM models,
or to Wilson coefficients of a more fundamental EFT above the electroweak scale. 
However, the momentum exchange in beta decays is far below the QCD scale, and thus we need to connect the quark-level Lagrangian above with the nucleon-level Lee-Yang Lagrangian in~\eref{INT_Lleeyang}, which can be rewritten in the following form:
\bea
\label{eq:TH_Lleeyang}
\cL_{\rm Lee-Yang} &=& 
-  \bar{p}\gamma^\mu n 
\left( C_V^+ \bar{e} \gamma_\mu  \nu_L 
+ C_V^- \bar{e} \gamma_\mu \nu_R  \right) 
-  \bar{p}\gamma^\mu \gamma_5 n 
\left( C_A^+ \bar{e} \gamma_\mu  \nu_L 
-  C_A^- \bar{e} \gamma_\mu  \nu_R   \right)  
\nonumber\\
 &-&  
   \bar{p}n \left( 
C_S^+ \bar{e}  \nu_L + C_S^- \bar{e} \nu_R  \right)
- \frac{1}{2}\bar{p}\sigma^{\mu\nu} n 
 \left( C_T^+ \bar{e} \sigma_{\mu\nu} \nu_L 
 + C_T^- \bar{e} \sigma_{\mu\nu} \nu_R  \right) 
\nonumber\\
& + &  \bar{p} \gamma_5 n  \left( C_P^+ \bar{e}\nu_L 
-  C_P^- \bar{e}\nu_R  \right)
+ \hc   
\eea
after a simple change of variables $C_X = (C_X^+ + C_X^-)/2$, $C_X' = (C_X^+ - C_X^-)/2$ that
separates the left-handed (via $C_X^+$) and right-handed (via $C_X^-$) neutrino couplings, which affect the nuclear observables in a different way due to the fact that only the former may interfere with the SM amplitudes.  
Consequently, in the $C_X^\pm$ variables the experimental constraints  are physically more transparent, and the matching to the quark-level Lagrangian  is more straightforward. 
At a more practical level, correlations in the global fit are hugely reduced  when using  the $C_X^\pm$ variables.  
The relation between the parameters in \eref{TH_Lrweft} and in \eref{TH_Lleeyang} is given by~\cite{Gonzalez-Alonso:2018omy}
\bea
\label{eq:TH_LYtoRWEFT}
C_V^+ & = & {V_{ud} \over  v^2} g_V \sqrt{1 + \Delta_R^V} \big ( 1+ \epsilon_L + \epsilon_R \big ) , 
\qquad 
C_V^-  =  {V_{ud} \over v^2} g_V  \sqrt{1 + \Delta_R^V}  \big ( \tilde \epsilon_L + \tilde \epsilon_R \big ) , 
\nnl 
C_A^+ & = & - {V_{ud} \over v^2} g_A  \sqrt{1 + \Delta_R^A}  \big ( 1+ \epsilon_L - \epsilon_R \big ) , 
\qquad 
C_A^-  =  {V_{ud} \over v^2} g_A  \sqrt{1 + \Delta_R^A}  \big ( \tilde \epsilon_L - \tilde \epsilon_R \big ) , 
\nnl
C_T^+ & = & {V_{ud} \over v^2} g_T \epsilon_T  , 
\qquad 
C_T^-  =  {V_{ud} \over v^2} g_T \tilde \epsilon_T, 
\nnl
C_S^+ & = & {V_{ud} \over v^2} g_S \epsilon_S  , 
\qquad 
C_S^-  =  {V_{ud} \over  v^2} g_S \tilde \epsilon_S,  
\nnl
C_P^+ & = & {V_{ud} \over v^2} g_P \epsilon_P  , 
\qquad 
C_P^-  =  - {V_{ud} \over  v^2} g_P \tilde \epsilon_P,  
\eea 
where $g_{V,A,S,P,T}$ are  vector, axial, scalar, pseudoscalar, and tensor charges of the nucleon~\cite{Herczeg:2001vk,Gonzalez-Alonso:2018omy}, which must be determined using lattice or other theoretical techniques. 
For the vector charge, one can prove that  $g_V = 1$ up to (negligible) quadratic corrections in isospin-symmetry breaking~\cite{Ademollo:1964sr}. 
We will use the FLAG'19 averages~\cite{Aoki:2019cca} for the axial, scalar and tensor charges: $g_A = 1.251(33)$, $g_S = 1.022(100)$ and $g_T = 0.989(33)$~\cite{Chang:2018uxx,Gupta:2018qil} (see also Ref.~\cite{Gonzalez-Alonso:2013ura}). 
Although the pseudoscalar charge is enhanced by the pion pole, namely $g_P=349(9)$~\cite{Gonzalez-Alonso:2013ura}, the suppression of the pseudoscalar contributions to the observables is larger and they will be neglected in the following.

The matching in \eref{TH_LYtoRWEFT} includes the short-distance (inner) radiative corrections $\Delta_R^V$ and $\Delta_R^A$. 
Especially the former is important, because it is necessary to extract $V_{ud}$ from nuclear data. 
Four recent calculations of this quantity are available~\cite{Seng:2018yzq,Czarnecki:2019mwq,Seng:2020wjq,Hayen:2020cxh}, all within $1\sigma$. In  this analysis we use the Seng et al. evaluation, $\Delta_R^V = 0.02467(22)$~\cite{Seng:2018yzq}, which has the smallest uncertainty. We will discuss the impact of this choice. 
Given  the assumption about the reality of $\epsilon_X$ and $\tilde \epsilon_X$, all $C_X^\pm$ are then also real.   

At leading order, nuclear effects are encapsulated in the so-called Fermi and GT matrix elements, $M_{F,GT}$. Leading-order expressions for beta decay observables in terms of the Lee-Yang Wilson coefficients $C_X^\pm$ can be found in Refs.~\cite{Jackson:1957zz,EBEL1957213}. 
However, given the experimental precision, subleading effects such as weak-magnetism and long-distance electromagnetic corrections have to be included in the SM terms. These small contributions can be calculated with large accuracy for the transitions that are included in this work~\cite{Holstein:1974zf,Cirigliano:2013xha,Hayen:2017pwg,Hayen:2020nej}.

The main results of this work are the constraints on the $C_X^\pm$ Wilson coefficients using neutron and nuclear physics data, with special attention to the role played by mirror beta decays which are included in a global fit for the first time. 
In the remainder of this section we review how beta decay observables included in this work depend on the $C_X^\pm$ Wilson coefficients.

\subsection{{\cal F}t values and neutron lifetime}
\label{sec:Ftvalues}

The total decay width of an allowed beta transition can be calculated from the formula
\beq
\label{eq:TH_gamma}
\Gamma_i   =  \big (1  + \delta_i \big )   
{M_F^2 m_e^5  \over 4 \pi^3} f^i_V   \hat \xi_i 
\bigg [1+ \gamma_i b_i \bigg \langle{m_e \over E_e} \bigg \rangle_i \bigg ]  \, . 
\eeq 
The index $i$ labels transition-dependent quantities.
Radiative corrections, 
other than the short-distance ones already included in $\Delta_R^{V,A}$, are encoded in   $\delta_i$. 
They are customarily split as 
$1+ \delta_i = (1 + \delta_R') (1 + \delta_{NS}^V - \delta_C^V)$, 
where $\delta_R'$ is the dominant piece of the long-distance (outer) radiative corrections that depends trivially (only via $Z$ and $\Delta$) on the nucleus, 
$\delta_{NS}^V$ is the nuclear-structure-dependent piece, 
and $\delta_C^V$ is the isospin-symmetry breaking correction.
$M_F$ is the Fermi matrix element in the isospin limit, which for a transition between two members of the same isospin multiplet is given by $M_F = \sqrt{j(j+1) - m(m \pm 1)}$, where $(j,m)$ are the isospin quantum numbers of the parent nucleus, 
The factor $f^i_V$ is the phase space integral given by 
\beq 
f^i_V = {1 \over m_e^5} \int_{m_e}^{\Delta_i} 
d E_e F_i (\Delta_i - E_e)^2 p_e E_e , 
\eeq 
where $m_e$ is the electron mass, 
$p_e = \sqrt{E_e^2 - m_e^2}$, 
$\Delta_i$ is the maximal total energy of the beta particle, 
and  $F_i \equiv 4 (2p_e\,R_i)^{2(\gamma_i -1)} e^{\pi \eta_i} {|\Gamma(\gamma_i + i \eta_i)|^2 
/ \Gamma(1 + 2 \gamma_i)^2}$ 
is the Fermi function describing the Coulomb corrections, with
$\gamma_i \equiv \sqrt{ 1 - (\alpha Z_i)^2}$,
$Z_i$ ($R_i$) the charge (radius) of the daughter nucleus, 
 $\alpha$ the electromagnetic structure constant, and
 $\eta_i \equiv \pm {\alpha Z_i E_e / p_e}$. 
Above and in all of the following, the upper (lower) sign applies to   $\beta^-$ ($\beta^+$) transitions. 
Further details about subleading corrections that have to be included in $f^i_V$ are discussed in
Refs.~\cite{Hardy:2004id,Hayen:2017pwg}. 
The second term in the square bracket in \eref{TH_gamma} is called the Fierz term. It is proportional to the weighted average of the electron mass over its energy: 
\beq
\label{eq:TH_meOverEe}
\bigg \langle { m_e \over E_e} \bigg \rangle_i \equiv {\int_{m_e}^{\Delta_i} d E_e F_i (\Delta_i - E_e)^2  p_e m_e 
\over  
\int_{m_e}^{\Delta_i} d E_e F_i (\Delta_i - E_e)^2  p_e E_e }. 
\eeq 

The dependence of the decay width on the Wilson coefficients of the Lee-Yang Lagrangian enters via the combinations $\hat \xi_i$ and $b_i$ defined as\footnote{%
The relation with the traditional notation~\cite{Jackson:1957auh} is simply $\xi_i = M_F^2 \,\hat{\xi}_i/2$.}
\bea 
\label{eq:TH_x}
\hat \xi_i   & \equiv  & (C_V^+)^2   + (C_S^+)^2  +  (C_V^-)^2   + (C_S^-)^2  +  {f_A^i \over f_V^i}  {(C_V^+)^2 \over (C_A^+)^2 } \tilde \rho_i^2 \bigg [  (C_A^+)^2  + (C_T^+)^2  +  (C_A^-)^2  + (C_T^-)^2 \bigg ]  ,
\nnl 
b_i \hat \xi_i   & \equiv & \pm 2   \bigg \{  C_V^+ C_S^+ +  C_V^- C_S^- +   {(C_V^+)^2 \over (C_A^+)^2 } \tilde \rho_i^2   \bigg [ C_A^+ C_T^+  +  C_A^- C_T^-   \bigg ] \bigg \},   
\eea
where the factors $f_A^i/f_V^i$ encode corrections to the (axial-vector) phase space integral.\footnote{%
The $f_A^i/f_V^i$ factors should be included in $\hat{\xi}_i$ for the calculations of the ${\cal F}t$ values, but not for the correlations.}
For the mirror transitions relevant to our analysis,   
$f_A^i/f_V^i$ can be estimated by theoretical methods, and have values close to unity~\cite{Hayen:2019nic}. 
The ``polluted'' mixing ratio $\tilde \rho_i$ in \eref{TH_x} is defined as 
\beq
\label{eq:TH_tildeRho}
\tilde \rho_i =  {C_A^+ \over C_V^+}  {M_{\rm GT}\over M_{\rm F} } 
{ \big ( 1 + \delta^A_{\rm NS} -  \delta^A_{\rm C} \big  )^{1/2} 
\over 
\big ( 1 + \delta^V_{\rm NS} -  \delta^V_{\rm C} \big  )^{1/2} } ~,
\eeq 
where $M_{\rm GT}/ M_{\rm F}$ is the ratio  between the GT and Fermi matrix elements in the isospin limit, 
and  $\delta^A_{\rm NS}$, $\delta^A_{\rm C}$ are the axial-vector equivalents to $\delta^V_{\rm NS}$, $\delta^V_{\rm C}$ mentioned earlier. 
Using \eref{TH_LYtoRWEFT} to re-express $C_A^+/C_V^+$  
one can see that, in the SM limit, 
$\tilde \rho_i$ reduces to the usual mixing ratio $\rho_i$  defined in Refs.~\cite{Severijns:2008ep,NaviliatCuncic:2008xt}. However, in the presence of new physics, 
$\tilde \rho_i$ depends on unknown parameters $\epsilon_{L,R}$, and it is not anymore a pure QCD/nuclear quantity. In practice, the distinction between $\rho_i$ and $\tilde \rho_i$ is not relevant because the mixing ratios $\rho_i$ cannot be calculated with  sufficient precision by current theoretical techniques for nuclei with $A>1$. 
Thus, the mixing ratios, whether $\rho_i$ or $\tilde \rho_i$, have to be treated as free parameters in the fits, together with $C_X^\pm$.

Rather than the observable decay width, experimental groups or theory compilations often communicate the corrected half-life ${\cal F}t$ defined as 
\beq
\label{eq:TH_ft}
{\cal F} t_i  \equiv {f^i_V (1+ \delta_i)\log 2 \over \Gamma_i }
= 
 {4 \pi^3 \log 2 \over M_F^2 m_e^5} 
\bigg [ \hat \xi_i + \gamma_i b_i\hat \xi_i \bigg \langle { m_e \over E_e} \bigg \rangle_i   
\bigg ]^{-1}  . 
\eeq 
For the {\em superallowed} ($0^+ \to 0^+$, $j=1$) beta decays,
one has $M_F= \sqrt 2$ and $\rho = 0$ in
\eref{TH_ft} and \eref{TH_x}. 
In the SM limit, ${\cal F} t$ is predicted to be universal for all superallowed transitions. 
However, beyond the SM  one can have $b_i \neq 0$ and then ${\cal F} t$ depends on the transition via the Fierz factor 
$b_i \langle m_e/E_e\rangle_i$.
For {\em mirror}  ($J^+ \to J^+$, $j=1/2$) beta decays one has $M_F = 1$, and ${\cal F} t_i$ is a function (via $\hat\xi_i$ and $b_i$) 
of the  transition-dependent mixing ratio $\rho_i$.
Consequently, for mirror decays ${\cal F} t_i$ is transition dependent even in the SM limit.  
For {\em neutron decay} one has $M_F = 1$ and $\rho_n = - \sqrt{3} g_A$, up to ${\cal O}(0.2\%)$ radiative corrections~\cite{Hayen:2020cxh}.
Rather than the corrected half-life ${\cal F} t$, 
 experiments customarily quote the neutron lifetime:  
\beq
\label{eq:TH_taun}
\tau_n \equiv {1\over \Gamma_n}  
= {4 \pi^3 \over (1+ \delta_n) m_e^5 f_n} \bigg   [ 
\hat \xi_n  + \gamma_n b_n \hat \xi_n  \bigg \langle { m_e \over E_e} \bigg \rangle_n  \bigg ]^{-1}~.
\eeq 
We use here $f_n = 1.6887(1)$~\cite{Czarnecki:2018okw} and 
$\delta_n = \delta_R'{}^n = 0.014902(2)$~\cite{Towner:2010zz}, 
and ignore the negligible~\cite{Wilkinson:1982hu} correction due to $f_A^n/f_V^n$ in \eref{TH_x}.

\subsection{Correlation measurements}

The lifetime measurements discussed in the previous subsection probe only a limited number of combinations of the Lee-Yang Wilson coefficients and the mixing ratios. 
To avoid degeneracy in a global fit, 
one needs to include a wider palette of observables.
Nuclear experiments measure various angular correlations between the decay products. 
Assuming CP conservation 
and summing over polarizations of the  daughter nucleus and $\beta$ particle, the differential decay width can be cast in the form\footnote{In the presence of CP-violating interactions, an additional term appears in~\eref{TH_correlations} multiplied by the so-called D-coefficient~\cite{Jackson:1957auh}. Likewise, the R-coefficient appears if the electron polarization is measured. We do not discuss them in this work because we assume real Wilson coefficients. 
}
\bea
\label{eq:TH_correlations}
{d \Gamma_i \over  dE_e d {\Omega_e}  d {\Omega_\nu} }&  = &  
 \hat \xi_i   {M_F^2   \over 64 \pi^5 } 
 F_i (\Delta_i - E_e)^2  p_e E_e
\bigg \{ 1+ b_i {m_e \over E_e} 
 + a_i {\vec p_e \over E_e} \cdot  {\vec p_\nu \over E_\nu}  
+  A_i   {\vec p_e \over E_e} \cdot {\langle \vec J \rangle \over J} 
 + B_i   {\vec p_\nu \over E_\nu} \cdot {\langle \vec J \rangle \over J}
  \nnl & + &
  c_i  \bigg [  
{\vec p_e \cdot \vec p_\nu   \over 3 E_e  E_\nu}   
- { (\vec p_e  \cdot \vec j ) 
(\vec p_\nu  \cdot \vec j )  \over  E_e  E_\nu }   \bigg ]   \bigg [ { J(J+1)  - 3 (\langle \vec J \rangle \cdot \vec j)^2  \over 
J (2 J - 1 ) } \bigg ]
 \bigg \}~, 
\eea
where the indices $e$ and $\nu$ refer to the $\beta$ particle and neutrino, respectively and the symbols $\Omega$,  $\vec p$ and $E$ denote the angular coordinates, momentum and energy of the leptons, 
$\langle \vec J \rangle$ is the polarization of the parent nucleus, 
$J$ is its spin, and $\vec j$ is the unit vector in the polarization direction.
Since no experiment measures the coefficient $c_i$,  we will not discuss it in the following. 
The remaining parameters in \eref{TH_correlations} are referred to as the  $\beta$-$\nu$ correlation ($a$), 
the $\beta$-asymmetry ($A$), 
and the neutrino asymmetry ($B$). 
For a mixed Fermi and GT $\beta$ transition  they can be expressed by the Lee-Yang Wilson coefficients as 
\bea
\label{eq:correlations}
\hat \xi_i a_i  & = &
(C_V^+)^2  - (C_S^+)^2 + (C_V^-)^2  -(C_S^-)^2
- {\tilde \rho_i^2 \over 3 } {(C_V^+)^2 \over (C_A^+)^2} \bigg  [  (C_A^+)^2 - (C_T^+)^2   + (C_A^-)^2   - (C_T^-)^2 \bigg ] , 
\nnl
\hat \xi_i A_i  & = &  - 2  \tilde \rho_i {C_V^+ \over C_A^+}  \sqrt{J \over J+1}  \bigg  \{  
 C_V^+ C_A^+   - C_S^+ C_T^+  
-    C_V^- C_A^-   +    C_S^- C_T^-
 \bigg \} 
 \nnl  & & 
\mp  {\tilde \rho_i^2 \over  J+1 } {(C_V^+)^2 \over (C_A^+)^2} \bigg \{  
(C_A^+)^2 -  (C_T^+)^2   - (C_A^-)^2   +  (C_T^-)^2   \bigg \}  , 
\nnl 
\hat \xi_i B_i  & = &
-2 \tilde \rho_i {C_V^+ \over C_A^+}   \sqrt{J \over J+1}   \bigg \{ 
 C_V^+ C_A^+     +  C_S^+ C_T^+   
-  C_V^- C_A^-   -   C_S^- C_T^- 
 \nnl  & & 
 \pm   {m_e \over E_e} \left [    
C_S^+ C_A^+   +  C_V^+ C_T^+   
 - C_S^- C_A^-    -  C_V^- C_T^- 
  \right ]    \bigg \} 
\\ & & 
\pm {\tilde \rho_i^2 \over  J+1 } {(C_V^+)^2 \over (C_A^+)^2} \bigg \{ 
(C_A^+)^2    +  (C_T^+)^2   - (C_A^-)^2   -  (C_T^-)^2 
\pm  {2 m_e \over E_e} \bigg  [   C_A^+ C_T^+  
 -  C_A^-  C_T^-  \bigg  ] 
\bigg \} . 
\nonumber
\eea
We also define the tilde correlation coefficients as
\beq
\label{eq:tildeprescription}
\tilde X_i \equiv  {X_i \over 
1 + b_i   \big \langle { m_e / E_e} \big \rangle_i} , 
\qquad 
\tilde X_i(E_e) \equiv  {X_i \over  1 + b_i { m_e / E_e} },  
\qquad X = a, A, B~,
\eeq
which are the quantities that can be determined from asymmetries (integrated over energies and for a fixed-energy respectively)~\cite{Gonzalez-Alonso:2016jzm}. 
The distinction between the $\tilde X_i$ and $X_i$ correlations is only relevant beyond the SM,  when the Fierz term is non-zero. 
Many experimental  results in the literature do not make this distinction explicit, because they operate under the SM hypothesis. Some care is needed when interpreting such results~\cite{Gonzalez-Alonso:2016jzm}.  
We discuss this issue in more detail in the following section. 

For completeness, in \aref{formulas} we collect the expressions for other observables used in  the present analysis: 
the ratio between longitudinal polarization of $\beta$ particles from Fermi and Gamow-Teller transitions 
and the correlation coefficients $a$ and $A$ in pure Gamow-Teller decays.

\section{Experimental data}
\label{sec:experiment}

\subsection{Mirror beta decays} 
The ${\cal F} t$ values have been extracted quite precisely for numerous mirror transitions from the measured lifetimes, branching ratios and $Q$-values and calculated $\delta_i$ corrections~\cite{Severijns:2008ep}. 
However, those alone do not provide any information about the fundamental SM parameters or new physics contributions, because the decay rates also depend on the mixing ratio $\tilde \rho_i$ (see~\eref{TH_gamma} and~\eref{TH_x}) which cannot be currently calculated from first principles with high precision. 
But when both ${\cal F} t$ and some correlation coefficient are determined for a given transition, then $V_{ud}$ and the  BSM coefficients can be probed. 
The dependence of observables on the Wilson coefficients $C_X^\pm$ is a function of the transition-dependent $\tilde\rho_i$ and nuclear spin $J$,  thus each mirror transition probes a distinct combination of  $C_X^\pm$.
This results in a strong interplay between different transitions, 
especially regarding the constraints on $C_X^\pm$ in the presence of new physics.  

There are currently six mirror transitions for which precise spectroscopic measurements exist to extract the ${\cal F} t$ values as well as measurements of a correlation coefficient.
These are summarized in \tref{mirror} and they are a crucial experimental input for the fits carried out in this work.

Since some of the measurements have a precision at the level of $1\%$ or below, linear recoil effects (mainly weak magnetism) have to be considered~\cite{Holstein:1974zf}. We follow the same prescription as in Ref.~\cite{NaviliatCuncic:2008xt} to take into account its contribution to the asymmetries in ${}^{19}$Ne, ${}^{21}$Na, and ${}^{35}$Ar. We use updated values for the nuclear magnetic moments from Ref.~\cite{IAEAdatabase}, though these changes do not have any impact in the analysis. For ${}^{37}$K, we do not include recoil effects because they were already subtracted from the value of $A$ in Ref.~\cite{Fenker:2017rcx}. Finally for ${}^{17}$F and ${}^{29}$P we neglect all subleading contributions because the experimental uncertainty is much larger for these transitions.

Other changes with respect to the experimental input used in the SM analyses carried out in Refs.~\cite{Severijns:2008ep,NaviliatCuncic:2008xt} are the following:
\begin{itemize}
    \item
    To allow for non-standard interactions, we reinterpret the extractions of $A$ and $B$ through the usual tilde prescription, \eref{tildeprescription}. The $a$ extraction in Ref.~\cite{Vetter:2008zz} is more complicated. In principle the tilde prescription is not valid because the correlation is not extracted from an asymmetry measurement  and thus the data should have been analyzed using two independent free parameters $a$ and $b$~\cite{Gonzalez-Alonso:2016jzm}. However, we have checked that, under the conditions of this measurement, the time of flight distribution is mainly sensitive to a specific combination of $a$ and $b$ that happens to be well approximated by $\tilde{a}$.
    \item The $f_A/f_V$ values are taken from Refs.~\cite{Hayen:2020cxh,Hayen:private}, where a small double-counting affecting previous values was pointed out and resolved. These corrections are all below $0.15\%$. Their uncertainties are very small and can be neglected  in our analysis.
    \item
    The analysis in Ref.~\cite{NaviliatCuncic:2008xt} included the measurement of the neutrino asymmetry ${B}_\nu$ in $^{37}$K decay, which has a relative uncertainty of 3.1\%~\cite{Melconian:2007zz}. We include here  as well the recent precise measurement of $\tilde A$ from Ref.~\cite{Fenker:2017rcx}. This $\beta$-asymmetry parameter has the smallest relative uncertainty from all mirror transitions.
    \item We include the recent result of Ref.~\cite{Combs:2020ttz} for the beta asymmetry in $^{19}$Ne.
    \item The results from Ref.~\cite{NaviliatCuncic:2008xt} have motivated several measurements of branching ratios, lifetimes and $Q$-values for the extraction of ${\cal F} t$ values~\cite{Shidling:2014ura,Rebeiro:2018lwo,Karthein:2019bss}. The review of spectroscopic data is not within the scope of the present analysis. 
    The ${\cal F} t$ values adopted here are indicated in \tref{mirror}.
    \item For completeness, for all these $\beta^+$ transitions, we update the maximal total energies $\Delta = Q_{\rm EC} - m_e$, using the transition energies $Q_{\rm EC}$ given in the 2016 Atomic Mass Evaluation (AME) Database~\cite{Wang_2017}, except for $^{21}$Na  for which we use a newer  measurement~\cite{Karthein:2019bss}. 
\end{itemize}

\begin{table}[tb]
\bc
\setlength{\tabcolsep}{3pt}
\setlength{\extrarowheight}{7pt}
\begin{tabular}{ccrcrrr}
\hline\hline
Parent & Spin 
& $\Delta$~[MeV]~~~
& $\langle m_e/E_e \rangle$
& $f_A/f_V$~~
& ${\cal F}t$ [s]~~~~~ 
& Correlation~~~~~~  
 \\  \hline
${}^{17}$F  & 5/2 
&  2.24947(25) 
& 0.447 & 1.0007(1)  & 2292.4(2.7)~\cite{Brodeur:2016spm}
&  $\tilde A=0.960(82)$~\cite{PhysRevLett.63.1050,Severijns:2006dr} 
 \\
${}^{19}$Ne  & 1/2 
& 2.72849(16)
& 0.386 & 1.0012(2) & 1721.44(92)~\cite{Rebeiro:2018lwo}
  &  $\tilde{A}_{0} =-0.0391(14)$~\cite{Calaprice:1975zz}  
  \\
  & & & & & &$\tilde A_0 = -0.03871(91)$~\cite{Combs:2020ttz}
   \\
${}^{21}$Na & 3/2 
&  3.035920(18) 
& 0.355 & 1.0019(4)   &  4071(4)~\cite{Karthein:2019bss} 
&  $\tilde a=0.5502(60)$~\cite{Vetter:2008zz}   
       \\
${}^{29}$P & 1/2 
& 4.4312(4) 
& 0.258 & 0.9992(1) & 4764.6(7.9)~\cite{Long:2020lby}
  &  $\tilde A=0.681(86)$~\cite{Masson:1990zz}
   \\
${}^{35}$Ar & 3/2 
& 5.4552(7)
&  0.215  & 0.9930(14)  & 5688.6(7.2)~\cite{Severijns:2008ep}     
& $\tilde A=0.430(22)$~\cite{Garnett:1987gw,Converse:1993ba,NaviliatCuncic:2008xt} 
\\
${}^{37}$K & 3/2 
&  5.63647(23)
&  0.209 & 0.9957(9)    & 4605.4(8.2) \cite{Shidling:2014ura}

& $\tilde A =-0.5707(19)$~\cite{Fenker:2017rcx}  \\
&&&&&&     
$\tilde B =  -0.755(24)$~\cite{Melconian:2007zz}
     \\  
\hline\hline
\end{tabular} 
\ec 
\caption{\label{tab:mirror}
Mirror beta decays used in this analysis. 
The quantity $\langle m_e/E_e \rangle$ is calculated via \eref{TH_meOverEe}, using  the endpoint energy
listed in the table.
The latter are taken from AME2016~\cite{Wang_2017}, except that of $^{21}$Na~\cite{Karthein:2019bss}.
The values of $f_A/f_V$ come from
Ref.~\cite{Hayen:2019nic,Hayen:private}.
We also used the notation $\tilde A_{0} \equiv \tilde A(m_e)$.}
\end{table}

The measurement of the {\it total} $\beta$-asymmetry (i.e. the asymmetry integrated over the energy of the beta particle) only gives us access to $\tilde A$. 
However, it is clear that measuring the energy dependence of the $\beta$-asymmetry makes possible to extract separately $A$ and the Fierz term $b$, {\it cf.}~\eref{tildeprescription}.
We encourage experimental groups to carry out such analyses in order to extract all the information contained in the data. Such measurements of the $\beta$-asymmetry as a function of the energy have already been performed, see e.g. Refs.~\cite{Fenker:2017rcx,Combs:2020ttz}, but not analyzed with a two-parameter fit.

\subsection{Fermi, Gamow-Teller and neutron decays}
\label{sec:nonmirrordata}
For pure Fermi, pure GT, and neutron decay, we use the same data set included in the global fit of Ref.~\cite{Gonzalez-Alonso:2018omy} (total rates and asymmetries) with some updates that we explain in this section. 
The complete list of observables and references is collected in~\aref{tables}.

The measurement of the $\beta$-asymmetry in neutron decay by the PERKEO-III collaboration~\cite{Markisch:2018ndu} represents a major change, not only because it is the most precise to date, but also because after its inclusion in the global data set and using the PDG criteria for averaging various measurements~\cite{Zyla:2020zbs}, the scale factor $S$ inflating the error has decreased considerably.
The numerical change is very significant:
\begin{eqnarray}
    \tilde{A}_n &=& -0.11869(99)~\text{($S=2.6$, pre PERKEO-III)}~,
    \\
    \tilde{A}_n &=& -0.11958(21)~\text{($S=1.2$, post PERKEO-III)}~.
\end{eqnarray}
We also include the aSPECT'19 measurement, $a_n=-0.10430(84)$~\cite{Beck:2019xye}. The new average of $a_n$ is
\bea
a_n = -0.10426(82)~,
\eea
which is a significant improvement compared with the previous average, $a_n = -0.1034(37)$ ~\cite{Gonzalez-Alonso:2018omy}.
In contrast to the situation in $^{21}\rm{Na}$ decay mentioned above, we have found that for neutron decay, the tilde prescription does not work in practice for the extraction of $a$ (as a numerical approximation) since the correlation between $a$ and $b$ from the analysis of the energy spectrum of the proton is strongly reduced. The same issue affects also the $a_n$ extractions carried out in Refs.~\cite{Stratowa:1978gq,Byrne:2002tx}. Since their of other observables included in the fits, we find that the inclusion of these values of $a_n$ have a negligible impact in the parameters of the fits. For this same reason, including a 1-parameter fit extraction of $a_n$ (instead of 2-parameter one) will not have practical consequences in the results. We stress that this is not the case in fits with less observables, and we encourage once again experimental groups to analyze the data including the Fierz term $b$, especially once higher precision is reached.

Apart from these changes in the experimental input, this analysis also takes into account important changes concerning theory input, which we discuss in the rest of this section. 
As mentioned in~\sref{Lagrangian}, we take into account the most recent calculations and averages of the nucleon charges $g_{A,S,T}$~\cite{Aoki:2019cca,Chang:2018uxx,Gupta:2018qil}, as well as the new calculations of the inner radiative corrections $\Delta_R^{V,A}$~\cite{Seng:2018yzq,Czarnecki:2019mwq,Seng:2020wjq,Hayen:2020cxh}. 
In addition to these developments at the nucleon level, Refs.~\cite{Seng:2018qru,Gorchtein:2018fxl} studied the $\gamma$-$W$ box correction in nuclei (which contributes to $\delta_{NS}$ in the usual notation, {\it cf.}~\sref{Ftvalues}) using a free Fermi-gas model. 
This effect is taken into account in the central values of ${\cal F} t$ provided in the most recent evaluation by Hardy and Towner~\cite{Hardy:2020qwl},  which we show in~\tref{superallowed}.
However, the errors displayed in~\tref{superallowed} do not include the theoretical uncertainty associated with the corrections of  Refs.~\cite{Seng:2018qru,Gorchtein:2018fxl} or $\delta'_R$, because they are strongly correlated between different transitions. To take them into account, we modify the ${\cal F}t_i^{HT}$ values in~\tref{superallowed} as follows:
\begin{equation}
    \label{eq:Seng2correction}
    {\cal F}t_i =  {\cal F}t_i^{HT} 
 \big (1 + 
 \eta_1\, \Delta\delta^{\prime \,i}_R
 + \eta_2\, \Delta\delta_{NS,A}
 + \eta_3\, \Delta\delta_{NS,E}^{i} \big )
\end{equation}
where $\eta_{1,2,3}$ are three independent nuisance parameters with zero central value and the $1\sigma$ confidence interval being $[-1,1]$, 
$\Delta\delta^{\prime\,i}_R=[\delta'_R]^i_{Z^2\alpha^3}/3$~\cite{Towner:2007np,Hardy:2014qxa,Gonzalez-Alonso:2018omy}, 
$\Delta\delta_{NS,A} = 0.00033$~\cite{Hardy:2020qwl}, 
and $\Delta\delta_{NS,E}^{i}=8\times 10^{-5}Q_{EC}/\rm{MeV}$~\cite{Gorchtein:2018fxl}.
This procedure makes it possible to understand the implications of these additional uncertainties not only in the SM case but also in the presence of a scalar current ($b\neq 0$).

The information contained in the 15 superallowed transitions can be conveniently encoded in 2 parameters if we write~\eref{TH_ft} as ${\cal F}t_i = {\cal F}t_0 (1 + b \langle m_e/E_e \rangle )^{-1}$. In~\tref{RC} we show the values of ${\cal F}t_0$ and $b$ that are obtained from the data using different inputs for the nuclear-structure dependent corrections. From this table we conclude that:
\bi
\item In the SM limit we reproduce the results of~\cite{Seng:2018qru,Gorchtein:2018fxl,Hardy:2020qwl}. The very small differences are not surprising, since there are minor differences between our approaches such as the inclusion of the $\delta'_R$ uncertainty and the most recent ${\cal F}t$ values~\cite{Hardy:2020qwl}.
The slightly smaller uncertainty of ${\cal F}t$ compared to Ref.~\cite{Hardy:2020qwl} is the consequence of the fact that we treat the effects of Ref.~\cite{Seng:2018qru} and Ref.~\cite{Gorchtein:2018fxl} as uncorrelated (i.e. $\eta_2$ and $\eta_3$ in \eref{Seng2correction} are independent nuisance parameters).
\item The impact of the new calculations is significant in the value of ${\cal F}t_0$ obtained for $b=0$,
and hence in the extraction of $V_{ud}$ in the SM limit  (\fref{VudSM} right panel). 
\item On the other hand, the impact on the bound obtained on the Fierz term is much smaller. Thus, we conclude that such bound is quite robust with respect to these corrections.
\item Although the results in Refs.~\cite{Seng:2018qru,Gorchtein:2018fxl} should be taken with caution because they are obtained using a free Fermi-gas model, their main effect (with respect to the previously standard approach~\cite{Seng:2018yzq}) is to increase the uncertainties, so one can consider their inclusion as a conservative approach.
\item The value of the $\chi^2$ function at the minimum is significantly lower than the degrees of freedom, namely $\chi_{min}^2$/dof$ \sim 0.5$. This is worth keeping in mind and investigating further, since it could be reflecting overestimated uncertainties or mis-calculated central values. We note that this low $\chi_{min}^2$/dof value (i) is not affected by the correlated uncertainties discussed above; and (ii) has been obtained in every survey since the 2002 evaluation of the nuclear-structure-dependent corrections were introduced~\cite{Towner:2002rg}.
\ei

\begin{table}[tb]
\centering
\begin{tabular}{lccc}
\hline\hline
&   Pre-2018 $\delta_{NS}$~\cite{Towner:1994mw,Towner:1992xm} &    
        Seng et al.~\cite{Seng:2018qru}  &   Gorchtein~\cite{Seng:2018qru,Gorchtein:2018fxl} \\
\hline
${\cal F}t_0$ [s] & $3071.7(1.9)$& $3070.1(2.1)$ & $3072.7(3.1)$ \\
$b\times 10^3$    & $-1.0(2.1)$ &   $-1.0(2.1)$ &  $0.4(2.5)$ \\
$\rho({\cal F}t_0,b)$  &  0.93  & 0.83 &  0.86 \\
\hline
${\cal F}t_0$ for $b\!=\!0$ [s]    & $3072.54(0.68)$ & $3071.0(1.2)$ & $3072.3(1.6)$\\
\hline\hline
\end{tabular}
\caption{Results of global fits to superallowed $0^+\to 0^+$ transitions using different nuclear-structure dependent corrections. The 2nd column uses standard pre-2018 values~\cite{Towner:1994mw,Towner:1992xm}. The 3rd column includes the correction pointed out in Seng et al.~\cite{Seng:2018qru}, and the 4th column adds Gorchtein~\cite{Gorchtein:2018fxl} as well. We find $\chi_{min}^2$/dof$\sim 0.5$ for all cases. 
$\rho$ gives the correlation between the two free parameters of the fit. 
The last row shows the result in the absence of a Fierz term $b$.}
\label{tab:RC}
\end{table}

\section{Numerical analysis}
\label{sec:fits}

\subsection{Standard Model scenario}
\label{sec:SMscenario}

If the Lee-Yang Lagrangian in \eref{TH_Lleeyang} is derived as the low-energy EFT for the SM, 
then the only non-zero Wilson coefficients at the leading order are 
$C_V^+$ and $C_A^+$. 
These parameterize the vector and axial interactions, 
descending from the $V$-$A$ quark-level 4-fermion terms predicted by the SM. 
The remaining Wilson coefficients: $C_{S,T,P}^+$ and all $C_{X}^-$ are set to zero in this subsection. 
We will refer to this set of assumptions as the {\em SM scenario}. 

A global fit to all beta decay data discussed in~\sref{experiment} (i.e. mirror and non-mirror data) gives the following $1\sigma$ confidence intervals:\footnote{%
For ${}^{19}$Na, since both $a$ and ${\cal F}$ depend on $\rho^2_{\rm{Na}}$, the sign of the mixing parameter is not fixed by the data. We use an input from shell model calculations to select the positive sign~\cite{Severijns:2008ep}.}
\beq
\label{eq:RES_CVArhoSM}
\bvec 
 v^2 C_V^+ \\ 
 v^2 C_A^+  \\ 
\rho_{\rm F} \\ 
\rho_{\rm Ne} \\ 
\rho_{\rm Na} \\
\rho_{\rm P} \\
\rho_{\rm Ar} \\
\rho_{\rm K} \\   
\evec 
= \bvec
0.98564(23)  \\ 
-1.25700(44) \\ 
-1.2958(13) \\
1.60183(76) \\ 
-0.7129(11) \\ 
-0.5383(21) \\ 
-0.2838(25) \\ 
0.5789(20) 
\evec , 
\eeq 
with $\chi^2$/dof=0.8 at the minimum and with the correlation matrix 
\beq \rho  = 
 \left(
\begin{array}{cccccccc}
 1. & -0.34 & 0.37 & -0.64 & 0.42 & 0.27 & 0.34 & -0.24 \\
 -0.34 & 1. & -0.12 & 0.22 & -0.14 & -0.09 & -0.11 & 0.08 \\
 0.37 & -0.12 & 1. & -0.24 & 0.15 & 0.1 & 0.12 & -0.09 \\
 -0.64 & 0.22 & -0.24 & 1. & -0.27 & -0.17 & -0.22 & 0.15 \\
 0.42 & -0.14 & 0.15 & -0.27 & 1. & 0.11 & 0.14 & -0.1 \\
 0.27 & -0.09 & 0.1 & -0.17 & 0.11 & 1. & 0.09 & -0.06 \\
 0.34 & -0.11 & 0.12 & -0.22 & 0.14 & 0.09 & 1. & -0.08 \\
 -0.24 & 0.08 & -0.09 & 0.15 & -0.1 & -0.06 & -0.08 & 1. \\
\end{array}
\right). 
\eeq 
Since the Wilson coefficients in the Lee-Yang Lagrangian are dimensionful, it is more transparent to display the corresponding confidence intervals in units of $\sqrt{2}G_F=1/v^2$, where $v = 246.219651(63)$~GeV~\cite{Zyla:2020zbs}.  
It is remarkable that, within the SM scenario,
both  $C_V^+$ and $C_A^+$ are  independently measured with an accuracy approaching $\cO(10^{-4})$. 

In addition to the results shown in~\eref{RES_CVArhoSM}, there is another global minimum where $C_{V,A}^+$ have the opposite sign. We work in the convention where $V_{ud}>0$, and thus such solution with $C_V^+ < 0$ is only possible in the presence of very large and very fine-tuned new physics contributions, cf. \eref{TH_LYtoRWEFT}. 
We ignore this ``shadow'' minimum in this section, as well as in the subsequent BSM fits (where the same argument holds flipping the signs of all Wilson coefficients, $C_X^\pm \to - C_X^\pm$).

Although the global quality of the fit is good ($\chi_{\rm{min}}^2$/dof=27/35), its breakdown in the various datasets reveals a quite heterogeneous situation. Superallowed decays, mirror transitions and the remaining nuclear input have very low $\chi^2$/dof values (7/13, 2/6 and 3/8, respectively), whereas the neutron dataset has a large value (16/4).

Using the dictionary in \eref{TH_LYtoRWEFT}, the Wilson coefficients $C_V^+$ and $C_{A}^+$ can be related to more fundamental parameters. In the SM scenario, from $C_V^+$ and $C_A^+$ one can extract the CKM element $V_{ud}$, and the axial coupling of the nucleon $g_A$. 
The confidence intervals for $C_{V,A}^+$ in \eref{RES_CVArhoSM} translate to 
\beq
\label{eq:RES_vudga}
V_{ud} = 0.97370(25), \qquad  g_A^{exp} = 1.27528(45) , 
\eeq
with the correlation coefficient $-0.29$.  
To ease the comparison with previous extractions we display the result for the quantity $g_A^{exp} \equiv g_A \left(1+ (\Delta_R^A -\Delta_R^V)/2\right)$, which is simply denoted $g_A$ in most of the past literature.\footnote{The PDG determination, $g_A=1.2756(13)$~\cite{Zyla:2020zbs}, is significantly less precise mainly because it does not use the neutron lifetime, which is the most sensitive observable.} 
Using the recent determination 
$\Delta_R^A -\Delta_R^V= 0.00060(5)$~\cite{Hayen:2020cxh}
we obtain\footnote{In the published version of this paper we used 
$\Delta_R^A -\Delta_R^V= 0.00407(8)$ quoted in versions 1-3 of Ref.~\cite{Hayen:2020cxh}.}
\beq
g_A = 1.27491(45) .  
\eeq 
These results for $g_A^{exp}$ and $g_A$ are the most precise obtained so far. This is, to a large extent, due to the inclusion of the recent accurate measurement of the $\beta$-asymmetry of the neutron by the PERKEO-III experiment~\cite{Markisch:2018ndu}. 
On the other hand, the values of $V_{ud}$ are {\em less precise} than those  presented in most of the previous  surveys, see e.g. \cite{Hardy:2018zsb}. 
This is so because we take into account the new sources of uncertainty in superallowed transitions discussed in Refs.~\cite{Seng:2018qru,Gorchtein:2018fxl}, {\it cf}.~\sref{nonmirrordata}.   
We remark that the $V_{ud}$ value depends on the inner radiative correction $\Delta_R^V$, {\it cf.}~\eref{TH_LYtoRWEFT}.
The results in~\eref{RES_vudga} are obtained using  
the Seng et al. evaluation:  
$\Delta_R^V = 0.02467(22)$~\cite{Seng:2018yzq}.
Using instead the Czarnecki et al. evaluation 
$\Delta_R^{V} = 0.02426(32)$~\cite{Czarnecki:2019mwq} one finds  
$V_{ud} = 0.97390(27)$, which has a similar error as the result in \eref{RES_vudga}, while the two central values differ by less than one standard deviation. 
The dependence of $V_{ud}$ on the different choices of radiative corrections is illustrated in~\fref{VudSM} (right panel), which also shows the CKM-unitarity $V_{ud}$ value obtained using current PDG values $V_{us}=0.2245(8)$ (S=2.0) and $V_{ub}=0.00382(24)$ (S=1.6)~\cite{Zyla:2020zbs}.
\begin{table}[tb]
\bc
\begin{tabular}{cccccc}
\hline \hline
Parameter & Mirror & Superallowed & Neutron & All but mirror & Global
\\ \hline
$V_{ud}$ & 
0.97424(95) & 0.97367(28)  & 0.97368(56) &  0.97367(25) & 
0.97370(25)
\\ 
$g_A^{exp}$ &
$-$ & $-$ &  1.27530(55) & 1.27531(45) &
1.27529(45)
\\ 
$\rho_{\rm Ne}$ & 
1.6007(21)  & $-$ & $-$ & $-$ &  
1.60183(76) 
\\  \hline\hline
\end{tabular}
\ec
\caption{
\label{tab:RES_sm_comparison}
Values for the CKM element $V_{ud}$,
the axial charge of the nucleon $g_A^{exp}$, 
and the mixing ratio of ${}^{19}$Ne, obtained from fits
assuming the SM as the fundamental theory. The columns indicate the different subsets of data used in the fits. The errors of $g_A^{exp}$ and $V_{ud}$ in the neutron column are the nominal errors (i.e. not inflated a posteriori), so that they reflect the weight that the neutron dataset has in the global fit, where no $S$ factor is needed since $\chi^2_{\rm{min}}/\rm{dof}=0.8$. The same holds for~\fref{VudSM}. 
Let us note for completeness that pion semileptonic decay gives $V_{ud}=0.9740(28)$~\cite{Feng:2020zdc,Pocanic:2003pf}.}
\end{table}

In~\fref{VudSM} (left panel) and~\tref{RES_sm_comparison} we show the confidence intervals for $V_{ud}$ and $g_A$ obtained  using various subsets of  the nuclear and neutron data. 
It is clear that the sensitivity to $V_{ud}$ is dominated by the superallowed decays, while $g_A$ is dominated by the neutron data. 
The impact of the mirror decays on these bounds is negligible in the SM scenario, although they represent valuable and nontrivial inputs that improve the robustness of the $V_{ud}$ extraction from beta decay, given they are sensitive to very different systematics. The error on $V_{ud}$ determined from the mirror data alone is currently 3.4 times larger than that determined from the superallowed data. 
The former error has decreased by a factor of two since the pioneering work of Ref.~\cite{NaviliatCuncic:2008xt}.

Another tangible effect of the mirror data is the very precise extraction of the corresponding mixing ratios $\rho$ in \eref{RES_CVArhoSM}.
These do not probe fundamental parameters, 
but may be nevertheless interesting from the point of view of nuclear theory. 
In this regard, we note that the errors on $\rho$ are a factor of $2$-$4$ smaller in the global fit,  
compared to the determination based on the mirror data only (due to significant correlations in the only-mirrors fit).
This is illustrated in \tref{RES_sm_comparison} by the comparison of the mixing ratio of ${}^{19}$Ne determined from the mirror and global data. 
Finally we note that the mixing ratios $\rho$ in \eref{RES_CVArhoSM} are significantly more precise than those obtained in Ref.~\cite{Severijns:2008ep}, mainly because of the improvement in the ${\cal F}t$ values and the $f_A/f_V$ factors. Some of these inputs have shifted from their former values, which is reflected in shifts in the extracted mixing ratios with respect to Ref.~\cite{Severijns:2008ep}. 

\begin{table}[tb]
\bc
\begin{tabular}{ccccccc}
\hline\hline
 & ${}^{17}$F & ${}^{19}$Ne &  ${}^{21}$Na 
 &  ${}^{29}$P &  ${}^{35}$Ar &  ${}^{37}$K
\\  \hline
$V_{ud}$ 
& 0.83(22) 
& 0.9742(11) 
& 0.9734(33) 
&  0.951(43)   
& 0.9749(39) 
& 0.9750(26) 
\\  \hline\hline
\end{tabular}
\ec
\caption{
\label{tab:RES_sm_compareMirror}
Values of $V_{ud}$ obtained from each individual mirror transition.}
\end{table}

In \tref{RES_sm_compareMirror} we also compare the sensitivity to $V_{ud}$ of various mirror transitions taken on its own. 
Currently, the determination using ${}^{19}$Ne is the most accurate one, dominating the mirror-decays extraction, followed by those based on ${}^{37}$K, ${}^{21}$Na, and ${}^{35}$Ar. 
For ${}^{29}$P, the weaker sensitivity is due to a relatively large error in the correlation measurement. 
The sensitivity of ${}^{17}$F looks abysmal, despite the fact that the correlation measurement has a similar error as that of  ${}^{29}$P.   
For $^{17}$F decay the experimentally measured   $\beta$-asymmetry is near the maximum of $A$ regarded as a function of $\rho_{F}$, which leads to a large uncertainty on $\rho_{F}$ when only the input from ${}^{17}$F is used. 
For this reason the ${}^{17}$F transition is rarely used in this context.  
However, for the sake of new physics searches the sensitivity of  ${}^{17}$F  and  ${}^{29}$P will be comparable, 
therefore we keep the former input in the analysis.

\begin{figure}[tb]
\centering
\includegraphics[width=0.45\textwidth]{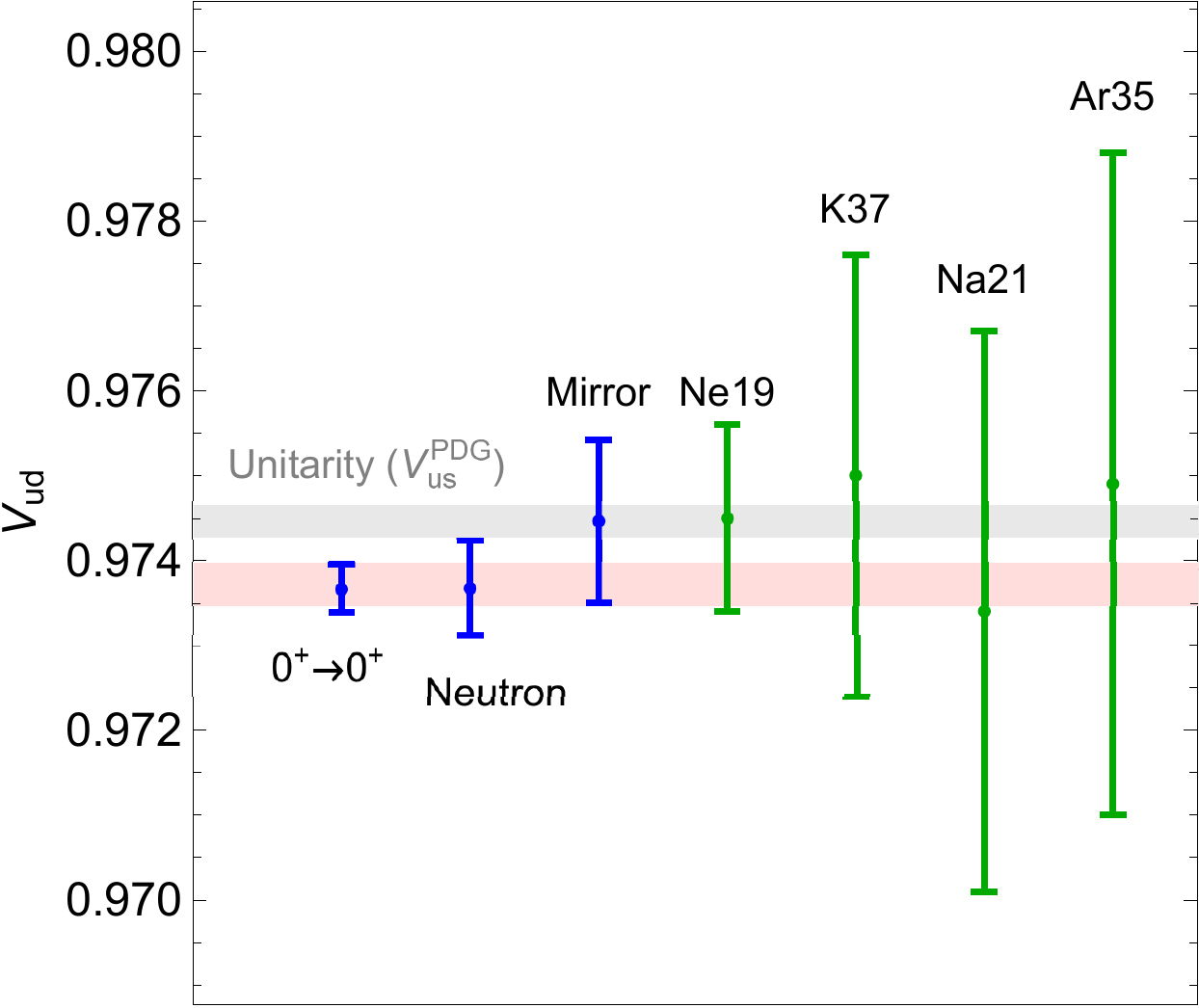}
\quad 
\includegraphics[width=0.45\textwidth]{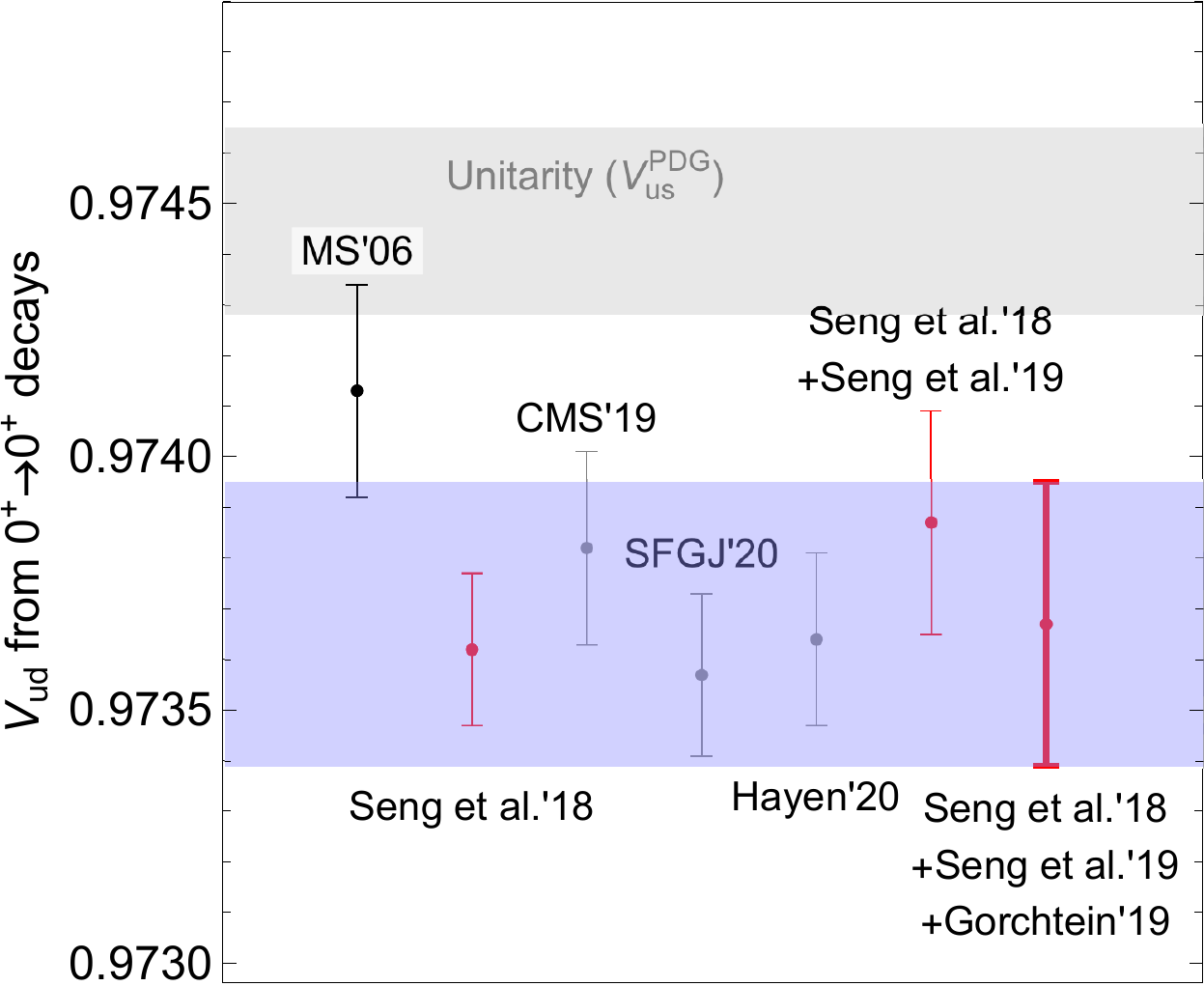}
\caption{\label{fig:VudSM}
{\em Left:} 
comparison between the $1\sigma$ confidence intervals for $V_{ud}$ extracted from various data sets, assuming the SM scenario.  
The blue error bars show the determination based on the superallowed, neutron, and mirror data. 
The green error bars display the four most precise determinations from individual mirror transitions.  
The salmon band corresponds to the value in \eref{RES_vudga} using all data included in this analysis.
{\em Right:} 
comparison between the $V_{ud}$ value obtained in this work from superallowed data (violet band) and alternative extractions (black error bars) using different values of the inner radiative correction $\Delta_R^V$ and of the nuclear-structure dependent corrections $\delta_{NS}$.  
The first four error bars correspond to
$\Delta_R^V = 0.02361(38)$~\cite{Marciano:2005ec}~(MS'06), 
$\Delta_R^V =0.02467(22)$~\cite{Seng:2018yzq}~(Seng et al.'18), $\Delta_R^V = 0.02426(32)$~\cite{Czarnecki:2019mwq} (CMS'19), 
$\Delta_R^V = 0.02477(24)$~\cite{Seng:2020wjq} (SFGJ'20), and $\Delta_R^V=0.02473(27)$~\cite{Hayen:2020cxh} (Hayen'20),
and they do  not take into account the nuclear-structure dependent corrections  pointed out in Seng et al.'19~\cite{Seng:2018qru} and Gorchtein'19~\cite{Gorchtein:2018fxl} ({\it cf.}~\sref{nonmirrordata}), which are included in the last two points. 
The $V_{ud}$ value obtained from CKM unitarity using current PDG values $V_{us}=0.2245(8)$ 
and $V_{ub}=0.00382(24)$
~\cite{Zyla:2020zbs} is shown in both panels (gray bands).
}
\end{figure}

\subsection{Non-standard interactions involving left-handed neutrinos}
\label{sec:WEFTfit}

We move to discussing constraints on physics beyond the SM.
Before attacking the general case, 
we first consider the scenario where the only non-zero Wilson coefficients in the Lee-Yang Lagrangian of \eref{TH_Lleeyang} are $C_X^+$, 
$X = V,A,S,T,P$. 
Recall that $C_X^-$ parametrize 4-fermion interactions of proton, neutrons, electrons, and the {\em right-handed} electron neutrino. 
Setting $C_X^- = 0$ thus corresponds to neglecting beta decays into the right-handed neutrino, 
either because this degree of freedom is simply absent in nature, 
or because it acquires a Majorana mass significantly larger than few MeV.
We recall also that the Wilson coefficient $C_P^+$ does not affect $\beta$-decay observables at the leading order in the recoil velocity expansion, 
thus it does not enter the fits. 
All in all, in this subsection we simultaneously fit 4 Wilson coefficients $C_{V,A,S,T}^+$, together with the 6 mixing ratios of mirror nuclei and the 3 nuisance parameters in \eref{Seng2correction}.

In this scenario we find the following $1\sigma$ confidence intervals and the correlation matrix:
\beq
\label{eq:RES_CXp}
v^2 \begin{pmatrix}
C_V^+ \\ C_A^+ \\ C_S^+ \\ C_T^+ 
\end{pmatrix}
=  \begin{pmatrix}
\phantom{-}0.98571(41)  \\ 
-1.25707(55)  \\ 
\phantom{-}0.0001(10) \\ 
\phantom{-}0.0004(12)
\end{pmatrix} , 
\qquad
\rho =\left(
\begin{array}{cccc}
 1. & -0.62 & 0.80 & 0.65 \\
 -0.62 & 1. & -0.49 & -0.56 \\
 0.80 & -0.49 & 1. & 0.59 \\
 0.65 & -0.56 & 0.59 & 1. \\
\end{array}
\right). 
\eeq
These results deserve a number of comments:
\begin{itemize}
\item  Nuclear observables depend on the Wilson coefficients in the Lee-Yang Lagrangian in a non-linear way and thus the likelihood function we constructed is in general non-Gaussian.
Nevertheless,  the central values, $1 \sigma$ errors, and  the correlation matrix displayed in \eref{RES_CXp} fully characterize the likelihood in the region of the parameter space  compatible with the data. 
This is a consequence of two facts. 
One is that the BSM Wilson coefficients $C_{S,T}^+$ interfere with the SM amplitudes, therefore they affect nuclear observables already at the linear level. 
The other is that all the parameters in \eref{RES_CXp} are stringently constrained, at the per-mille level or better. 
These two facts ensure that, near the maximum of the likelihood,  $\chi^2 = - 2 \log L$ can be very well approximated by a quadratic form: 
$\chi^2 \approx  \chi^2_{\rm min} + 
{1 \over 2}(\vec x - \vec x_0) \Delta^{-1} (\vec x - \vec x_0)$, 
where $\vec x$ is a 10-dimensional vector of $C_{X}^+$ and mixing ratios $\rho_i$, 
$\vec x_0$ is its central value, 
and $\Delta$ is the error matrix. 
\item The SM makes two predictions about the Wilson coefficients in \eref{RES_CXp}. 
One is that scalar and tensor currents are absent:  
$C_S^+ = C_T^+ = 0$. 
The other is that $C_V^+$ and $-C_A^+/C_V^+$ are respectively equal to $V_{ud}$ $\leq 1$ and to the axial charge of the nucleon $g_A$ up to small radiative corrections. 
Both of these predictions are in perfect agreement with the fit results in \eref{RES_CXp}. Thus, in this scenario, there is no slightest hint of physics beyond the SM affecting the nuclear observables.
It is remarkable that data require $C_S^+$ and $C_T^+$ to vanish within the per-mille precision.
\item In~\fref{CSvsCT} we show the marginalized bounds on scalar and tensor coefficients using different subsets of data. The plot shows the complementarity of the different subsets, and the strong bounds obtained using only mirror decays. 
The constraints on the scalar currents are dominated by the superallowed data: a non-zero $C_S^+$  would lead to a non-universal shift of the ${\cal F} t$ values, cf.~\eref{TH_ft}. 
We note that there are significant correlations in this fit with ten free parameters that cannot be shown in a 2D plot. These correlations explain, e.g., that the combination of neutron and superallowed data, provides bounds almost as strong as the entire dataset. 
In other words, given the input from the superallowed transitions, tensor currents are strongly constrained by the neutron data, where $C_T^+$ would affect the precisely measured $\tau_n$ and $A_n$. We will come back to this later (see~\tref{RES_weft_comparison}). 
\item The neutron ellipse in~\fref{CSvsCT} shows a mild tension with the other ellipses, but it is not statistically significant in the global fit, where we find $\chi^2_{\rm min}/{\rm dof} <1$. It has its origin in a few measurements (mainly $a_n$ from aSPECT~\cite{Beck:2019xye} and $B_n$) that are in some tension with the most precise ones ($A_n$ and $\tau_n$). We will come back to this in~\sref{globalfit}.
\item Allowing for the possibility of per-mille level $C_{S,T}^+$ contributions to the nuclear observables  somewhat relaxes the constraints on $C_{V,A}^+$, 
compared to the constraints obtained within the SM scenario.  
\item 
In order to obtain \eref{RES_CXp} we fit the four Wilson coefficients $C_X^+$ together with the six ``polluted'' mixing ratios $\tilde \rho_i$ for the mirror transitions used in the analysis, 
cf. \eref{TH_tildeRho}. 
The results for  $\tilde \rho_i$ are not displayed in \eref{RES_CXp} because we do not consider them to be of interest.
\end{itemize} 

\begin{figure}[tb]
\centering
\includegraphics[width=0.6\textwidth]{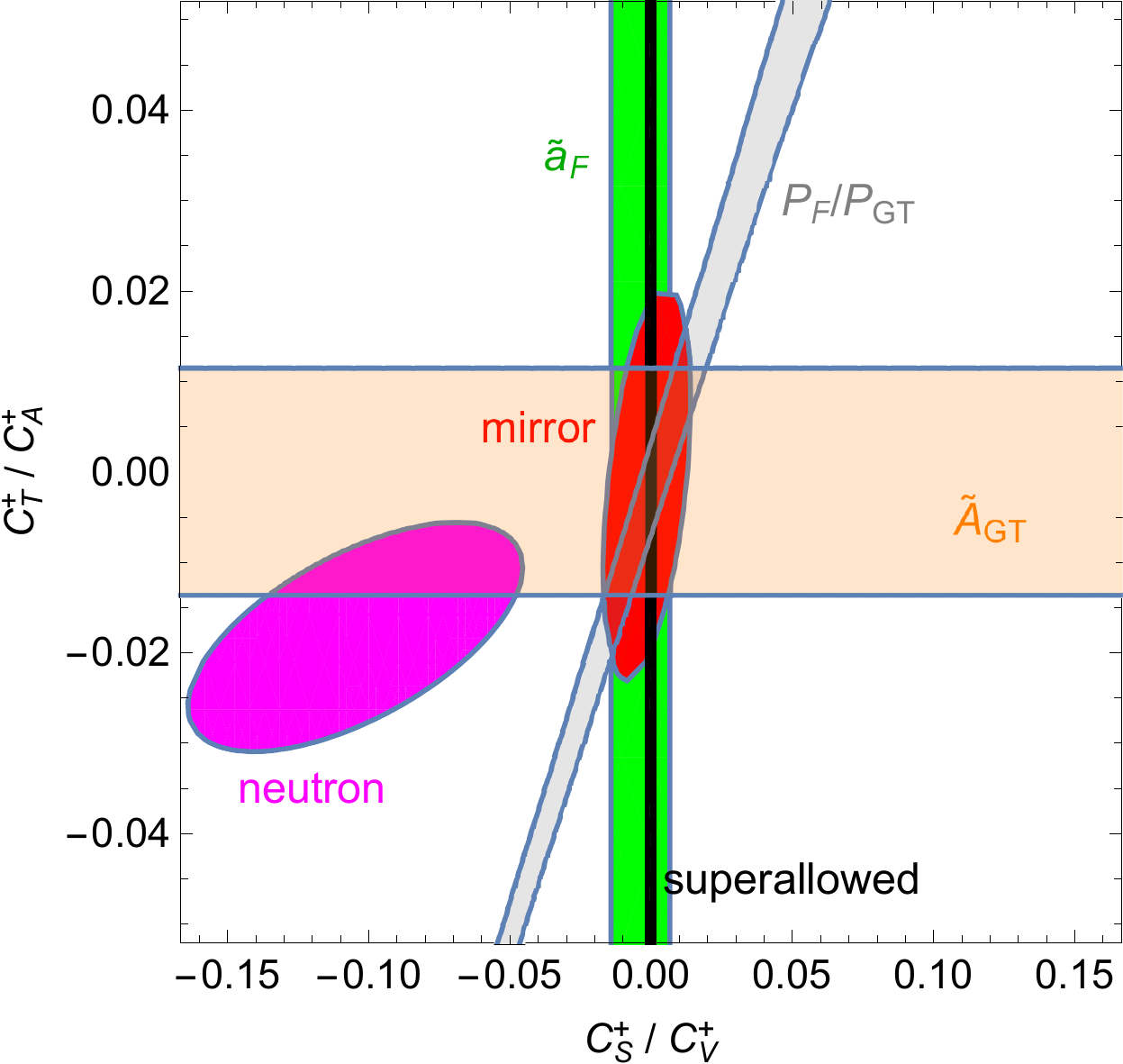}
\caption{\label{fig:CSvsCT}
1$\sigma$ constraints on scalar and tensor coefficients using different subsets of beta decay data. The measurement of $\tilde{a}(^6\rm{He})$ does not exclude any portion of the region shown in this plot. 
See~\aref{tables} for details about the various data sets.}
\end{figure}

It is instructive to translate the results in \eref{RES_CXp} into constraints on the parameters of the quark-level effective Lagrangian in \eref{TH_Lrweft}. 
The assumption $C_X^- = 0$ translates into $\tilde \epsilon_X = 0$ for all $X$, 
so that the free parameters are $V_{ud}$ and $\epsilon_X$, 
$X = L,R,T,S$. 
Given the absence of right-handed neutrinos, 
\eref{TH_Lrweft} is a part of the weak EFT (WEFT) Lagrangian~\cite{Jenkins:2017jig} valid between the hadronic scale and $m_W$.   
At the latter scale it can be matched to another EFT, called the SMEFT, whose degrees of freedom are those of the SM. 
The matching equations are known at one loop~\cite{Dekens:2019ept}, and the anomalous dimensions  describing the running of $\epsilon_X$ between the hadronic and  $m_W$ scales in the WEFT have also been written down~\cite{Gonzalez-Alonso:2017iyc}.  
Using those, the results below can be easily translated into constraints on the Wilson coefficients in the SMEFT.

Since the data require $|\epsilon_X| \ll 1$, we will work to linear order in $\epsilon_X$.
Then the dictionary in \eref{TH_LYtoRWEFT} reduces to 
\bea 
\label{eq:RES_LYtoWEFT}
C_V^+  &=&  {\hat V_{ud} \over  v^2}\sqrt{1 + \Delta_R^V} \,g_V, 
\qquad 
C_A^+ \approx  - {\hat V_{ud} \over v^2} \sqrt{1 + \Delta_R^A} \,g_A \big ( 1 - 2 \epsilon_R \big ),
\nnl 
C_T^+  &\approx &  {\hat V_{ud} \over v^2} g_T \epsilon_T  , 
\qquad \qquad 
C_S^+  \approx  {\hat V_{ud} \over v^2} g_S \epsilon_S , 
\eea 
where we defined  the ``polluted'' CKM element 
$\hat V_{ud} \equiv  V_{ud} \big ( 1+ \epsilon_L + \epsilon_R \big )$. It is important to realize that, using the nuclear data alone, 
it is {\em not} possible to disentangle the true CKM element $V_{ud}$ from the new physics corrections parameterized by $\epsilon_L + \epsilon_R$. 
Indeed, the data independently constrain four Wilson coefficients 
$C_{V,A,S,T}^+$, which however depend on five quark-level parameters  
$V_{ud}$ and $\epsilon_{L,R,S,T}$, leaving one flat direction. 
Note that, for $\epsilon_L + \epsilon_R \neq 0$, 
$\hat V_{ud}$ is {\em not} an element of a unitary matrix, 
and thus it is not tied by the unitarity relation to $V_{us}$ measured in kaon decays.
Conversely, a conclusive proof that $\hat V_{ud}^2 +  V_{us}^2 \neq 1$ would be an evidence for the existence of new physics,  manifesting as  $\epsilon_L + \epsilon_R \neq 0$ in the quark-level effective Lagrangian.\footnote{%
In reality the issue is slightly more complicated, because kaon decays also probe a ``polluted" $\hat V_{us}$ rather than the original CKM element  $V_{us}$. 
Thus, evidence for $\hat V_{ud}^2 +  \hat V_{us}^2 \neq 1$ can be interpreted as new physics in the $ud$ sectors, {\em or} in the $us$ sector, or both~\cite{Cirigliano:2009wk,Gonzalez-Alonso:2016etj}.} 
Furthermore, in the presence of new physics, the nuclear data can no longer disentangle $g_A$ from effects of the new physics parameter $\epsilon_R$, which encodes non-standard $V$+$A$ interactions in the quark-level Lagrangian.
Instead, a lattice determination of  $g_A$ has to be used to disentangle $g_A$ and $\epsilon_R$~\cite{Bhattacharya:2011qm,Gonzalez-Alonso:2016etj}. 
In this analysis we will use the FLAG'19 average $g_A = 1.251\pm 0.033$~\cite{Aoki:2019cca}. 
With this additional lattice input to the dictionary in \eref{RES_LYtoWEFT}, 
the fit in \eref{RES_CXp} translates into\footnote{We stress that $\epsilon_T$ was defined in this work with a different normalization (by a factor of 4) than in previous works, {\it cf.}~\eref{TH_Lrweft}.}
\beq
\begin{pmatrix}
\hat V_{ud} \\ \epsilon_R \\\epsilon_S \\ \epsilon_T
\end{pmatrix}
= \begin{pmatrix}
0.97377(41)  \\ -0.010(13)  \\ 0.0001(10) \\ 0.0005(13)
\end{pmatrix}, 
\qquad \rho = 
\left(
\begin{array}{cccc}
 1. & 0.01 & 0.77 & 0.62 \\
 0.01 & 1. & 0. & 0. \\
 0.77 & 0. & 1. & 0.59 \\
 0.62 & 0. & 0.59 & 1. \\
\end{array}
\right).
\eeq

Per-mille-level constraints on $C_{S,T}^+$ translate into per-mille-level constraints on $\epsilon_{S,T}$ in the quark-level Lagrangian.
On the other hand, for  $\epsilon_{R}$ the constraint is only at the percent level, 
due to the the percent-level accuracy of the lattice determination of $g_A$. 
We note here that if, instead of the FLAG average, we use the CalLat determination $g_A = 1.271 \pm 0.013$~\cite{Chang:2018uxx}
then we find $\epsilon_R = -0.0015(51)$ - an improvement by a factor of 2.5!  
Future improvements of the lattice determination of $g_A$~\cite{Walker-Loud:2019cif} will immediately translate into more stringent constraints on the non-standard $V$+$A$ currents encoded in $\epsilon_R$. 
\begin{table}[tb]
\bc
\begin{tabular}{cccccc}
\hline\hline
Parameter & Mirror & All but mirror & Global
\\  \hline
$\hat V_{ud}$ &  0.9744(36) &  0.97369(46) &
0.97377(41)
\\  
$\epsilon_S$ & -0.002(10) & 0.0000(11) & 
0.0001(10)
\\   
$\epsilon_T$ & 0.002(19) &  0.0002(14) &  
0.0005(13)
\\  \hline\hline
\end{tabular}
\ec
\caption{
\label{tab:RES_weft_comparison}
Comparison of the sensitivity of mirror and other data sets to the parameters of the quark-level Lagrangian in \eref{TH_Lrweft}, 
in the scenario where all $\tilde \epsilon_X = 0$. 
}
\end{table}

\begin{figure}[tb]
\centering
\includegraphics[width=0.6\textwidth]{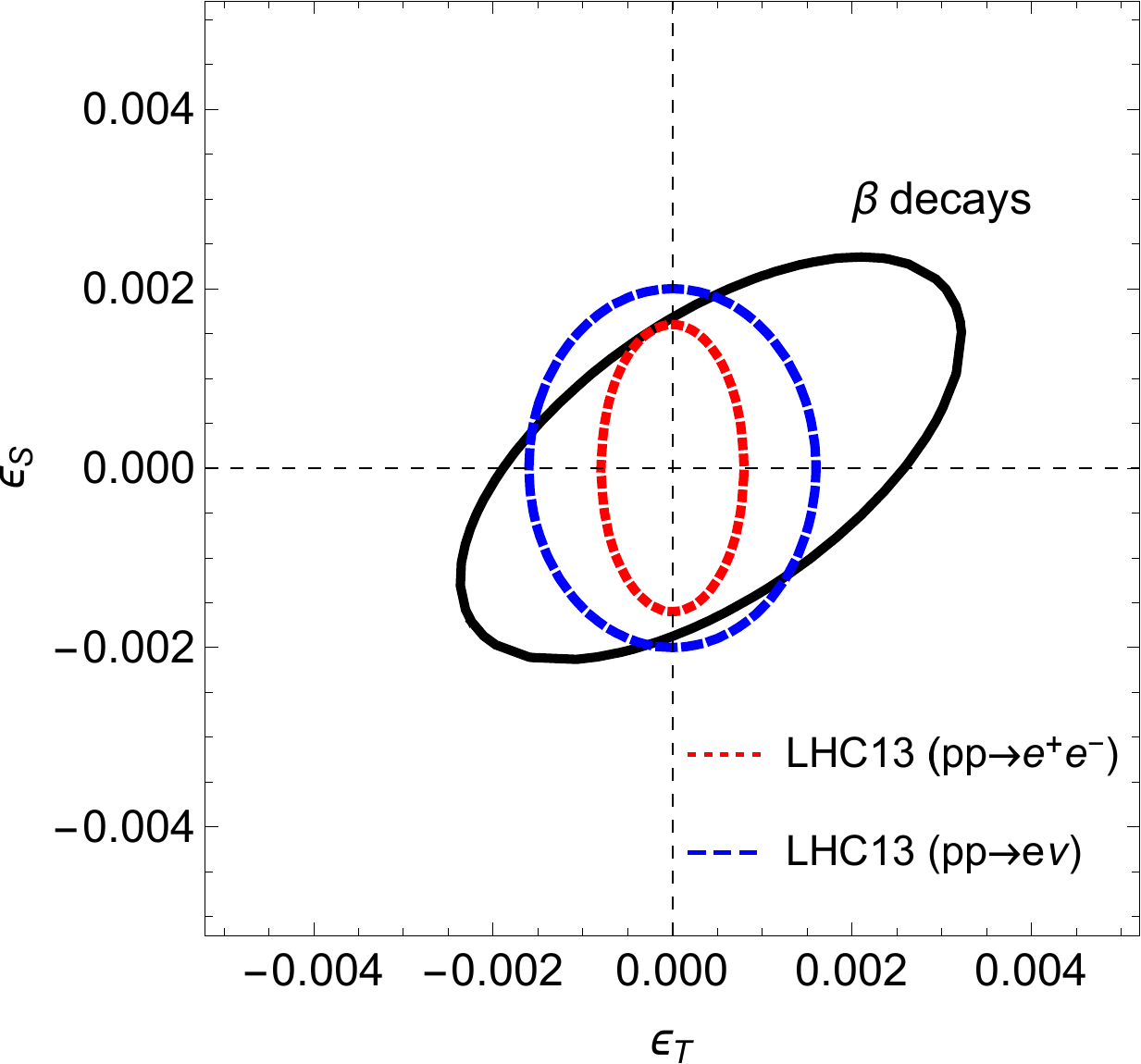}
\caption{\label{fig:LHCcomparison}
90\% CL constraints on scalar and tensor coefficients obtained from beta decays in this work (solid black line) and from LHC data (dashed blue and dotted red lines)~\cite{Gupta:2018qil}. We  stress here again that $\epsilon_T$ is defined in this work with a different normalization (by a factor of 4) than in Ref.~\cite{Gupta:2018qil}. 
}
\end{figure}

Other processes are sensitive to the same effective operators, see Ref.~\cite{Gonzalez-Alonso:2018omy} for a detailed review. 
For instance, assuming the so-called SMEFT as the underlying theory valid at LHC scales, one can relate the Wilson coefficients of dimension-6 SMEFT operators to the low-energy EFT parameters, and translate LHC constraints on the former into constraints on $\epsilon_X$.  
Let us stress that this implicitly involves non-trivial assumptions that new physics is heavier than a few TeV and that dimension-8 and higher SMEFT operators can be neglected.    
Moreover, due to the humongous number of the SMEFT operators, LHC analyses often involve simplifying assumptions that only a small subset of dimension-6 operators is simultaneously present. 
Given these caveats, one can for example set bounds on $\epsilon_{S,T}$ using high-energy $pp\to e\nu$ and $pp\to e^+e^-$ processes~\cite{Cirigliano:2012ab}.
Fig.~\ref{fig:LHCcomparison} shows the comparison of the bounds obtained from beta decays in this work and the latest bounds from LHC data~\cite{Gupta:2018qil} assuming two particular dimension-6 operators present at the TeV scale.
It is remarkable that the beta decay and the LHC constraints are comparable, which indicates that precision measurements of beta decays can effectively probe similarly high scales as the LHC, even though they involve merely MeV energy transfers!

To close this subsection, we come back to the comparison of the constraining power of the mirror transitions and other nuclear observables. 
In \tref{RES_weft_comparison} we compare the $1\sigma$ confidence intervals obtained with and without including the mirror data. 
The mirror data alone, without any other input,  are capable of simultaneously constraining $\hat V_{ud}$,  $\epsilon_{S}$, $\epsilon_{T}$, together with the six relevant mixing ratios $\rho$. 
This shows that the mirror transitions can potentially play an important role in probing new physics beyond the SM,
in addition to measuring the CKM element $V_{ud}$ within the SM scenario.  
However, much as in the SM case,
the impact of the mirror transitions  is  currently limited in the scenario with only left-handed neutrinos. 
As anticipated above, the reason is that 
$\hat V_{ud}$, $\epsilon_{S}$, $\epsilon_{T}$ are already well constrained by a combination of superallowed and neutron data,  without leaving flat directions in the parameter space.  
Compared to the superallowed and neutron data, the uncertainties of correlation measurements in mirror transitions is still too large by a factor of few, therefore mirror data does not improve the constraints in this scenario. 
Still, and much like in the SM scenario, mirror decays improve the robustness of beta decay constraints since they come from different experiments and are subject to different systematics.

\subsection{Non-standard interactions involving left- and right-handed neutrinos}
\label{sec:globalfit}

Finally, we discuss the constraints on the Wilson coefficients of the Lee-Yang Lagrangian in \eref{TH_Lleeyang} when all of them are allowed to be simultaneously present. 
In particular, the Wilson coefficients $C_X^-$, 
which characterize the interaction strength of {\em right-handed} neutrinos, are allowed to be non-zero.    
For the Wilson coefficients we find the 1$\sigma$ confidence intervals
\beq
\label{eq:RES_CXpm}
v^2 \begin{pmatrix}
C_V^+ \\ C_A^+ \\ C_S^+ \\ C_T^+ 
\end{pmatrix}
= \begin{pmatrix}
0.98501^{(+75)}_{(-114)}   \\
-1.2544^{(+14)}_{(-11)}   \\
-0.0007^{(+29)}_{(-14)} \\
-0.0010^{(+33)}_{(-22)} 
\end{pmatrix}, 
\qquad
 \begin{pmatrix}
\hspace{-1.2cm}v^2 |C_V^-|  < 0.053  \\ 
\hspace{-1.2cm} v^2 |C_A^-|  <  0.063   \\
\hspace{-1.2cm}v^2|C_S^-| < 0.050 \\ 
v^2 |C_T^-| \in [0.072,0.099] 
\end{pmatrix}. 
\eeq
Simultaneously with the 8 Wilson coefficients in \eref{RES_CXpm}, we also fit the 6 mixing ratios $\tilde\rho_i$ of the mirror nuclei. 
Thus, we perform a 14-parameter minimization of a highly non-linear likelihood with over 40 distinct experimental inputs. 
To obtain the confidence intervals in \eref{RES_CXpm}, for each Wilson coefficient we construct a one-dimensional likelihood marginalized over the 13 remaining parameters. 
In spite of these technical challenges, we obtain a smooth likelihood for each Wilson coefficients and a stable fit.  
In fact, adding the mirror data improves the stability, even though it necessitates including 6 additional free parameters ($\tilde\rho_i$) in the fit. 
This indicates that the mirror data are vital for lifting degeneracies in the space of the Wilson coefficients $C_X^\pm$. 
The marginalized likelihoods for some of the Wilson coefficients, with and without including the mirror data,  are displayed in \fref{MirrorNoMirror} and \fref{MirrorNoMirror2}. 

 These results deserve also several comments:
\begin{itemize}
\item 
One should keep in mind that the likelihood, being highly non-Gaussian, contains more than one global minimum. First, the confidence intervals displayed in \eref{RES_CXpm} encompass {\em two} degenerate global minima of $\chi^2 = - 2 \log L$, related by $C_X^- \to - C_X^-$~\cite{Boothroyd:1984fz,Severijns:2006dr}. 
In addition, as discussed in~\sref{SMscenario}, we have the ``shadow minima" where $C_V^+<0$, which we ignore.
Finally,  the likelihood contains several shallow local minima with $\cO(1)$ difference in $\chi^2$ compared to the global one, in close vicinity in the parameter space to the global minima. 
These are responsible for the wiggles in the marginalized likelihood for some Wilson coefficients like $C_T^+$, (\fref{MirrorNoMirror} right). 
\item Given that the likelihood is symmetric under  $C_X^- \to - C_X^-$, the marginalized likelihood functions for all $C_X^-$ are always symmetric around zero: $\chi_{\rm marg}^2(C_X^-) = \chi_{\rm marg}^2(-C_X^-)$, as can be seen in \fref{MirrorNoMirror2}. 
\item For the Wilson coefficients $C_X^+$ associated with left-handed neutrinos, the constraints become ${\cal O}(2-3)$ weaker than in the $C_X^+$-only fit (cf. \eref{RES_CXp}).
This is a limited increase of uncertainties, given that we have introduced 4 additional free parameters into the fit!   
The robustness of the constraints is  an evidence of the power of the precision data on beta decays.
\item The results in \eref{RES_CXpm}  provide model-independent constraints on the Wilson coefficients  $C_X^-$ associated with right-handed neutrinos. 
This is the first time such general constraints have been extracted from nuclear observables while keeping all 8 Wilson coefficients $C_X^\pm$ in the fit.
As expected, the constraints on $C_X^-$ are much less stringent than those on $C_X^+$, since the former do no interfere with the SM contributions. 
As a consequence, they enter the nuclear observables only at the quadratic order (in contrast to $C_X^+$, which enter at the linear order).  
\item 
One consequence of the relaxed constraints is that the constraints on the CKM element $V_{ud}$ are also less stringent in the scenario with right-handed neutrinos. 
More precisely, as in the previous scenario in \sref{WEFTfit}, we can only constrain the ``polluted'' matrix element $\hat V_{ud} \equiv V_{ud} (1 + \epsilon_L + \epsilon_R)$.
We find $\hat V_{ud} = 0.97308^{(+75)}_{(-113)}$
to be compared with $V_{ud} = 0.97370(25)$ in the SM, 
and $\hat V_{ud} = 0.97377(41)$ in the BSM scenario with only left-handed neutrinos. 

\item Unlike in the previous two subsections, the likelihood function for the Wilson coefficients is highly non-Gaussian.
As we mentioned above,  $C_X^-$ enter at the quadratic order, and thus the marginalized likelihood for  $C_X^-$ cannot be approximated by a quadratic function (e.g. in a fit with a single $C_X^-$, the $\chi^2$ would be a quartic polynomial in $C_X^-$).
The departure from Gaussianity is clearly visible in \fref{MirrorNoMirror2}. 
As for $C_X^+$, while they enter at the linear order, the asymmetric errors in \eref{RES_CXpm}  demonstrate that the marginalized likelihood for $C_X^+$ cannot be  well approximated by a quadratic function either. 
We remark however that these latter non-Gaussianities are somewhat reduced  due to the inclusion of the mirror decay data in the global fit (\fref{MirrorNoMirror}).
Because of the non-Gaussianities, we do not quote the correlation matrix for this fit. 
Unlike in the Gaussian case, confidence intervals together with the correlation matrix  do {\em not} allow one to reconstruct a non-Gaussian likelihood.
The full likelihood is instead available in the numerical form, as a {\tt Mathematica} code. 
\item For the SM fit, and for the BSM fit with only left-handed neutrinos, the impact of the mirror decay data was found to be negligible. 
The situation changes in the present fit because the parameter space is enlarged to include the $C_X^-$, pertinent to right-handed neutrinos.
Indeed,  including the mirror data shrinks the confidence intervals significantly, typically by an $\cO(2)$ factor, 
as shown in~\tref{MirrorNoMirror}. 
\item 
The results from the fit in \eref{RES_CXpm} show a striking preference for a non-zero value of the new physics parameter $C_T^-$. 
The significance of the anomaly is $3.2~\sigma$, given the $\chi^2$ displayed in the right-panel of \fref{MirrorNoMirror2}.
This tension appears because non-zero values of  $C_T^-$ allow one to improve the fit to the neutron data, especially to eliminate the tension between the aSPECT measurement of $a_n$~\cite{Beck:2019xye} and other inputs. 
Indeed, removing this single measurement from the fit the ``anomaly'' is reduced to $1.8\sigma$.
The mirror data are not an eminent player in this anomaly,  however adding them to the global fit slightly strengthens (by one unit of $\chi^2$) the hint for new physics. 
\end{itemize}

\begin{table}[tb]
\setlength{\tabcolsep}{10pt}
\renewcommand{\arraystretch}{1.5}
    \centering
\begin{tabular}{ccccccccc}
\hline \hline
Parameter & 
$C_V^+$ & $C_A^+$ & $C_S^+$ & $C_T^+$ & 
$C_V^-$ & $C_A^-$ & $C_S^-$ & $C_T^-$ 
\\     \hline  
Improvement factor & 
2.8 & 2.8 & 1.6 & 2.3 & 
1.8 & 1.7 & 1.0 & 2.0 
\\     \hline\hline
\end{tabular}
\caption{Improvement of the marginalized constraints on the Wilson coefficients of the Lee-Yang effective Lagrangian in \eref{TH_Lleeyang}, in the scenario where both left- and right-handed neutrinos are present. 
The improvement factor is defined as the ratio between the widths of the 68\% CL intervals in the fit without and with the mirror data. 
\label{tab:MirrorNoMirror}
}
\end{table}

\begin{figure}[tb]
\centering
\includegraphics[width=0.45\textwidth]{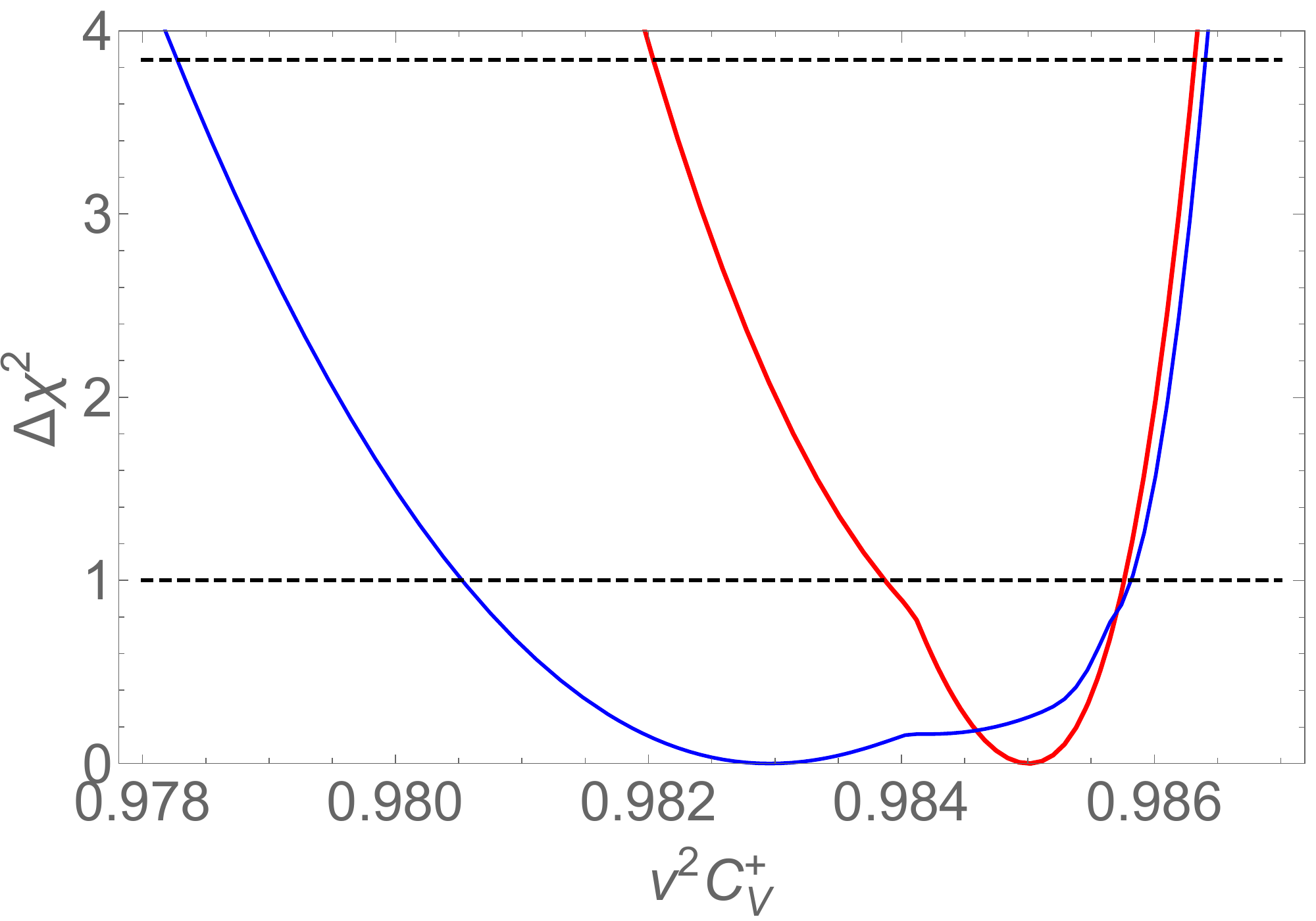}
\quad 
\includegraphics[width=0.45\textwidth]{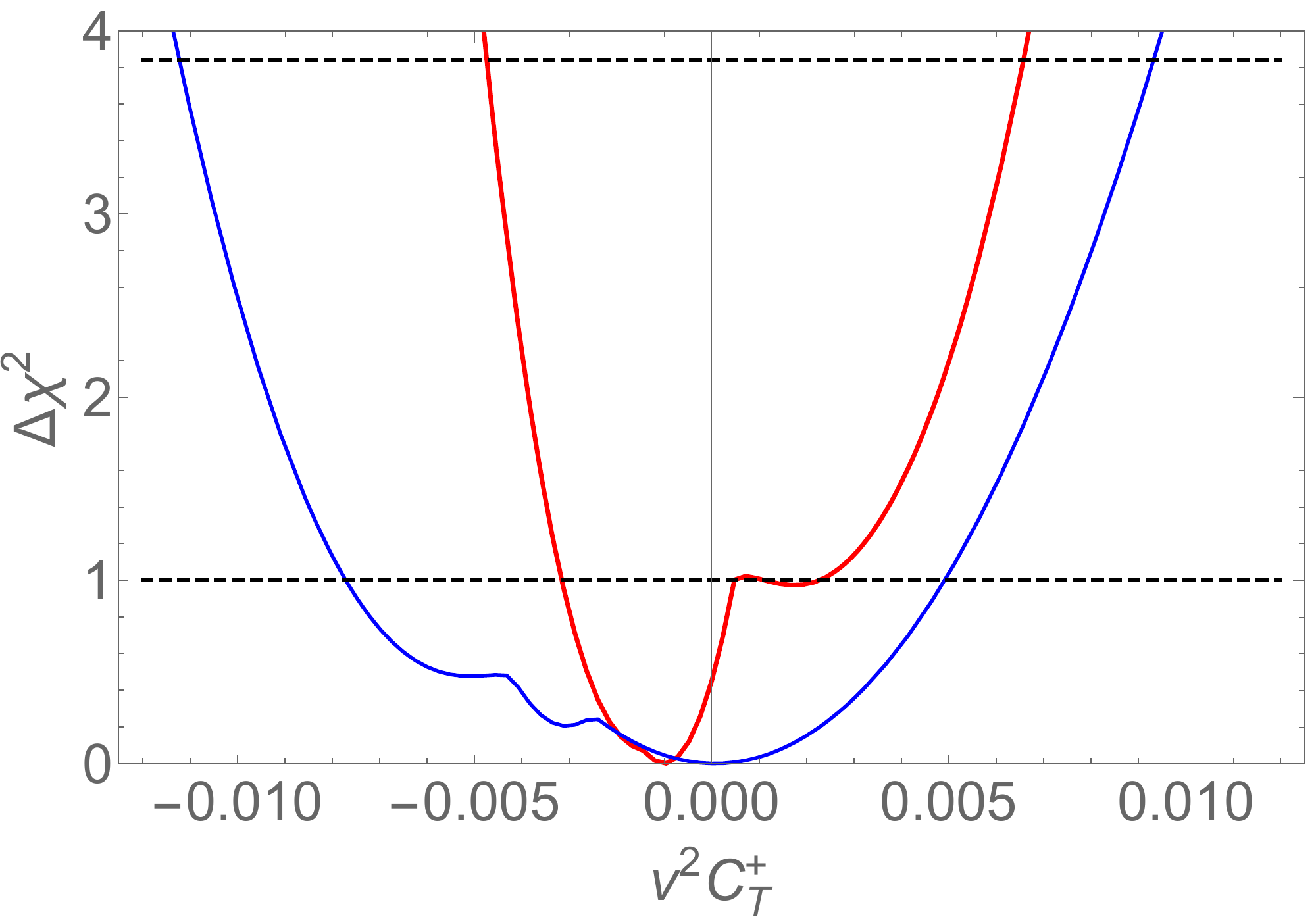}
\caption{\label{fig:MirrorNoMirror}
Marginalized $\Delta \chi^2 \equiv \chi^2 - \chi^2_{\rm min}$ distributions for the Wilson coefficients $C_{V}^+$ (left) and $C_T^+$ (right),  with (red) and without (blue) taking account the input from  mirror beta decay. 
}
\end{figure}

\begin{figure}[tb]
\centering
\includegraphics[width=0.45\textwidth]{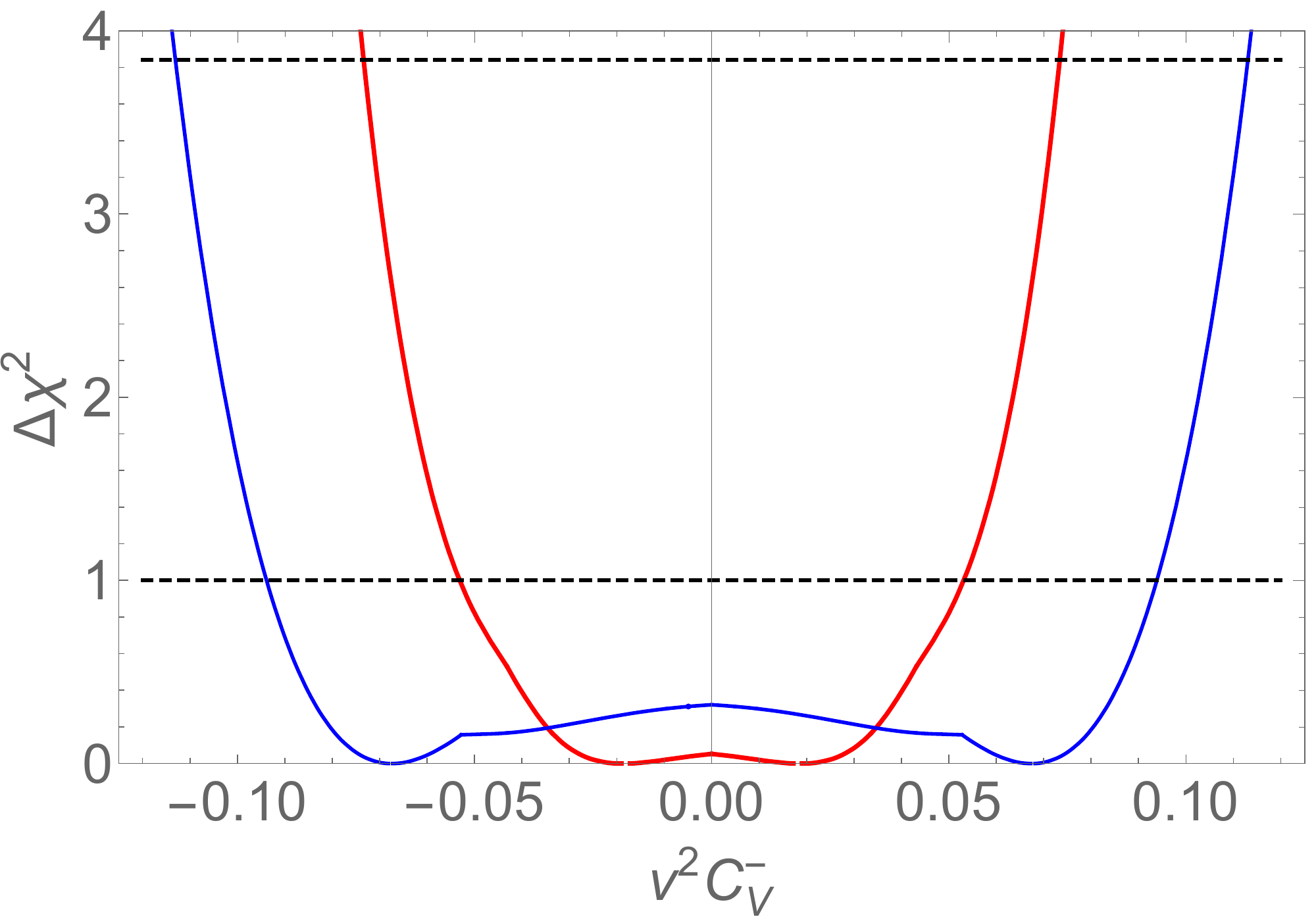}
\quad 
\includegraphics[width=0.45\textwidth]{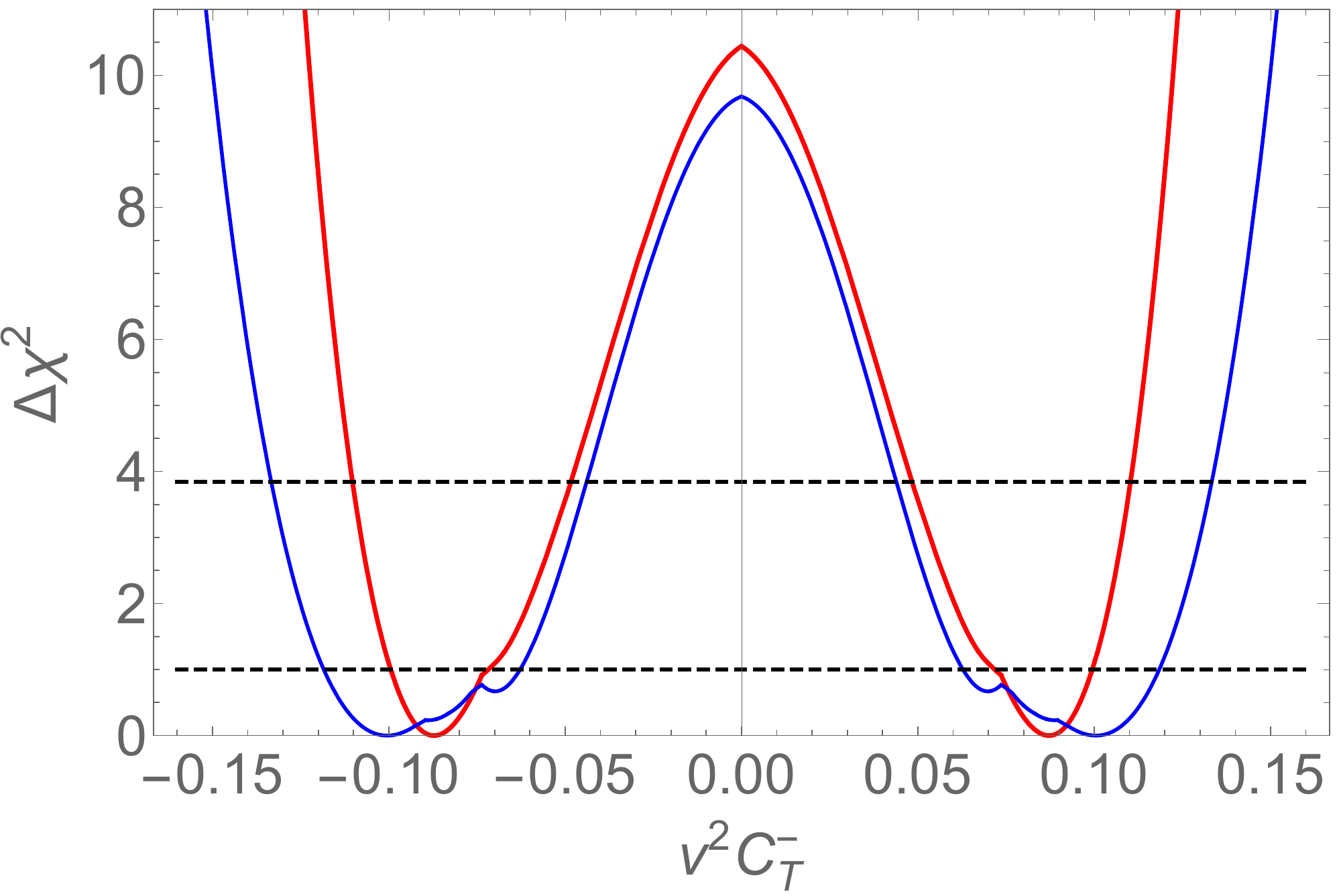}
\caption{\label{fig:MirrorNoMirror2}
Marginalized $\Delta \chi^2 \equiv \chi^2 - \chi^2_{\rm min}$ distributions for the Wilson coefficients $C_V^-$ (left) and $C_T^-$ (right),  with (red) and without (blue) taking account the input from  mirror beta decay. 
}
\end{figure}

We close this section with a historical comment.
One of the central questions in the 1950's was whether the weak interactions were mediated by  vector-axial or  scalar-tensor currents. After some initial confusion, experiments settled on the former possibility, paving the way to the discovery of the SM.
In fact, that conclusion has always hinged on simplifying assumptions that only a couple of Wilson coefficients of the Lee-Yang Lagrangian were present at the same time.
The present analysis is the first complete and model-independent demonstration that the weak interactions involving the lightest quarks are of the vector-axial type. In contrast to the global fits performed in Refs.~\cite{Boothroyd:1984fz,Severijns:2006dr} the present work includes also the extraction of the overall strength, keeping track of all correlations in a consistent way.  
At this point in history, such an observation has only an anecdotal value. 
Of more practical interest is that the present analysis provides the most up-to-date precise quantitative limits on possible departures from the vector-axial picture. 
We find that scalar and tensor currents associated with the {\em left-hand}  neutrino have to be below a percent level at 95\% CL.  
On the other hand, corrections from scalar and tensor currents associated with the {\em right-hand}  neutrino can be larger,  $\cO(10)$\%. 

\section{Conclusions and Future Directions}
\label{sec:conclusions}

In this paper we have discussed the constraints from nuclear beta decays on the parameters of the effective Lee-Yang Lagrangian describing the weak interactions between nucleons, electrons, and neutrinos, paying special attention to the role played by mirror decays.  
The Wilson coefficients of this Lagrangian carry precious information about physics occurring at higher-energies compared to the nuclear scale.  
First, they are sensitive to the fundamental SM parameters, 
specifically to the combination $V_{ud}/v^2 = \sqrt{2} G_F V_{ud}$. 
Second, they are affected by non-perturbative QCD entering via the nucleon charges $g_{V,A,S,T}$, and also by loop corrections which in part have to be evaluated in the non-perturbative regime as well. 
Finally, the Wilson coefficient are sensitive to physics beyond the SM, that is to masses and interaction strength of new hypothetical particles ($Z'$ bosons, leptoquarks, etc.) with masses in the 1 GeV - 100 TeV  range. 
We constructed a global likelihood function, using the latest experimental and theoretical input for a broad selection of allowed beta transitions. 
This allows us to extract the most up-to-date constraints on the Wilson coefficients in the  Lee-Yang Lagrangian in \eqref{eq:TH_Lleeyang}.

Two hallmarks stand out in the present analysis as compared to  previous works.
One is that we perform a completely general model-independent fit, allowing all the leading order Wilson coefficients in the Lee-Yang Lagrangian to be {\em simultaneously} present. 
This follows the lines of analyses in Refs.~\cite{Boothroyd:1984fz,Severijns:2006dr} but allowing in addition the extraction of the overall interaction strength in a consistent fashion. 
Therefore, the results obtained in this work can be applied to constrain generic new physics models that, at low energies, lead to an arbitrary pattern of vector, axial, scalar, and tensor currents. 
Moreover, the present results are valid for models with only left-handed neutrinos (as in the SM), 
as well as for models with both left- and right-handed neutrinos in the low-energy spectrum. 
The other hallmark is that we take into account measurements of half-lives and correlations in mirror beta transitions. 
Previously, mirror transitions have been employed to extract a value of the $V_{ud}$ matrix element in the SM scenario~\cite{NaviliatCuncic:2008xt}.
However, it is the first time that they are used in a consistent and global way to constrain new physics beyond the SM.
In our analysis we include the data on beta decays of 
${}^{17}$F, ${}^{19}$Ne, ${}^{21}$Na, ${}^{29}$P, ${}^{35}$Ar,  and 
${}^{37}$K.  
The distinguishing feature of these nuclei is that not only their half-life but also some correlation parameter ($a$, $A$, or $B$) has been measured with a decent accuracy. 
This allows one, simultaneously, to determine the mixing parameters $\rho$ for these transition and to constrain several new combinations of the Lee-Yang Wilson coefficients. 

It is also worth stressing that  the present analysis incorporates for the first time some recent developments in non-mirror decays. They encompass theoretical aspects, such as the inclusion of the nuclear-structure dependent corrections pointed out in Refs.~\cite{Seng:2018yzq,Gorchtein:2018fxl}, and experimental developments, such as the PERKEO-III measurement of the neutron beta asymmetry~\cite{Markisch:2018ndu}.

 The results for the confidence intervals are given in Eqs. \eqref{eq:RES_CVArhoSM}, \eqref{eq:RES_CXp}, and \eqref{eq:RES_CXpm}, 
depending on the assumptions about the pattern of Wilson coefficients present in the Lagrangian.
For the SM scenario and the BSM scenario involving only left-handed neutrinos, the present results supersede therefore those of Ref.~\cite{Gonzalez-Alonso:2018omy}.
\eref{RES_CVArhoSM} assumes the SM as the underlying theory  and thus the only Wilson coefficients present are $C_V^+$ and $C_A^+$,  as a consequence of the $V$-$A$ structure of the weak interactions.  
\eref{RES_CXp} assumes that the particle spectrum at low energies of order ~2 GeV is that predicted by the SM, in particular that  right-handed neutrinos are absent. 
However, it allows for generic new physics contributions, which manifest as scalar and tensor currents parameterized by the Wilson coefficients $C_S^+$ and $C_T^+$.  
Finally, \eref{RES_CXpm} is the most general scenario, as it allows for the presence of right-handed neutrinos in the low-energy spectrum. 
 Here, the relevant weak interactions are described by 8 Wilson coefficients: $C_{V,A,S,T}^+$ characterizing the vector, axial, scalar, and tensor currents with a left-handed neutrino, 
and $C_{V,A,S,T}^-$ characterizing the same currents but with a right-handed neutrino. 
In all these 3 scenarios we obtain a stable fit and stringent constraints on the Wilson coefficients. 
At the quantitative level, our most important findings are: 
\bi 
\item 
The relative uncertainty on $C_V^+$ and $C_A^+$ is $\cO(10^{-4})$ in the SM scenario. 
Via \eref{TH_LYtoRWEFT} it translates to an $\cO(10^{-4})$ precision measurement of $V_{ud}$ and $g_A$. 
The uncertainty on these parameters relaxes to $\cO(10^{-3})$ in the general new physics scenario; 
\item  
In the scenario with left-handed neutrinos the uncertainty on $C_S^+$ and $C_T^+$ is $\cO(10^{-3})$ in the units of $1/v^2$, where $v \approx 246$~GeV. 
This implies that nuclear observables are sensitive to new particles with $\cO(10)$~TeV masses and order one couplings to the SM, 
or with  $\cO(100)$~TeV masses and maximally strong coupling to the SM. 
In the presence of right-handed neutrinos these bounds are relaxed only by an $\cO(2)$ factor; 
\item 
The uncertainty on the Wilson coefficients $C_X^-$ characterizing weak interactions  of the right-handed neutrino is much larger, $\cO(10^{-1})$ in the units of $1/v^2$. 
That is because the contributions from right-handed neutrinos to the nuclear observables enter only at the quadratic level in $C_X^-$. 
\item 
In the scenario with only left-handed neutrinos, the current data do not show any hint of new physics effects, 
that is to stay, $C_S^+$ and $C_T^+$ are compatible with zero  within $1\sigma$ confidence level.
On the other hand, in the general scenario the data show a $3.2\sigma$ indication\footnote{The tension is reduced to $3.0~\sigma$ if one takes into account the updated measurement of neutron's $\tilde a$ by the aCORN experiment~\cite{Hassan:2020hrj}, which appeared after our paper.} for new physics manifesting via right-handed tensor currents with  $|C_T^-| \sim 0.1/v^2$. 
That Wilson coefficient could be generated e.g. via a leptoquark  with the mass in the TeV  range and $\cO(1)$ couplings to the light quarks, electron and right-handed neutrino. 
It is not clear, however, whether a phenomenologically viable model of this kind can be constructed, given the constraints from direct and Drell-Yan searches  at the LHC~\cite{Cirigliano:2012ab}. 
\ei 
The inclusion of mirror transitions in the global fit has little effect in the SM scenario and in the BSM scenario with only left-handed neutrinos,
where the precise data from the superallowed and neutron decays dominate the final constraints. 
On the other hand, the mirror transitions are {\em vital} to lift approximate degeneracies in the parameter space of the more general scenario with right-handed neutrinos.  
 Here, the confidence intervals for the Wilson coefficients shrink by a factor of two due to the inclusion of the mirror data. 
Incidentally, the hint for a non-zero $C_T^-$ is slightly enhanced (by one unit of $\chi^2$) after including the mirror data. 
All this demonstrates the potential importance of mirror transitions as a means to search for new physics. 

\begin{table}[tb]
\renewcommand{\arraystretch}{1.5}
\bc
\begin{tabular}{ccccccc}
\hline\hline
Measurement & ${}^{17}$F & ${}^{19}$Ne &  ${}^{21}$Na 
 &  ${}^{29}$P &  ${}^{35}$Ar &  ${}^{37}$K 
 \\  \hline
 $\Delta a = 10^{-3}$ &
1.3 &  1.2 & 1.6 &  1.4 &  1.6 &  1.3 
 \\  
$\Delta \tilde A  = 10^{-3}$ & 
1.0 & 1.2 & 1.4 & 1.4  &  1.7 &  1.1 
 \\  
 $\Delta \tilde A = 10^{-3}$,
${\Delta {\cal F}t / {\cal F}t} = 5 \times 10^{-4}$& 
1.0&  1.2 & 1.5 & 2.1  & 3.2  & 1.2 
 \\  
  $\Delta \tilde A  = 10^{-4}$ & 
1.4 & 3.1 &  2.1 &  1.5 & 1.8 & 1.4 
 \\  
  $\Delta \tilde A = 10^{-4}$,
${\Delta {\cal F}t / {\cal F}t} = 5 \times 10^{-4}$& 
1.5 & 3.8 & 3.8 & 3.8  & 3.9  & 3.4
 \\  \hline\hline
\end{tabular}
\ec
\caption{
\label{tab:RES_futureSM}
Potential improvements of the determination of the CKM matrix element $V_{ud}$ in the SM using only mirror decays with respect to the current extraction, $V_{ud}^{mirror}=0.97424(95)$. 
A given row and column indicate the observable and transition, respectively, that is added to the current dataset. 
We remind the reader that an improvement of a 3.4 (1.8) factor is required to reach the current $V_{ud}$ extractions from superallowed (neutron) data.
}
\end{table}

We close with a comment about the prospects for improving the constraints presented in this work. 
As shown in~\tref{mirror}, past measurements of correlations in mirror decays are typically at the $10^{-2}$ level, with the notable exception of $\tilde{A}(^{37}{\rm K})$ which has a $3\times 10^{-3}$ uncertainty. On the other hand, ${\cal F}t$ values in mirror decays are much better known, with typical uncertainties in the $(0.5-1.8)\times 10^{-3}$ range.
The expected sensitivity of  improved beta decay experiments, many of which involve mirror nuclei, was recently discussed in Ref.~\cite{Cirgiliano:2019nyn}.
Our goal here is merely to establish a number of goalposts, to serve as a simple reference for planning future experiments. 
Potential improvements of the $V_{ud}$ determination from mirror decays are shown in~\tref{RES_futureSM} for the SM scenario.
We see for instance that a new per-mille-accuracy $A$ measurement in ${}^{35}$Ar will shrink the error bars by a 1.7 factor, producing a $V_{ud}^{mirror}$ extraction as precise as the current one from neutron decay.
A more significant improvement would require knowledge of some correlation and the corresponding ${\cal F} t$ value with an $\cO(10^{-4})$ accuracy.
Beyond the SM, the impact on mirror-only fits will be far more consequential. 
As one can see in \tref{RES_weft_comparison}, 
the existing mirror data provide loose constraints on the SM and new physics parameters in the BSM scenario with only left-handed neutrinos.  
Any new per-mille-level correlation measurement will help lifting approximately flat directions in the parameter space, leading to a significant strengthening of the mirror-only constraints. 
On the other hand, it is less trivial for the new mirror measurements to make an impact in the {\em global} fit, where they have to compete with very precise superallowed and neutron data. 
We would need an $\cO(10^{-4})$-level correlation measurement for a lighter mirror nuclei (${}^{17}$F or ${}^{19}$Ne) to have a non-negligible improvement in the BSM bounds. 
Finally, in the general BSM scenario with right-handed neutrinos the future mirror data will continue playing an outstanding role, 
whether in the mirror-only or in the global fit. 
In this case, because of the many-dimensional parameter space and the existence of multiple quasi-degenerate minima, any new per-mille level mirror measurement will be invaluable for reducing the error bars on both the SM and new physics parameters.

\section*{Acknowledgments}
We thank M.~Gorchtein, L.~Hayen, and N.~Severijns for useful discussions and correspondence.
AF is partially supported by the Agence Nationale de la Recherche (ANR) under grant ANR-19-CE31-0012 (project MORA). MGA is supported by the {\it Generalitat Valenciana} (Spain) through the {\it plan GenT} program (CIDEGENT/2018/014).
\appendix 

\section{Gamow-Teller decays}
\label{app:formulas}

For completeness, in this appendix we summarize the theoretical expressions for the correlation parameters in Gamow-Teller (GT) beta decays used in the global fit.   
We define 
\beq
\hat \xi_{\rm GT}   \equiv   
(C_A^+)^2 + (C_T^+)^2  +  (C_A^-)^2+   (C_T^-)^2 . 
\eeq 
The $\beta$-$\nu$ correlation is then expressed by the Wilson coefficient of the Lee-Yang Lagrangian in \eref{TH_Lleeyang} as  \beq 
\hat \xi_{\rm GT}   a_{\rm GT}    =  
 - {1 \over 3 } \bigg [  (C_A^+)^2 - (C_T^+)^2 +  (C_A^-)^2 - (C_T^-)^2 \bigg ] . 
\eeq 
The expression for the $\beta$-asymmetry depends on relative spins of the initial (J) and final (J') state nuclei. 
For  $J' = J-1$ we have 
\beq
\hat \xi_{\rm GT} A_{\rm GT}  =   \mp  \bigg \{ 
(C_A^+)^2  -   (C_T^+)^2    - (C_A^-)^2  +  (C_T^-)^2   \bigg \}  ,
\eeq 
while for $J' = J+1$ we have 
\beq
\hat \xi_{\rm GT} A_{\rm GT}   =  
 \pm  {J \over J+1}  \bigg \{ 
 (C_A^+)^2  -   (C_T^+)^2    - (C_A^-)^2  +   (C_T^-)^2  \bigg \} .
\eeq 
Finally, the ratio of beta polarizations in pure Fermi and GT transitions is given by the expression:  
\bea
{P_{\rm F} \over P_{\rm GT} } & = &  
{ (C_V^+)^2   -  (C_S^+)^2   -   (C_V^-)^2  +    (C_S^-)^2 \over 
 (C_V^+)^2   +  (C_S^+)^2   +   (C_V^-)^2  +    (C_S^-)^2 
- 2 \langle m_e/E_e \rangle_{F} \big [  C_V^+ C_S^+  +   C_V^- C_S^-  \big ] 
  }
  \nnl & \times & 
 { (C_A^+)^2   +  (C_T^+)^2   +   (C_A^-)^2  +    (C_T^-)^2
- 2 \langle m_e/E_e \rangle_{GT} \big  [  C_A^+ C_T^+  +   C_A^- C_T^-  \big  ]  
 \over 
 (C_A^+)^2   -  (C_T^+)^2   -   (C_A^-)^2  +    (C_T^-)^2   }.
\eea

\section{Data for non-mirror beta decays}
\label{app:tables}

In this section we collect the additional experimental values for observables used in the analysis. The values for mirror beta decays are listed in~\tref{mirror}. 

\begin{table}[h]
\caption{
\label{tab:superallowed}
Data from superallowed decays used in the fits~\cite{Hardy:2020qwl}.}
\begin{center}
\begin{tabular}{r@{\hspace{8mm}} r@{\hspace{8mm}}  c}
\hline\hline
Parent & ${\cal F}t$ [s] ~~~ & $\langle m_e/E_e \rangle$\\
\hline
$^{10}$C   & $3075.7 \pm 4.4$ &  0.619  \\
$^{14}$O   & $3070.2\pm 1.9$ &   0.438  \\
$^{22}$Mg  & $3076.2 \pm 7.0$ &  0.308  \\
$^{26m}$Al & $3072.4 \pm 1.1$ &  0.300  \\
$^{26}$Si & $3075.4 \pm 5.7$ &   0.264  \\
$^{34}$Cl  & $3071.6 \pm 1.8$ &  0.234  \\
$^{34}$Ar  & $3075.1\pm 3.1$ &   0.212  \\
$^{38m}$K  & $3072.9\pm 2.0$ &   0.213  \\
$^{38}$Ca  & $3077.8\pm 6.2$ &   0.195  \\
$^{42}$Sc  & $3071.7\pm 2.0$ &   0.201  \\
$^{46}$V   & $3074.3\pm 2.0$ &   0.183  \\
$^{50}$Mn  & $3071.1\pm 1.6$ &   0.169  \\
$^{54}$Co  & $3070.4\pm 2.5$ &   0.157  \\
$^{62}$Ga  & $3072.4\pm 6.7$ &   0.142  \\
$^{74}$Rb  & $3077\pm 11$ &      0.125  \\
\hline\hline
\end{tabular}
\end{center}
\end{table}

\tref{superallowed} lists the ${\cal F}t$ values for the superallowed decays used in the fits. 
These are all $\beta^+$ transitions of the Fermi type between nuclei of spin~0 and positive parity.  
The ${\cal F} t$ values are copied from Table~XVI of Ref.~\cite{Hardy:2020qwl}.
The central values take into account both the $\delta_R'$ correction and the effects pointed out in Refs.~\cite{Seng:2018yzq,Gorchtein:2018fxl}, 
however the errors do not include the associated theoretical uncertainties as they are strongly correlated between the decays. 
The fits carried out in the present work do take into account those correlated errors following~\eref{Seng2correction}. 
The $\langle m_e/E_e \rangle$ values are calculated using \eref{TH_meOverEe}.

\begin{table}[h]
\caption{
\label{tab:NeutronData}
Inputs from neutron decay  used in the fits.
}
\begin{center}
\begin{tabular}{
c@{\hspace{3mm}}  r@{\hspace{5mm}}  
c@{\hspace{3mm}}
c@{\hspace{3mm}}
c }
\hline\hline
Observable		& Value~~ 	& S factor		
& $\langle m_e/E_e \rangle$	& References \\
\hline
$\tau_n$~(s)	& 879.75(76) & 1.9
& 0.655					&\cite{Mampe:1993an,Byrne:1996zz,Serebrov:2004zf,Pichlmaier:2010zz,Steyerl:2012zz,Yue:2013qrc,Ezhov:2014tna,Arzumanov:2015tea,Pattie:2017vsj,Serebrov:2017bzo} 
\\
$\tilde{A}_n$	&  $-0.11958(21)$ & 1.2 		
& 0.569					& \cite{Bopp:1986rt,Liaud:1997vu,Erozolimsky:1997wi,Mund:2012fq,Brown:2017mhw,Markisch:2018ndu,Zyla:2020zbs}\\
$\tilde{B}_n$	&  0.9805(30)	&			
& 0.591					& \cite{Kuznetsov:1995sk,Serebrov:1998aj,Kreuz:2005jz,Schumann:2007qe}\\
$\lambda_{AB}$&  $-1.2686(47)$ & 
& 0.581					& \cite{Mostovoi:2001ye}\\
$a_n$		& $-0.10426(82)$	& 		
& &\cite{Stratowa:1978gq,Byrne:2002tx,Beck:2019xye}  
\\
$\tilde{a}_n$	& $-0.1090(41)$	&  		
& 0.695					& \cite{Darius:2017arh}
\\
\hline\hline
\end{tabular}
\end{center}
\end{table}

The input from neutron decay used in the fits is shown in \tref{NeutronData}.
When multiple references are given, the value is a Gaussian average of several experimental results. 
For the neutron lifetime, due to mutually inconsistent measurements,  the error is inflated by the scale factor $S=1.9$ following the standard PDG procedure~\cite{Zyla:2020zbs}.
Contrary to the latest PDG edition~\cite{Zyla:2020zbs}, 
we do not discard the beam measurements~\cite{Byrne:1996zz,Yue:2013qrc} following the arguments of Ref.~\cite{Czarnecki:2018okw}, since these arguments are valid only in the SM context. 
In fact, as shown in~\fref{Neutronlifetime}, allowing for scalar and tensor currents in the effective Lagrangian, the global fit actually predicts a central value of $\tau_n$ closer to the beam than to the bottle results~\cite{Mampe:1993an,Serebrov:2004zf,Pichlmaier:2010zz,Steyerl:2012zz,Ezhov:2014tna,Arzumanov:2015tea,Pattie:2017vsj,Serebrov:2017bzo}, although compatible with both of them.
\begin{figure}[tb]
\centering
\includegraphics[width=0.6\textwidth]{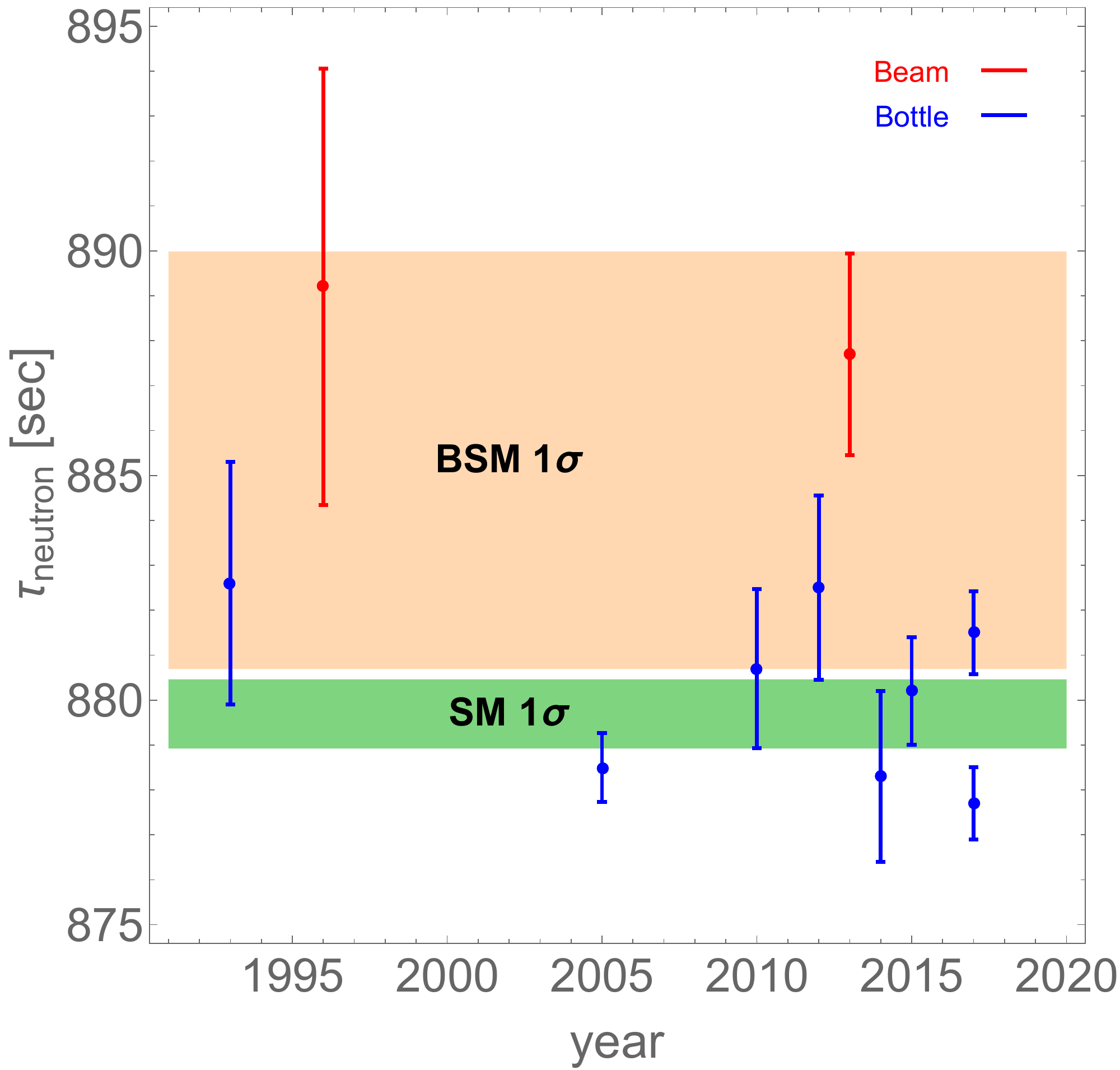}
\caption{\label{fig:Neutronlifetime}
Neutron lifetime prediction in a global fit that includes all other measurements listed in~\aref{tables}. Bottle measurements are favored in the SM fit (green band) but not in the BSM scenario with  only left-handed neutrinos (salmon band) (\emph{cf.}~\sref{WEFTfit}).  
}
\end{figure}

In the combination of the $\beta$-asymmetry measurements $\tilde A_n$ we inflate the error by the scale factor $S=1.2$, 
again following PDG~\cite{Zyla:2020zbs}. 
For the $\tau_n$ measurement, $\langle m_e/E_e \rangle$ is calculated using \eref{TH_meOverEe}; 
for the remaining measurements we use the effective  $\langle m_e/E_e \rangle$ values provided in Ref.~\cite{Gonzalez-Alonso:2018omy}, which take into account the experimental conditions.

\begin{table}[h]
\caption{
\label{tab:NuclearData}
Data from correlation measurements in pure Fermi and pure Gamow-Teller decays used in the fits. 
}
\begin{center}
\begin{tabular}{
l@{\hspace{5mm}} c @{\hspace{3mm}}  c@{\hspace{3mm}}  c@{\hspace{3mm}}  c@{\hspace{3mm}}
 r@{\hspace{5mm}}  
c@{\hspace{3mm}}  c 
}
\hline\hline
Parent		& $J_i$	& $J_f$	& Type	& Observable	& Value~~
& $\langle m_e/E_e \rangle$	& Ref.
\\ \hline
$^6$He		& 0		& 1		& GT/$\beta^-$		& $\tilde{a}$		& $-0.3308(30)$ 
&	0.286		& \cite{Johnson:1963zza}
\\
$^{32}$Ar		& 0		& 0		&  F/$\beta^+$		& $\tilde{a}$	& 0.9989(65)
&  0.210	& \cite{Adelberger:1999ud} 
\\
$^{38m}$K	& 0		& 0		&  F/$\beta^+$		& $\tilde{a}$	& 0.9981(48)	
&  0.161  & \cite{Gorelov:2004hv}
\\
$^{60}$Co		& 5		& 4		& GT/$\beta^-$		& $\tilde{A}$	& $-1.014(20)$			
& 0.704			& \cite{Wauters:2010gh} \\
$^{67}$Cu		& 3/2		& 5/2		& GT/$\beta^-$		& $\tilde{A}$	& 0.587(14)			
& 0.395			& \cite{Soti:2014xua} 
\\
$^{114}$In	& 1		& 0		& GT/$\beta^-$		& $\tilde{A}$	& $-0.994(14)$			
& 0.209			& \cite{Wauters:2009jw}
\\
$^{14}$O/$^{10}$C &	&		& F-GT/$\beta^+$	& $P_F/P_{GT}$& 0.9996(37)			
&  0.292			&  \cite{Carnoy:1991jd}
\\
$^{26}$Al/$^{30}$P &	&		& F-GT/$\beta^+$	& $P_F/P_{GT}$& 1.0030	(40) 		
&  0.216			& \cite{Wichers:1986es}
\\ \hline\hline
\end{tabular}
\end{center}
\end{table}
Finally, the present analysis includes an input from various correlation measurements in pure Fermi and pure Gamow-Teller decays. 
These are collected in \tref{NuclearData}.

\bibliographystyle{apsrev4-1}
\bibliography{mirrormirror}

\begin{thebibliography}{108}%
\makeatletter
\providecommand \@ifxundefined [1]{%
 \@ifx{#1\undefined}
}%
\providecommand \@ifnum [1]{%
 \ifnum #1\expandafter \@firstoftwo
 \else \expandafter \@secondoftwo
 \fi
}%
\providecommand \@ifx [1]{%
 \ifx #1\expandafter \@firstoftwo
 \else \expandafter \@secondoftwo
 \fi
}%
\providecommand \natexlab [1]{#1}%
\providecommand \enquote  [1]{``#1''}%
\providecommand \bibnamefont  [1]{#1}%
\providecommand \bibfnamefont [1]{#1}%
\providecommand \citenamefont [1]{#1}%
\providecommand \href@noop [0]{\@secondoftwo}%
\providecommand \href [0]{\begingroup \@sanitize@url \@href}%
\providecommand \@href[1]{\@@startlink{#1}\@@href}%
\providecommand \@@href[1]{\endgroup#1\@@endlink}%
\providecommand \@sanitize@url [0]{\catcode `\\12\catcode `\$12\catcode
  `\&12\catcode `\#12\catcode `\^12\catcode `\_12\catcode `\%12\relax}%
\providecommand \@@startlink[1]{}%
\providecommand \@@endlink[0]{}%
\providecommand \url  [0]{\begingroup\@sanitize@url \@url }%
\providecommand \@url [1]{\endgroup\@href {#1}{\urlprefix }}%
\providecommand \urlprefix  [0]{URL }%
\providecommand \Eprint [0]{\href }%
\providecommand \doibase [0]{http://dx.doi.org/}%
\providecommand \selectlanguage [0]{\@gobble}%
\providecommand \bibinfo  [0]{\@secondoftwo}%
\providecommand \bibfield  [0]{\@secondoftwo}%
\providecommand \translation [1]{[#1]}%
\providecommand \BibitemOpen [0]{}%
\providecommand \bibitemStop [0]{}%
\providecommand \bibitemNoStop [0]{.\EOS\space}%
\providecommand \EOS [0]{\spacefactor3000\relax}%
\providecommand \BibitemShut  [1]{\csname bibitem#1\endcsname}%
\let\auto@bib@innerbib\@empty
\bibitem [{\citenamefont {Cirigliano}\ \emph
  {et~al.}(2013{\natexlab{a}})\citenamefont {Cirigliano}, \citenamefont
  {Gardner},\ and\ \citenamefont {Holstein}}]{Cirigliano:2013xha}%
  \BibitemOpen
  \bibfield  {author} {\bibinfo {author} {\bibfnamefont {V.}~\bibnamefont
  {Cirigliano}}, \bibinfo {author} {\bibfnamefont {S.}~\bibnamefont {Gardner}},
  \ and\ \bibinfo {author} {\bibfnamefont {B.}~\bibnamefont {Holstein}},\
  }\href {\doibase 10.1016/j.ppnp.2013.03.005} {\bibfield  {journal} {\bibinfo
  {journal} {Prog. Part. Nucl. Phys.}\ }\textbf {\bibinfo {volume} {71}},\
  \bibinfo {pages} {93} (\bibinfo {year} {2013}{\natexlab{a}})},\ \Eprint
  {http://arxiv.org/abs/1303.6953} {arXiv:1303.6953 [hep-ph]} \BibitemShut
  {NoStop}%
\bibitem [{\citenamefont {Naviliat-Cuncic}\ and\ \citenamefont
  {Gonz\'alez-Alonso}(2013)}]{Gonzalez-Alonso:2013uqa}%
  \BibitemOpen
  \bibfield  {author} {\bibinfo {author} {\bibfnamefont {O.}~\bibnamefont
  {Naviliat-Cuncic}}\ and\ \bibinfo {author} {\bibfnamefont {M.}~\bibnamefont
  {Gonz\'alez-Alonso}},\ }\href {\doibase 10.1002/andp.201300072} {\bibfield
  {journal} {\bibinfo  {journal} {Annalen Phys.}\ }\textbf {\bibinfo {volume}
  {525}},\ \bibinfo {pages} {600} (\bibinfo {year} {2013})},\ \Eprint
  {http://arxiv.org/abs/1304.1759} {arXiv:1304.1759 [hep-ph]} \BibitemShut
  {NoStop}%
\bibitem [{\citenamefont {Vos}\ \emph {et~al.}(2015)\citenamefont {Vos},
  \citenamefont {Wilschut},\ and\ \citenamefont {Timmermans}}]{Vos:2015eba}%
  \BibitemOpen
  \bibfield  {author} {\bibinfo {author} {\bibfnamefont {K.~K.}\ \bibnamefont
  {Vos}}, \bibinfo {author} {\bibfnamefont {H.~W.}\ \bibnamefont {Wilschut}}, \
  and\ \bibinfo {author} {\bibfnamefont {R.~G.~E.}\ \bibnamefont
  {Timmermans}},\ }\href {\doibase 10.1103/RevModPhys.87.1483} {\bibfield
  {journal} {\bibinfo  {journal} {Rev. Mod. Phys.}\ }\textbf {\bibinfo {volume}
  {87}},\ \bibinfo {pages} {1483} (\bibinfo {year} {2015})},\ \Eprint
  {http://arxiv.org/abs/1509.04007} {arXiv:1509.04007 [hep-ph]} \BibitemShut
  {NoStop}%
\bibitem [{\citenamefont {Gonzalez-Alonso}\ \emph {et~al.}(2019)\citenamefont
  {Gonzalez-Alonso}, \citenamefont {Naviliat-Cuncic},\ and\ \citenamefont
  {Severijns}}]{Gonzalez-Alonso:2018omy}%
  \BibitemOpen
  \bibfield  {author} {\bibinfo {author} {\bibfnamefont {M.}~\bibnamefont
  {Gonzalez-Alonso}}, \bibinfo {author} {\bibfnamefont {O.}~\bibnamefont
  {Naviliat-Cuncic}}, \ and\ \bibinfo {author} {\bibfnamefont {N.}~\bibnamefont
  {Severijns}},\ }\href {\doibase 10.1016/j.ppnp.2018.08.002} {\bibfield
  {journal} {\bibinfo  {journal} {Prog. Part. Nucl. Phys.}\ }\textbf {\bibinfo
  {volume} {104}},\ \bibinfo {pages} {165} (\bibinfo {year} {2019})},\ \Eprint
  {http://arxiv.org/abs/1803.08732} {arXiv:1803.08732 [hep-ph]} \BibitemShut
  {NoStop}%
\bibitem [{\citenamefont {Zyla}\ \emph {et~al.}(2020)\citenamefont {Zyla} \emph
  {et~al.}}]{Zyla:2020zbs}%
  \BibitemOpen
  \bibfield  {author} {\bibinfo {author} {\bibfnamefont {P.}~\bibnamefont
  {Zyla}} \emph {et~al.} (\bibinfo {collaboration} {Particle Data Group}),\
  }\href {\doibase 10.1093/ptep/ptaa104} {\bibfield  {journal} {\bibinfo
  {journal} {PTEP}\ }\textbf {\bibinfo {volume} {2020}},\ \bibinfo {pages}
  {083C01} (\bibinfo {year} {2020})}\BibitemShut {NoStop}%
\bibitem [{\citenamefont {Lee}\ and\ \citenamefont {Yang}(1956)}]{Lee:1956qn}%
  \BibitemOpen
  \bibfield  {author} {\bibinfo {author} {\bibfnamefont {T.~D.}\ \bibnamefont
  {Lee}}\ and\ \bibinfo {author} {\bibfnamefont {C.-N.}\ \bibnamefont {Yang}},\
  }\href {\doibase 10.1103/PhysRev.104.254} {\bibfield  {journal} {\bibinfo
  {journal} {Phys. Rev.}\ }\textbf {\bibinfo {volume} {104}},\ \bibinfo {pages}
  {254} (\bibinfo {year} {1956})}\BibitemShut {NoStop}%
\bibitem [{\citenamefont {Jackson}\ \emph
  {et~al.}(1957{\natexlab{a}})\citenamefont {Jackson}, \citenamefont
  {Treiman},\ and\ \citenamefont {Wyld}}]{Jackson:1957auh}%
  \BibitemOpen
  \bibfield  {author} {\bibinfo {author} {\bibfnamefont {J.~D.}\ \bibnamefont
  {Jackson}}, \bibinfo {author} {\bibfnamefont {S.~B.}\ \bibnamefont
  {Treiman}}, \ and\ \bibinfo {author} {\bibfnamefont {H.~W.}\ \bibnamefont
  {Wyld}},\ }\href {\doibase 10.1016/0029-5582(87)90019-8} {\bibfield
  {journal} {\bibinfo  {journal} {Nucl. Phys.}\ }\textbf {\bibinfo {volume}
  {4}},\ \bibinfo {pages} {206} (\bibinfo {year}
  {1957}{\natexlab{a}})}\BibitemShut {NoStop}%
\bibitem [{\citenamefont {Ebel}\ and\ \citenamefont
  {Feldman}(1957)}]{EBEL1957213}%
  \BibitemOpen
  \bibfield  {author} {\bibinfo {author} {\bibfnamefont {M.}~\bibnamefont
  {Ebel}}\ and\ \bibinfo {author} {\bibfnamefont {G.}~\bibnamefont {Feldman}},\
  }\href {\doibase https://doi.org/10.1016/0029-5582(87)90020-4} {\bibfield
  {journal} {\bibinfo  {journal} {Nuclear Physics}\ }\textbf {\bibinfo {volume}
  {4}},\ \bibinfo {pages} {213 } (\bibinfo {year} {1957})}\BibitemShut
  {NoStop}%
\bibitem [{\citenamefont {Pich}(1998)}]{Pich:1998xt}%
  \BibitemOpen
  \bibfield  {author} {\bibinfo {author} {\bibfnamefont {A.}~\bibnamefont
  {Pich}},\ }in\ \href@noop {} {\emph {\bibinfo {booktitle} {{Les Houches
  Summer School in Theoretical Physics, Session 68: Probing the Standard Model
  of Particle Interactions}}}}\ (\bibinfo {year} {1998})\ pp.\ \bibinfo {pages}
  {949--1049},\ \Eprint {http://arxiv.org/abs/hep-ph/9806303}
  {arXiv:hep-ph/9806303} \BibitemShut {NoStop}%
\bibitem [{\citenamefont {Paul}(1970)}]{Paul:1970gfz}%
  \BibitemOpen
  \bibfield  {author} {\bibinfo {author} {\bibfnamefont {H.}~\bibnamefont
  {Paul}},\ }\href {\doibase 10.1016/0375-9474(70)91077-8} {\bibfield
  {journal} {\bibinfo  {journal} {Nucl. Phys. A}\ }\textbf {\bibinfo {volume}
  {154}},\ \bibinfo {pages} {160} (\bibinfo {year} {1970})}\BibitemShut
  {NoStop}%
\bibitem [{\citenamefont {Boothroyd}\ \emph {et~al.}(1984)\citenamefont
  {Boothroyd}, \citenamefont {Markey},\ and\ \citenamefont
  {Vogel}}]{Boothroyd:1984fz}%
  \BibitemOpen
  \bibfield  {author} {\bibinfo {author} {\bibfnamefont {A.}~\bibnamefont
  {Boothroyd}}, \bibinfo {author} {\bibfnamefont {J.}~\bibnamefont {Markey}}, \
  and\ \bibinfo {author} {\bibfnamefont {P.}~\bibnamefont {Vogel}},\ }\href
  {\doibase 10.1103/PhysRevC.29.603} {\bibfield  {journal} {\bibinfo  {journal}
  {Phys. Rev. C}\ }\textbf {\bibinfo {volume} {29}},\ \bibinfo {pages} {603}
  (\bibinfo {year} {1984})}\BibitemShut {NoStop}%
\bibitem [{\citenamefont {Severijns}\ \emph {et~al.}(2006)\citenamefont
  {Severijns}, \citenamefont {Beck},\ and\ \citenamefont
  {Naviliat-Cuncic}}]{Severijns:2006dr}%
  \BibitemOpen
  \bibfield  {author} {\bibinfo {author} {\bibfnamefont {N.}~\bibnamefont
  {Severijns}}, \bibinfo {author} {\bibfnamefont {M.}~\bibnamefont {Beck}}, \
  and\ \bibinfo {author} {\bibfnamefont {O.}~\bibnamefont {Naviliat-Cuncic}},\
  }\href {\doibase 10.1103/RevModPhys.78.991} {\bibfield  {journal} {\bibinfo
  {journal} {Rev. Mod. Phys.}\ }\textbf {\bibinfo {volume} {78}},\ \bibinfo
  {pages} {991} (\bibinfo {year} {2006})},\ \Eprint
  {http://arxiv.org/abs/nucl-ex/0605029} {arXiv:nucl-ex/0605029} \BibitemShut
  {NoStop}%
\bibitem [{\citenamefont {Severijns}\ \emph {et~al.}(2008)\citenamefont
  {Severijns}, \citenamefont {Tandecki}, \citenamefont {Phalet},\ and\
  \citenamefont {Towner}}]{Severijns:2008ep}%
  \BibitemOpen
  \bibfield  {author} {\bibinfo {author} {\bibfnamefont {N.}~\bibnamefont
  {Severijns}}, \bibinfo {author} {\bibfnamefont {M.}~\bibnamefont {Tandecki}},
  \bibinfo {author} {\bibfnamefont {T.}~\bibnamefont {Phalet}}, \ and\ \bibinfo
  {author} {\bibfnamefont {I.~S.}\ \bibnamefont {Towner}},\ }\href {\doibase
  10.1103/PhysRevC.78.055501} {\bibfield  {journal} {\bibinfo  {journal} {Phys.
  Rev.}\ }\textbf {\bibinfo {volume} {C78}},\ \bibinfo {pages} {055501}
  (\bibinfo {year} {2008})},\ \Eprint {http://arxiv.org/abs/0807.2201}
  {arXiv:0807.2201 [nucl-ex]} \BibitemShut {NoStop}%
\bibitem [{\citenamefont {Naviliat-Cuncic}\ and\ \citenamefont
  {Severijns}(2009)}]{NaviliatCuncic:2008xt}%
  \BibitemOpen
  \bibfield  {author} {\bibinfo {author} {\bibfnamefont {O.}~\bibnamefont
  {Naviliat-Cuncic}}\ and\ \bibinfo {author} {\bibfnamefont {N.}~\bibnamefont
  {Severijns}},\ }\href {\doibase 10.1103/PhysRevLett.102.142302} {\bibfield
  {journal} {\bibinfo  {journal} {Phys. Rev. Lett.}\ }\textbf {\bibinfo
  {volume} {102}},\ \bibinfo {pages} {142302} (\bibinfo {year} {2009})},\
  \Eprint {http://arxiv.org/abs/0809.0994} {arXiv:0809.0994 [nucl-ex]}
  \BibitemShut {NoStop}%
\bibitem [{\citenamefont {Beck}\ \emph {et~al.}(2020)\citenamefont {Beck} \emph
  {et~al.}}]{Beck:2019xye}%
  \BibitemOpen
  \bibfield  {author} {\bibinfo {author} {\bibfnamefont {M.}~\bibnamefont
  {Beck}} \emph {et~al.},\ }\href {\doibase 10.1103/PhysRevC.101.055506}
  {\bibfield  {journal} {\bibinfo  {journal} {Phys. Rev. C}\ }\textbf {\bibinfo
  {volume} {101}},\ \bibinfo {pages} {055506} (\bibinfo {year} {2020})},\
  \Eprint {http://arxiv.org/abs/1908.04785} {arXiv:1908.04785 [nucl-ex]}
  \BibitemShut {NoStop}%
\bibitem [{\citenamefont {Aghanim}\ \emph {et~al.}(2020)\citenamefont {Aghanim}
  \emph {et~al.}}]{Aghanim:2018eyx}%
  \BibitemOpen
  \bibfield  {author} {\bibinfo {author} {\bibfnamefont {N.}~\bibnamefont
  {Aghanim}} \emph {et~al.} (\bibinfo {collaboration} {Planck}),\ }\href
  {\doibase 10.1051/0004-6361/201833910} {\bibfield  {journal} {\bibinfo
  {journal} {Astron. Astrophys.}\ }\textbf {\bibinfo {volume} {641}},\ \bibinfo
  {pages} {A6} (\bibinfo {year} {2020})},\ \Eprint
  {http://arxiv.org/abs/1807.06209} {arXiv:1807.06209 [astro-ph.CO]}
  \BibitemShut {NoStop}%
\bibitem [{\citenamefont {de~Blas}\ \emph {et~al.}(2018)\citenamefont
  {de~Blas}, \citenamefont {Criado}, \citenamefont {Perez-Victoria},\ and\
  \citenamefont {Santiago}}]{deBlas:2017xtg}%
  \BibitemOpen
  \bibfield  {author} {\bibinfo {author} {\bibfnamefont {J.}~\bibnamefont
  {de~Blas}}, \bibinfo {author} {\bibfnamefont {J.}~\bibnamefont {Criado}},
  \bibinfo {author} {\bibfnamefont {M.}~\bibnamefont {Perez-Victoria}}, \ and\
  \bibinfo {author} {\bibfnamefont {J.}~\bibnamefont {Santiago}},\ }\href
  {\doibase 10.1007/JHEP03(2018)109} {\bibfield  {journal} {\bibinfo  {journal}
  {JHEP}\ }\textbf {\bibinfo {volume} {03}},\ \bibinfo {pages} {109} (\bibinfo
  {year} {2018})},\ \Eprint {http://arxiv.org/abs/1711.10391} {arXiv:1711.10391
  [hep-ph]} \BibitemShut {NoStop}%
\bibitem [{\citenamefont {Herczeg}(2001)}]{Herczeg:2001vk}%
  \BibitemOpen
  \bibfield  {author} {\bibinfo {author} {\bibfnamefont {P.}~\bibnamefont
  {Herczeg}},\ }\href {\doibase 10.1016/S0146-6410(01)00149-1} {\bibfield
  {journal} {\bibinfo  {journal} {Prog. Part. Nucl. Phys.}\ }\textbf {\bibinfo
  {volume} {46}},\ \bibinfo {pages} {413} (\bibinfo {year} {2001})}\BibitemShut
  {NoStop}%
\bibitem [{\citenamefont {Ademollo}\ and\ \citenamefont
  {Gatto}(1964)}]{Ademollo:1964sr}%
  \BibitemOpen
  \bibfield  {author} {\bibinfo {author} {\bibfnamefont {M.}~\bibnamefont
  {Ademollo}}\ and\ \bibinfo {author} {\bibfnamefont {R.}~\bibnamefont
  {Gatto}},\ }\href {\doibase 10.1103/PhysRevLett.13.264} {\bibfield  {journal}
  {\bibinfo  {journal} {Phys. Rev. Lett.}\ }\textbf {\bibinfo {volume} {13}},\
  \bibinfo {pages} {264} (\bibinfo {year} {1964})}\BibitemShut {NoStop}%
\bibitem [{\citenamefont {Aoki}\ \emph {et~al.}(2020)\citenamefont {Aoki} \emph
  {et~al.}}]{Aoki:2019cca}%
  \BibitemOpen
  \bibfield  {author} {\bibinfo {author} {\bibfnamefont {S.}~\bibnamefont
  {Aoki}} \emph {et~al.} (\bibinfo {collaboration} {Flavour Lattice Averaging
  Group}),\ }\href {\doibase 10.1140/epjc/s10052-019-7354-7} {\bibfield
  {journal} {\bibinfo  {journal} {Eur. Phys. J. C}\ }\textbf {\bibinfo {volume}
  {80}},\ \bibinfo {pages} {113} (\bibinfo {year} {2020})},\ \Eprint
  {http://arxiv.org/abs/1902.08191} {arXiv:1902.08191 [hep-lat]} \BibitemShut
  {NoStop}%
\bibitem [{\citenamefont {Chang}\ \emph {et~al.}(2018)\citenamefont {Chang}
  \emph {et~al.}}]{Chang:2018uxx}%
  \BibitemOpen
  \bibfield  {author} {\bibinfo {author} {\bibfnamefont {C.}~\bibnamefont
  {Chang}} \emph {et~al.},\ }\href {\doibase 10.1038/s41586-018-0161-8}
  {\bibfield  {journal} {\bibinfo  {journal} {Nature}\ }\textbf {\bibinfo
  {volume} {558}},\ \bibinfo {pages} {91} (\bibinfo {year} {2018})},\ \Eprint
  {http://arxiv.org/abs/1805.12130} {arXiv:1805.12130 [hep-lat]} \BibitemShut
  {NoStop}%
\bibitem [{\citenamefont {Gupta}\ \emph {et~al.}(2018)\citenamefont {Gupta},
  \citenamefont {Jang}, \citenamefont {Yoon}, \citenamefont {Lin},
  \citenamefont {Cirigliano},\ and\ \citenamefont
  {Bhattacharya}}]{Gupta:2018qil}%
  \BibitemOpen
  \bibfield  {author} {\bibinfo {author} {\bibfnamefont {R.}~\bibnamefont
  {Gupta}}, \bibinfo {author} {\bibfnamefont {Y.-C.}\ \bibnamefont {Jang}},
  \bibinfo {author} {\bibfnamefont {B.}~\bibnamefont {Yoon}}, \bibinfo {author}
  {\bibfnamefont {H.-W.}\ \bibnamefont {Lin}}, \bibinfo {author} {\bibfnamefont
  {V.}~\bibnamefont {Cirigliano}}, \ and\ \bibinfo {author} {\bibfnamefont
  {T.}~\bibnamefont {Bhattacharya}},\ }\href {\doibase
  10.1103/PhysRevD.98.034503} {\bibfield  {journal} {\bibinfo  {journal} {Phys.
  Rev.}\ }\textbf {\bibinfo {volume} {D98}},\ \bibinfo {pages} {034503}
  (\bibinfo {year} {2018})},\ \Eprint {http://arxiv.org/abs/1806.09006}
  {arXiv:1806.09006 [hep-lat]} \BibitemShut {NoStop}%
\bibitem [{\citenamefont {Gonzalez-Alonso}\ and\ \citenamefont
  {Martin~Camalich}(2014)}]{Gonzalez-Alonso:2013ura}%
  \BibitemOpen
  \bibfield  {author} {\bibinfo {author} {\bibfnamefont {M.}~\bibnamefont
  {Gonzalez-Alonso}}\ and\ \bibinfo {author} {\bibfnamefont {J.}~\bibnamefont
  {Martin~Camalich}},\ }\href {\doibase 10.1103/PhysRevLett.112.042501}
  {\bibfield  {journal} {\bibinfo  {journal} {Phys. Rev. Lett.}\ }\textbf
  {\bibinfo {volume} {112}},\ \bibinfo {pages} {042501} (\bibinfo {year}
  {2014})},\ \Eprint {http://arxiv.org/abs/1309.4434} {arXiv:1309.4434
  [hep-ph]} \BibitemShut {NoStop}%
\bibitem [{\citenamefont {Seng}\ \emph {et~al.}(2018)\citenamefont {Seng},
  \citenamefont {Gorchtein}, \citenamefont {Patel},\ and\ \citenamefont
  {Ramsey-Musolf}}]{Seng:2018yzq}%
  \BibitemOpen
  \bibfield  {author} {\bibinfo {author} {\bibfnamefont {C.-Y.}\ \bibnamefont
  {Seng}}, \bibinfo {author} {\bibfnamefont {M.}~\bibnamefont {Gorchtein}},
  \bibinfo {author} {\bibfnamefont {H.~H.}\ \bibnamefont {Patel}}, \ and\
  \bibinfo {author} {\bibfnamefont {M.~J.}\ \bibnamefont {Ramsey-Musolf}},\
  }\href {\doibase 10.1103/PhysRevLett.121.241804} {\bibfield  {journal}
  {\bibinfo  {journal} {Phys. Rev. Lett.}\ }\textbf {\bibinfo {volume} {121}},\
  \bibinfo {pages} {241804} (\bibinfo {year} {2018})},\ \Eprint
  {http://arxiv.org/abs/1807.10197} {arXiv:1807.10197 [hep-ph]} \BibitemShut
  {NoStop}%
\bibitem [{\citenamefont {Czarnecki}\ \emph {et~al.}(2019)\citenamefont
  {Czarnecki}, \citenamefont {Marciano},\ and\ \citenamefont
  {Sirlin}}]{Czarnecki:2019mwq}%
  \BibitemOpen
  \bibfield  {author} {\bibinfo {author} {\bibfnamefont {A.}~\bibnamefont
  {Czarnecki}}, \bibinfo {author} {\bibfnamefont {W.~J.}\ \bibnamefont
  {Marciano}}, \ and\ \bibinfo {author} {\bibfnamefont {A.}~\bibnamefont
  {Sirlin}},\ }\href {\doibase 10.1103/PhysRevD.100.073008} {\bibfield
  {journal} {\bibinfo  {journal} {Phys. Rev. D}\ }\textbf {\bibinfo {volume}
  {100}},\ \bibinfo {pages} {073008} (\bibinfo {year} {2019})},\ \Eprint
  {http://arxiv.org/abs/1907.06737} {arXiv:1907.06737 [hep-ph]} \BibitemShut
  {NoStop}%
\bibitem [{\citenamefont {Seng}\ \emph {et~al.}(2020)\citenamefont {Seng},
  \citenamefont {Feng}, \citenamefont {Gorchtein},\ and\ \citenamefont
  {Jin}}]{Seng:2020wjq}%
  \BibitemOpen
  \bibfield  {author} {\bibinfo {author} {\bibfnamefont {C.-Y.}\ \bibnamefont
  {Seng}}, \bibinfo {author} {\bibfnamefont {X.}~\bibnamefont {Feng}}, \bibinfo
  {author} {\bibfnamefont {M.}~\bibnamefont {Gorchtein}}, \ and\ \bibinfo
  {author} {\bibfnamefont {L.-C.}\ \bibnamefont {Jin}},\ }\href {\doibase
  10.1103/PhysRevD.101.111301} {\bibfield  {journal} {\bibinfo  {journal}
  {Phys. Rev. D}\ }\textbf {\bibinfo {volume} {101}},\ \bibinfo {pages}
  {111301} (\bibinfo {year} {2020})},\ \Eprint
  {http://arxiv.org/abs/2003.11264} {arXiv:2003.11264 [hep-ph]} \BibitemShut
  {NoStop}%
\bibitem [{\citenamefont {Hayen}(2021)}]{Hayen:2020cxh}%
  \BibitemOpen
  \bibfield  {author} {\bibinfo {author} {\bibfnamefont {L.}~\bibnamefont
  {Hayen}},\ }\href {\doibase 10.1103/PhysRevD.103.113001} {\bibfield
  {journal} {\bibinfo  {journal} {Phys. Rev. D}\ }\textbf {\bibinfo {volume}
  {103}},\ \bibinfo {pages} {113001} (\bibinfo {year} {2021})},\ \Eprint
  {http://arxiv.org/abs/2010.07262} {arXiv:2010.07262 [hep-ph]} \BibitemShut
  {NoStop}%
\bibitem [{\citenamefont {Jackson}\ \emph
  {et~al.}(1957{\natexlab{b}})\citenamefont {Jackson}, \citenamefont
  {Treiman},\ and\ \citenamefont {Wyld}}]{Jackson:1957zz}%
  \BibitemOpen
  \bibfield  {author} {\bibinfo {author} {\bibfnamefont {J.~D.}\ \bibnamefont
  {Jackson}}, \bibinfo {author} {\bibfnamefont {S.~B.}\ \bibnamefont
  {Treiman}}, \ and\ \bibinfo {author} {\bibfnamefont {H.~W.}\ \bibnamefont
  {Wyld}},\ }\href {\doibase 10.1103/PhysRev.106.517} {\bibfield  {journal}
  {\bibinfo  {journal} {Phys. Rev.}\ }\textbf {\bibinfo {volume} {106}},\
  \bibinfo {pages} {517} (\bibinfo {year} {1957}{\natexlab{b}})}\BibitemShut
  {NoStop}%
\bibitem [{\citenamefont {Holstein}(1974)}]{Holstein:1974zf}%
  \BibitemOpen
  \bibfield  {author} {\bibinfo {author} {\bibfnamefont {B.~R.}\ \bibnamefont
  {Holstein}},\ }\href {\doibase 10.1103/RevModPhys.46.789} {\bibfield
  {journal} {\bibinfo  {journal} {Rev. Mod. Phys.}\ }\textbf {\bibinfo {volume}
  {46}},\ \bibinfo {pages} {789} (\bibinfo {year} {1974})},\ \bibinfo {note}
  {[Erratum: Rev. Mod. Phys.48,673(1976)]}\BibitemShut {NoStop}%
\bibitem [{\citenamefont {Hayen}\ \emph {et~al.}(2018)\citenamefont {Hayen},
  \citenamefont {Severijns}, \citenamefont {Bodek}, \citenamefont {Rozpedzik},\
  and\ \citenamefont {Mougeot}}]{Hayen:2017pwg}%
  \BibitemOpen
  \bibfield  {author} {\bibinfo {author} {\bibfnamefont {L.}~\bibnamefont
  {Hayen}}, \bibinfo {author} {\bibfnamefont {N.}~\bibnamefont {Severijns}},
  \bibinfo {author} {\bibfnamefont {K.}~\bibnamefont {Bodek}}, \bibinfo
  {author} {\bibfnamefont {D.}~\bibnamefont {Rozpedzik}}, \ and\ \bibinfo
  {author} {\bibfnamefont {X.}~\bibnamefont {Mougeot}},\ }\href {\doibase
  10.1103/RevModPhys.90.015008} {\bibfield  {journal} {\bibinfo  {journal}
  {Rev. Mod. Phys.}\ }\textbf {\bibinfo {volume} {90}},\ \bibinfo {pages}
  {015008} (\bibinfo {year} {2018})},\ \Eprint
  {http://arxiv.org/abs/1709.07530} {arXiv:1709.07530 [nucl-th]} \BibitemShut
  {NoStop}%
\bibitem [{\citenamefont {Hayen}\ and\ \citenamefont
  {Young}(2020)}]{Hayen:2020nej}%
  \BibitemOpen
  \bibfield  {author} {\bibinfo {author} {\bibfnamefont {L.}~\bibnamefont
  {Hayen}}\ and\ \bibinfo {author} {\bibfnamefont {A.~R.}\ \bibnamefont
  {Young}},\ }\href@noop {} {\  (\bibinfo {year} {2020})},\ \Eprint
  {http://arxiv.org/abs/2009.11364} {arXiv:2009.11364 [nucl-th]} \BibitemShut
  {NoStop}%
\bibitem [{\citenamefont {Hardy}\ and\ \citenamefont
  {Towner}(2005)}]{Hardy:2004id}%
  \BibitemOpen
  \bibfield  {author} {\bibinfo {author} {\bibfnamefont {J.~C.}\ \bibnamefont
  {Hardy}}\ and\ \bibinfo {author} {\bibfnamefont {I.~S.}\ \bibnamefont
  {Towner}},\ }\href {\doibase 10.1103/PhysRevC.71.055501} {\bibfield
  {journal} {\bibinfo  {journal} {Phys. Rev.}\ }\textbf {\bibinfo {volume}
  {C71}},\ \bibinfo {pages} {055501} (\bibinfo {year} {2005})},\ \Eprint
  {http://arxiv.org/abs/nucl-th/0412056} {arXiv:nucl-th/0412056} \BibitemShut
  {NoStop}%
\bibitem [{\citenamefont {Hayen}\ and\ \citenamefont
  {Severijns}(2019)}]{Hayen:2019nic}%
  \BibitemOpen
  \bibfield  {author} {\bibinfo {author} {\bibfnamefont {L.}~\bibnamefont
  {Hayen}}\ and\ \bibinfo {author} {\bibfnamefont {N.}~\bibnamefont
  {Severijns}},\ }\href@noop {} {\  (\bibinfo {year} {2019})},\ \Eprint
  {http://arxiv.org/abs/1906.09870} {arXiv:1906.09870 [nucl-th]} \BibitemShut
  {NoStop}%
\bibitem [{\citenamefont {Czarnecki}\ \emph {et~al.}(2018)\citenamefont
  {Czarnecki}, \citenamefont {Marciano},\ and\ \citenamefont
  {Sirlin}}]{Czarnecki:2018okw}%
  \BibitemOpen
  \bibfield  {author} {\bibinfo {author} {\bibfnamefont {A.}~\bibnamefont
  {Czarnecki}}, \bibinfo {author} {\bibfnamefont {W.~J.}\ \bibnamefont
  {Marciano}}, \ and\ \bibinfo {author} {\bibfnamefont {A.}~\bibnamefont
  {Sirlin}},\ }\href {\doibase 10.1103/PhysRevLett.120.202002} {\bibfield
  {journal} {\bibinfo  {journal} {Phys. Rev. Lett.}\ }\textbf {\bibinfo
  {volume} {120}},\ \bibinfo {pages} {202002} (\bibinfo {year} {2018})},\
  \Eprint {http://arxiv.org/abs/1802.01804} {arXiv:1802.01804 [hep-ph]}
  \BibitemShut {NoStop}%
\bibitem [{\citenamefont {Towner}\ and\ \citenamefont
  {Hardy}(2010)}]{Towner:2010zz}%
  \BibitemOpen
  \bibfield  {author} {\bibinfo {author} {\bibfnamefont {I.~S.}\ \bibnamefont
  {Towner}}\ and\ \bibinfo {author} {\bibfnamefont {J.~C.}\ \bibnamefont
  {Hardy}},\ }\href {\doibase 10.1088/0034-4885/73/4/046301} {\bibfield
  {journal} {\bibinfo  {journal} {Rept. Prog. Phys.}\ }\textbf {\bibinfo
  {volume} {73}},\ \bibinfo {pages} {046301} (\bibinfo {year}
  {2010})}\BibitemShut {NoStop}%
\bibitem [{\citenamefont {Wilkinson}(1982)}]{Wilkinson:1982hu}%
  \BibitemOpen
  \bibfield  {author} {\bibinfo {author} {\bibfnamefont {D.~H.}\ \bibnamefont
  {Wilkinson}},\ }\href {\doibase 10.1016/0375-9474(82)90051-3} {\bibfield
  {journal} {\bibinfo  {journal} {Nucl. Phys.}\ }\textbf {\bibinfo {volume}
  {A377}},\ \bibinfo {pages} {474} (\bibinfo {year} {1982})}\BibitemShut
  {NoStop}%
\bibitem [{\citenamefont {Gonzalez-Alonso}\ and\ \citenamefont
  {Naviliat-Cuncic}(2016)}]{Gonzalez-Alonso:2016jzm}%
  \BibitemOpen
  \bibfield  {author} {\bibinfo {author} {\bibfnamefont {M.}~\bibnamefont
  {Gonzalez-Alonso}}\ and\ \bibinfo {author} {\bibfnamefont {O.}~\bibnamefont
  {Naviliat-Cuncic}},\ }\href {\doibase 10.1103/PhysRevC.94.035503} {\bibfield
  {journal} {\bibinfo  {journal} {Phys. Rev.}\ }\textbf {\bibinfo {volume}
  {C94}},\ \bibinfo {pages} {035503} (\bibinfo {year} {2016})},\ \Eprint
  {http://arxiv.org/abs/1607.08347} {arXiv:1607.08347 [nucl-ex]} \BibitemShut
  {NoStop}%
\bibitem [{\citenamefont {IAEA-database}()}]{IAEAdatabase}%
  \BibitemOpen
  \bibfield  {author} {\bibinfo {author} {\bibnamefont {IAEA-database}},\
  }\href@noop {} {}\bibinfo {note} {{\tt
  https://www-nds.iaea.org/nuclearmoments/}}\BibitemShut {NoStop}%
\bibitem [{\citenamefont {Fenker}\ \emph {et~al.}(2018)\citenamefont {Fenker}
  \emph {et~al.}}]{Fenker:2017rcx}%
  \BibitemOpen
  \bibfield  {author} {\bibinfo {author} {\bibfnamefont {B.}~\bibnamefont
  {Fenker}} \emph {et~al.},\ }\href {\doibase 10.1103/PhysRevLett.120.062502}
  {\bibfield  {journal} {\bibinfo  {journal} {Phys. Rev. Lett.}\ }\textbf
  {\bibinfo {volume} {120}},\ \bibinfo {pages} {062502} (\bibinfo {year}
  {2018})},\ \Eprint {http://arxiv.org/abs/1706.00414} {arXiv:1706.00414
  [nucl-ex]} \BibitemShut {NoStop}%
\bibitem [{\citenamefont {Vetter}\ \emph {et~al.}(2008)\citenamefont {Vetter},
  \citenamefont {Abo-Shaeer}, \citenamefont {Freedman},\ and\ \citenamefont
  {Maruyama}}]{Vetter:2008zz}%
  \BibitemOpen
  \bibfield  {author} {\bibinfo {author} {\bibfnamefont {P.}~\bibnamefont
  {Vetter}}, \bibinfo {author} {\bibfnamefont {J.}~\bibnamefont {Abo-Shaeer}},
  \bibinfo {author} {\bibfnamefont {S.}~\bibnamefont {Freedman}}, \ and\
  \bibinfo {author} {\bibfnamefont {R.}~\bibnamefont {Maruyama}},\ }\href
  {\doibase 10.1103/PhysRevC.77.035502} {\bibfield  {journal} {\bibinfo
  {journal} {Phys. Rev. C}\ }\textbf {\bibinfo {volume} {77}},\ \bibinfo
  {pages} {035502} (\bibinfo {year} {2008})},\ \Eprint
  {http://arxiv.org/abs/0805.1212} {arXiv:0805.1212 [nucl-ex]} \BibitemShut
  {NoStop}%
\bibitem [{\citenamefont {Hayen}()}]{Hayen:private}%
  \BibitemOpen
  \bibfield  {author} {\bibinfo {author} {\bibfnamefont {L.}~\bibnamefont
  {Hayen}},\ }\href@noop {} {}\bibinfo {howpublished} {Private
  communication}\BibitemShut {NoStop}%
\bibitem [{\citenamefont {Melconian}\ \emph {et~al.}(2007)\citenamefont
  {Melconian} \emph {et~al.}}]{Melconian:2007zz}%
  \BibitemOpen
  \bibfield  {author} {\bibinfo {author} {\bibfnamefont {D.}~\bibnamefont
  {Melconian}} \emph {et~al.},\ }\href {\doibase
  10.1016/j.physletb.2007.04.047} {\bibfield  {journal} {\bibinfo  {journal}
  {Phys. Lett. B}\ }\textbf {\bibinfo {volume} {649}},\ \bibinfo {pages} {370}
  (\bibinfo {year} {2007})}\BibitemShut {NoStop}%
\bibitem [{\citenamefont {Combs}\ \emph {et~al.}(2020)\citenamefont {Combs},
  \citenamefont {Jones}, \citenamefont {Anderson}, \citenamefont {Calaprice},
  \citenamefont {Hayen},\ and\ \citenamefont {Young}}]{Combs:2020ttz}%
  \BibitemOpen
  \bibfield  {author} {\bibinfo {author} {\bibfnamefont {D.}~\bibnamefont
  {Combs}}, \bibinfo {author} {\bibfnamefont {G.}~\bibnamefont {Jones}},
  \bibinfo {author} {\bibfnamefont {W.}~\bibnamefont {Anderson}}, \bibinfo
  {author} {\bibfnamefont {F.}~\bibnamefont {Calaprice}}, \bibinfo {author}
  {\bibfnamefont {L.}~\bibnamefont {Hayen}}, \ and\ \bibinfo {author}
  {\bibfnamefont {A.}~\bibnamefont {Young}},\ }\href@noop {} {\  (\bibinfo
  {year} {2020})},\ \Eprint {http://arxiv.org/abs/2009.13700} {arXiv:2009.13700
  [nucl-ex]} \BibitemShut {NoStop}%
\bibitem [{\citenamefont {Shidling}\ \emph {et~al.}(2014)\citenamefont
  {Shidling} \emph {et~al.}}]{Shidling:2014ura}%
  \BibitemOpen
  \bibfield  {author} {\bibinfo {author} {\bibfnamefont {P.~D.}\ \bibnamefont
  {Shidling}} \emph {et~al.},\ }\href {\doibase 10.1103/PhysRevC.90.032501}
  {\bibfield  {journal} {\bibinfo  {journal} {Phys. Rev.}\ }\textbf {\bibinfo
  {volume} {C90}},\ \bibinfo {pages} {032501} (\bibinfo {year} {2014})},\
  \Eprint {http://arxiv.org/abs/1407.1742} {arXiv:1407.1742 [nucl-ex]}
  \BibitemShut {NoStop}%
\bibitem [{\citenamefont {Rebeiro}\ \emph {et~al.}(2019)\citenamefont {Rebeiro}
  \emph {et~al.}}]{Rebeiro:2018lwo}%
  \BibitemOpen
  \bibfield  {author} {\bibinfo {author} {\bibfnamefont {B.}~\bibnamefont
  {Rebeiro}} \emph {et~al.},\ }\href {\doibase 10.1103/PhysRevC.99.065502}
  {\bibfield  {journal} {\bibinfo  {journal} {Phys. Rev. C}\ }\textbf {\bibinfo
  {volume} {99}},\ \bibinfo {pages} {065502} (\bibinfo {year} {2019})},\
  \Eprint {http://arxiv.org/abs/1810.02331} {arXiv:1810.02331 [nucl-ex]}
  \BibitemShut {NoStop}%
\bibitem [{\citenamefont {Karthein}\ \emph {et~al.}(2019)\citenamefont
  {Karthein} \emph {et~al.}}]{Karthein:2019bss}%
  \BibitemOpen
  \bibfield  {author} {\bibinfo {author} {\bibfnamefont {J.}~\bibnamefont
  {Karthein}} \emph {et~al.},\ }\href {\doibase 10.1103/PhysRevC.100.015502}
  {\bibfield  {journal} {\bibinfo  {journal} {Phys. Rev. C}\ }\textbf {\bibinfo
  {volume} {100}},\ \bibinfo {pages} {015502} (\bibinfo {year} {2019})},\
  \bibinfo {note} {[Erratum: Phys.Rev.C 101, 049901 (2020)]},\ \Eprint
  {http://arxiv.org/abs/1906.01538} {arXiv:1906.01538 [nucl-ex]} \BibitemShut
  {NoStop}%
\bibitem [{\citenamefont {Wang}\ \emph {et~al.}(2017)\citenamefont {Wang},
  \citenamefont {Audi}, \citenamefont {Kondev}, \citenamefont {Huang},
  \citenamefont {Naimi},\ and\ \citenamefont {Xu}}]{Wang_2017}%
  \BibitemOpen
  \bibfield  {author} {\bibinfo {author} {\bibfnamefont {M.}~\bibnamefont
  {Wang}}, \bibinfo {author} {\bibfnamefont {G.}~\bibnamefont {Audi}}, \bibinfo
  {author} {\bibfnamefont {F.~G.}\ \bibnamefont {Kondev}}, \bibinfo {author}
  {\bibfnamefont {W.}~\bibnamefont {Huang}}, \bibinfo {author} {\bibfnamefont
  {S.}~\bibnamefont {Naimi}}, \ and\ \bibinfo {author} {\bibfnamefont
  {X.}~\bibnamefont {Xu}},\ }\href {\doibase 10.1088/1674-1137/41/3/030003}
  {\bibfield  {journal} {\bibinfo  {journal} {Chinese Physics C}\ }\textbf
  {\bibinfo {volume} {41}},\ \bibinfo {pages} {030003} (\bibinfo {year}
  {2017})}\BibitemShut {NoStop}%
\bibitem [{\citenamefont {Brodeur}\ \emph {et~al.}(2016)\citenamefont {Brodeur}
  \emph {et~al.}}]{Brodeur:2016spm}%
  \BibitemOpen
  \bibfield  {author} {\bibinfo {author} {\bibfnamefont {M.}~\bibnamefont
  {Brodeur}} \emph {et~al.},\ }\href {\doibase 10.1103/PhysRevC.93.025503}
  {\bibfield  {journal} {\bibinfo  {journal} {Phys. Rev. C}\ }\textbf {\bibinfo
  {volume} {93}},\ \bibinfo {pages} {025503} (\bibinfo {year}
  {2016})}\BibitemShut {NoStop}%
\bibitem [{\citenamefont {Severijns}\ \emph {et~al.}(1989)\citenamefont
  {Severijns}, \citenamefont {Wouters}, \citenamefont {Vanhaverbeke},\ and\
  \citenamefont {Vanneste}}]{PhysRevLett.63.1050}%
  \BibitemOpen
  \bibfield  {author} {\bibinfo {author} {\bibfnamefont {N.}~\bibnamefont
  {Severijns}}, \bibinfo {author} {\bibfnamefont {J.}~\bibnamefont {Wouters}},
  \bibinfo {author} {\bibfnamefont {J.}~\bibnamefont {Vanhaverbeke}}, \ and\
  \bibinfo {author} {\bibfnamefont {L.}~\bibnamefont {Vanneste}},\ }\href
  {\doibase 10.1103/PhysRevLett.63.1050} {\bibfield  {journal} {\bibinfo
  {journal} {Phys. Rev. Lett.}\ }\textbf {\bibinfo {volume} {63}},\ \bibinfo
  {pages} {1050} (\bibinfo {year} {1989})}\BibitemShut {NoStop}%
\bibitem [{\citenamefont {Calaprice}\ \emph {et~al.}(1975)\citenamefont
  {Calaprice}, \citenamefont {Freedman}, \citenamefont {Mead},\ and\
  \citenamefont {Vantine}}]{Calaprice:1975zz}%
  \BibitemOpen
  \bibfield  {author} {\bibinfo {author} {\bibfnamefont {F.}~\bibnamefont
  {Calaprice}}, \bibinfo {author} {\bibfnamefont {S.}~\bibnamefont {Freedman}},
  \bibinfo {author} {\bibfnamefont {W.}~\bibnamefont {Mead}}, \ and\ \bibinfo
  {author} {\bibfnamefont {H.}~\bibnamefont {Vantine}},\ }\href {\doibase
  10.1103/PhysRevLett.35.1566} {\bibfield  {journal} {\bibinfo  {journal}
  {Phys. Rev. Lett.}\ }\textbf {\bibinfo {volume} {35}},\ \bibinfo {pages}
  {1566} (\bibinfo {year} {1975})}\BibitemShut {NoStop}%
\bibitem [{\citenamefont {Long}\ \emph {et~al.}(2020)\citenamefont {Long} \emph
  {et~al.}}]{Long:2020lby}%
  \BibitemOpen
  \bibfield  {author} {\bibinfo {author} {\bibfnamefont {J.}~\bibnamefont
  {Long}} \emph {et~al.},\ }\href {\doibase 10.1103/PhysRevC.101.015501}
  {\bibfield  {journal} {\bibinfo  {journal} {Phys. Rev. C}\ }\textbf {\bibinfo
  {volume} {101}},\ \bibinfo {pages} {015501} (\bibinfo {year}
  {2020})}\BibitemShut {NoStop}%
\bibitem [{\citenamefont {Masson}\ and\ \citenamefont
  {Quin}(1990)}]{Masson:1990zz}%
  \BibitemOpen
  \bibfield  {author} {\bibinfo {author} {\bibfnamefont {G.}~\bibnamefont
  {Masson}}\ and\ \bibinfo {author} {\bibfnamefont {P.}~\bibnamefont {Quin}},\
  }\href {\doibase 10.1103/PhysRevC.42.1110} {\bibfield  {journal} {\bibinfo
  {journal} {Phys. Rev. C}\ }\textbf {\bibinfo {volume} {42}},\ \bibinfo
  {pages} {1110} (\bibinfo {year} {1990})}\BibitemShut {NoStop}%
\bibitem [{\citenamefont {Garnett}\ \emph {et~al.}(1988)\citenamefont
  {Garnett}, \citenamefont {Commins}, \citenamefont {Lesko},\ and\
  \citenamefont {Norman}}]{Garnett:1987gw}%
  \BibitemOpen
  \bibfield  {author} {\bibinfo {author} {\bibfnamefont {J.}~\bibnamefont
  {Garnett}}, \bibinfo {author} {\bibfnamefont {E.}~\bibnamefont {Commins}},
  \bibinfo {author} {\bibfnamefont {K.}~\bibnamefont {Lesko}}, \ and\ \bibinfo
  {author} {\bibfnamefont {E.}~\bibnamefont {Norman}},\ }\href {\doibase
  10.1103/PhysRevLett.60.499} {\bibfield  {journal} {\bibinfo  {journal} {Phys.
  Rev. Lett.}\ }\textbf {\bibinfo {volume} {60}},\ \bibinfo {pages} {499}
  (\bibinfo {year} {1988})}\BibitemShut {NoStop}%
\bibitem [{\citenamefont {Converse}\ \emph {et~al.}(1993)\citenamefont
  {Converse} \emph {et~al.}}]{Converse:1993ba}%
  \BibitemOpen
  \bibfield  {author} {\bibinfo {author} {\bibfnamefont {A.}~\bibnamefont
  {Converse}} \emph {et~al.},\ }\href {\doibase 10.1016/0370-2693(93)91400-H}
  {\bibfield  {journal} {\bibinfo  {journal} {Phys. Lett. B}\ }\textbf
  {\bibinfo {volume} {304}},\ \bibinfo {pages} {60} (\bibinfo {year}
  {1993})}\BibitemShut {NoStop}%
\bibitem [{\citenamefont {Markisch}\ \emph {et~al.}(2019)\citenamefont
  {Markisch} \emph {et~al.}}]{Markisch:2018ndu}%
  \BibitemOpen
  \bibfield  {author} {\bibinfo {author} {\bibfnamefont {B.}~\bibnamefont
  {Markisch}} \emph {et~al.},\ }\href {\doibase 10.1103/PhysRevLett.122.242501}
  {\bibfield  {journal} {\bibinfo  {journal} {Phys. Rev. Lett.}\ }\textbf
  {\bibinfo {volume} {122}},\ \bibinfo {pages} {242501} (\bibinfo {year}
  {2019})},\ \Eprint {http://arxiv.org/abs/1812.04666} {arXiv:1812.04666
  [nucl-ex]} \BibitemShut {NoStop}%
\bibitem [{\citenamefont {Stratowa}\ \emph {et~al.}(1978)\citenamefont
  {Stratowa}, \citenamefont {Dobrozemsky},\ and\ \citenamefont
  {Weinzierl}}]{Stratowa:1978gq}%
  \BibitemOpen
  \bibfield  {author} {\bibinfo {author} {\bibfnamefont {C.}~\bibnamefont
  {Stratowa}}, \bibinfo {author} {\bibfnamefont {R.}~\bibnamefont
  {Dobrozemsky}}, \ and\ \bibinfo {author} {\bibfnamefont {P.}~\bibnamefont
  {Weinzierl}},\ }\href {\doibase 10.1103/PhysRevD.18.3970} {\bibfield
  {journal} {\bibinfo  {journal} {Phys. Rev. D}\ }\textbf {\bibinfo {volume}
  {18}},\ \bibinfo {pages} {3970} (\bibinfo {year} {1978})}\BibitemShut
  {NoStop}%
\bibitem [{\citenamefont {Byrne}\ \emph {et~al.}(2002)\citenamefont {Byrne},
  \citenamefont {Dawber}, \citenamefont {van~der Grinten}, \citenamefont
  {Habeck}, \citenamefont {Shaikh}, \citenamefont {Spain}, \citenamefont
  {Scott}, \citenamefont {Baker}, \citenamefont {Green},\ and\ \citenamefont
  {Zimmer}}]{Byrne:2002tx}%
  \BibitemOpen
  \bibfield  {author} {\bibinfo {author} {\bibfnamefont {J.}~\bibnamefont
  {Byrne}}, \bibinfo {author} {\bibfnamefont {P.}~\bibnamefont {Dawber}},
  \bibinfo {author} {\bibfnamefont {M.~D.}\ \bibnamefont {van~der Grinten}},
  \bibinfo {author} {\bibfnamefont {C.}~\bibnamefont {Habeck}}, \bibinfo
  {author} {\bibfnamefont {F.}~\bibnamefont {Shaikh}}, \bibinfo {author}
  {\bibfnamefont {J.}~\bibnamefont {Spain}}, \bibinfo {author} {\bibfnamefont
  {R.}~\bibnamefont {Scott}}, \bibinfo {author} {\bibfnamefont
  {C.}~\bibnamefont {Baker}}, \bibinfo {author} {\bibfnamefont
  {K.}~\bibnamefont {Green}}, \ and\ \bibinfo {author} {\bibfnamefont
  {O.}~\bibnamefont {Zimmer}},\ }\href {\doibase 10.1088/0954-3899/28/6/314}
  {\bibfield  {journal} {\bibinfo  {journal} {J. Phys. G}\ }\textbf {\bibinfo
  {volume} {28}},\ \bibinfo {pages} {1325} (\bibinfo {year}
  {2002})}\BibitemShut {NoStop}%
\bibitem [{\citenamefont {Seng}\ \emph {et~al.}(2019)\citenamefont {Seng},
  \citenamefont {Gorchtein},\ and\ \citenamefont
  {Ramsey-Musolf}}]{Seng:2018qru}%
  \BibitemOpen
  \bibfield  {author} {\bibinfo {author} {\bibfnamefont {C.~Y.}\ \bibnamefont
  {Seng}}, \bibinfo {author} {\bibfnamefont {M.}~\bibnamefont {Gorchtein}}, \
  and\ \bibinfo {author} {\bibfnamefont {M.~J.}\ \bibnamefont
  {Ramsey-Musolf}},\ }\href {\doibase 10.1103/PhysRevD.100.013001} {\bibfield
  {journal} {\bibinfo  {journal} {Phys. Rev. D}\ }\textbf {\bibinfo {volume}
  {100}},\ \bibinfo {pages} {013001} (\bibinfo {year} {2019})},\ \Eprint
  {http://arxiv.org/abs/1812.03352} {arXiv:1812.03352 [nucl-th]} \BibitemShut
  {NoStop}%
\bibitem [{\citenamefont {Gorchtein}(2019)}]{Gorchtein:2018fxl}%
  \BibitemOpen
  \bibfield  {author} {\bibinfo {author} {\bibfnamefont {M.}~\bibnamefont
  {Gorchtein}},\ }\href {\doibase 10.1103/PhysRevLett.123.042503} {\bibfield
  {journal} {\bibinfo  {journal} {Phys. Rev. Lett.}\ }\textbf {\bibinfo
  {volume} {123}},\ \bibinfo {pages} {042503} (\bibinfo {year} {2019})},\
  \Eprint {http://arxiv.org/abs/1812.04229} {arXiv:1812.04229 [nucl-th]}
  \BibitemShut {NoStop}%
\bibitem [{\citenamefont {Hardy}\ and\ \citenamefont
  {Towner}(2020)}]{Hardy:2020qwl}%
  \BibitemOpen
  \bibfield  {author} {\bibinfo {author} {\bibfnamefont {J.}~\bibnamefont
  {Hardy}}\ and\ \bibinfo {author} {\bibfnamefont {I.}~\bibnamefont {Towner}},\
  }\href {\doibase 10.1103/PhysRevC.102.045501} {\bibfield  {journal} {\bibinfo
   {journal} {Phys. Rev. C}\ }\textbf {\bibinfo {volume} {102}},\ \bibinfo
  {pages} {045501} (\bibinfo {year} {2020})}\BibitemShut {NoStop}%
\bibitem [{\citenamefont {Towner}\ and\ \citenamefont
  {Hardy}(2008)}]{Towner:2007np}%
  \BibitemOpen
  \bibfield  {author} {\bibinfo {author} {\bibfnamefont {I.~S.}\ \bibnamefont
  {Towner}}\ and\ \bibinfo {author} {\bibfnamefont {J.~C.}\ \bibnamefont
  {Hardy}},\ }\href {\doibase 10.1103/PhysRevC.77.025501} {\bibfield  {journal}
  {\bibinfo  {journal} {Phys. Rev.}\ }\textbf {\bibinfo {volume} {C77}},\
  \bibinfo {pages} {025501} (\bibinfo {year} {2008})},\ \Eprint
  {http://arxiv.org/abs/0710.3181} {arXiv:0710.3181 [nucl-th]} \BibitemShut
  {NoStop}%
\bibitem [{\citenamefont {Hardy}\ and\ \citenamefont
  {Towner}(2015)}]{Hardy:2014qxa}%
  \BibitemOpen
  \bibfield  {author} {\bibinfo {author} {\bibfnamefont {J.~C.}\ \bibnamefont
  {Hardy}}\ and\ \bibinfo {author} {\bibfnamefont {I.~S.}\ \bibnamefont
  {Towner}},\ }\href {\doibase 10.1103/PhysRevC.91.025501} {\bibfield
  {journal} {\bibinfo  {journal} {Phys. Rev.}\ }\textbf {\bibinfo {volume}
  {C91}},\ \bibinfo {pages} {025501} (\bibinfo {year} {2015})},\ \Eprint
  {http://arxiv.org/abs/1411.5987} {arXiv:1411.5987 [nucl-ex]} \BibitemShut
  {NoStop}%
\bibitem [{\citenamefont {Towner}\ and\ \citenamefont
  {Hardy}(2002)}]{Towner:2002rg}%
  \BibitemOpen
  \bibfield  {author} {\bibinfo {author} {\bibfnamefont {I.}~\bibnamefont
  {Towner}}\ and\ \bibinfo {author} {\bibfnamefont {J.}~\bibnamefont {Hardy}},\
  }\href {\doibase 10.1103/PhysRevC.66.035501} {\bibfield  {journal} {\bibinfo
  {journal} {Phys. Rev. C}\ }\textbf {\bibinfo {volume} {66}},\ \bibinfo
  {pages} {035501} (\bibinfo {year} {2002})},\ \Eprint
  {http://arxiv.org/abs/nucl-th/0209014} {arXiv:nucl-th/0209014} \BibitemShut
  {NoStop}%
\bibitem [{\citenamefont {Towner}(1994)}]{Towner:1994mw}%
  \BibitemOpen
  \bibfield  {author} {\bibinfo {author} {\bibfnamefont {I.}~\bibnamefont
  {Towner}},\ }\href {\doibase 10.1016/0370-2693(94)91000-6} {\bibfield
  {journal} {\bibinfo  {journal} {Phys. Lett. B}\ }\textbf {\bibinfo {volume}
  {333}},\ \bibinfo {pages} {13} (\bibinfo {year} {1994})},\ \Eprint
  {http://arxiv.org/abs/nucl-th/9405031} {arXiv:nucl-th/9405031} \BibitemShut
  {NoStop}%
\bibitem [{\citenamefont {Towner}(1992)}]{Towner:1992xm}%
  \BibitemOpen
  \bibfield  {author} {\bibinfo {author} {\bibfnamefont {I.}~\bibnamefont
  {Towner}},\ }\href {\doibase 10.1016/0375-9474(92)90170-O} {\bibfield
  {journal} {\bibinfo  {journal} {Nucl. Phys. A}\ }\textbf {\bibinfo {volume}
  {540}},\ \bibinfo {pages} {478} (\bibinfo {year} {1992})}\BibitemShut
  {NoStop}%
\bibitem [{\citenamefont {Hardy}\ and\ \citenamefont
  {Towner}(2018)}]{Hardy:2018zsb}%
  \BibitemOpen
  \bibfield  {author} {\bibinfo {author} {\bibfnamefont {J.}~\bibnamefont
  {Hardy}}\ and\ \bibinfo {author} {\bibfnamefont {I.}~\bibnamefont {Towner}},\
  }in\ \href@noop {} {\emph {\bibinfo {booktitle} {{13th Conference on the
  Intersections of Particle and Nuclear Physics}}}}\ (\bibinfo {year} {2018})\
  \Eprint {http://arxiv.org/abs/1807.01146} {arXiv:1807.01146 [nucl-ex]}
  \BibitemShut {NoStop}%
\bibitem [{\citenamefont {Feng}\ \emph {et~al.}(2020)\citenamefont {Feng},
  \citenamefont {Gorchtein}, \citenamefont {Jin}, \citenamefont {Ma},\ and\
  \citenamefont {Seng}}]{Feng:2020zdc}%
  \BibitemOpen
  \bibfield  {author} {\bibinfo {author} {\bibfnamefont {X.}~\bibnamefont
  {Feng}}, \bibinfo {author} {\bibfnamefont {M.}~\bibnamefont {Gorchtein}},
  \bibinfo {author} {\bibfnamefont {L.-C.}\ \bibnamefont {Jin}}, \bibinfo
  {author} {\bibfnamefont {P.-X.}\ \bibnamefont {Ma}}, \ and\ \bibinfo {author}
  {\bibfnamefont {C.-Y.}\ \bibnamefont {Seng}},\ }\href {\doibase
  10.1103/PhysRevLett.124.192002} {\bibfield  {journal} {\bibinfo  {journal}
  {Phys. Rev. Lett.}\ }\textbf {\bibinfo {volume} {124}},\ \bibinfo {pages}
  {192002} (\bibinfo {year} {2020})},\ \Eprint
  {http://arxiv.org/abs/2003.09798} {arXiv:2003.09798 [hep-lat]} \BibitemShut
  {NoStop}%
\bibitem [{\citenamefont {Pocanic}\ \emph {et~al.}(2004)\citenamefont {Pocanic}
  \emph {et~al.}}]{Pocanic:2003pf}%
  \BibitemOpen
  \bibfield  {author} {\bibinfo {author} {\bibfnamefont {D.}~\bibnamefont
  {Pocanic}} \emph {et~al.},\ }\href {\doibase 10.1103/PhysRevLett.93.181803}
  {\bibfield  {journal} {\bibinfo  {journal} {Phys. Rev. Lett.}\ }\textbf
  {\bibinfo {volume} {93}},\ \bibinfo {pages} {181803} (\bibinfo {year}
  {2004})},\ \Eprint {http://arxiv.org/abs/hep-ex/0312030}
  {arXiv:hep-ex/0312030 [hep-ex]} \BibitemShut {NoStop}%
\bibitem [{\citenamefont {Marciano}\ and\ \citenamefont
  {Sirlin}(2006)}]{Marciano:2005ec}%
  \BibitemOpen
  \bibfield  {author} {\bibinfo {author} {\bibfnamefont {W.~J.}\ \bibnamefont
  {Marciano}}\ and\ \bibinfo {author} {\bibfnamefont {A.}~\bibnamefont
  {Sirlin}},\ }\href {\doibase 10.1103/PhysRevLett.96.032002} {\bibfield
  {journal} {\bibinfo  {journal} {Phys. Rev. Lett.}\ }\textbf {\bibinfo
  {volume} {96}},\ \bibinfo {pages} {032002} (\bibinfo {year} {2006})},\
  \Eprint {http://arxiv.org/abs/hep-ph/0510099} {arXiv:hep-ph/0510099 [hep-ph]}
  \BibitemShut {NoStop}%
\bibitem [{\citenamefont {Jenkins}\ \emph {et~al.}(2018)\citenamefont
  {Jenkins}, \citenamefont {Manohar},\ and\ \citenamefont
  {Stoffer}}]{Jenkins:2017jig}%
  \BibitemOpen
  \bibfield  {author} {\bibinfo {author} {\bibfnamefont {E.~E.}\ \bibnamefont
  {Jenkins}}, \bibinfo {author} {\bibfnamefont {A.~V.}\ \bibnamefont
  {Manohar}}, \ and\ \bibinfo {author} {\bibfnamefont {P.}~\bibnamefont
  {Stoffer}},\ }\href {\doibase 10.1007/JHEP03(2018)016} {\bibfield  {journal}
  {\bibinfo  {journal} {JHEP}\ }\textbf {\bibinfo {volume} {03}},\ \bibinfo
  {pages} {016} (\bibinfo {year} {2018})},\ \Eprint
  {http://arxiv.org/abs/1709.04486} {arXiv:1709.04486 [hep-ph]} \BibitemShut
  {NoStop}%
\bibitem [{\citenamefont {Dekens}\ and\ \citenamefont
  {Stoffer}(2019)}]{Dekens:2019ept}%
  \BibitemOpen
  \bibfield  {author} {\bibinfo {author} {\bibfnamefont {W.}~\bibnamefont
  {Dekens}}\ and\ \bibinfo {author} {\bibfnamefont {P.}~\bibnamefont
  {Stoffer}},\ }\href {\doibase 10.1007/JHEP10(2019)197} {\bibfield  {journal}
  {\bibinfo  {journal} {JHEP}\ }\textbf {\bibinfo {volume} {10}},\ \bibinfo
  {pages} {197} (\bibinfo {year} {2019})},\ \Eprint
  {http://arxiv.org/abs/1908.05295} {arXiv:1908.05295 [hep-ph]} \BibitemShut
  {NoStop}%
\bibitem [{\citenamefont {Gonz\'alez-Alonso}\ \emph {et~al.}(2017)\citenamefont
  {Gonz\'alez-Alonso}, \citenamefont {Martin~Camalich},\ and\ \citenamefont
  {Mimouni}}]{Gonzalez-Alonso:2017iyc}%
  \BibitemOpen
  \bibfield  {author} {\bibinfo {author} {\bibfnamefont {M.}~\bibnamefont
  {Gonz\'alez-Alonso}}, \bibinfo {author} {\bibfnamefont {J.}~\bibnamefont
  {Martin~Camalich}}, \ and\ \bibinfo {author} {\bibfnamefont {K.}~\bibnamefont
  {Mimouni}},\ }\href {\doibase 10.1016/j.physletb.2017.07.003} {\bibfield
  {journal} {\bibinfo  {journal} {Phys. Lett. B}\ }\textbf {\bibinfo {volume}
  {772}},\ \bibinfo {pages} {777} (\bibinfo {year} {2017})},\ \Eprint
  {http://arxiv.org/abs/1706.00410} {arXiv:1706.00410 [hep-ph]} \BibitemShut
  {NoStop}%
\bibitem [{\citenamefont {Cirigliano}\ \emph {et~al.}(2010)\citenamefont
  {Cirigliano}, \citenamefont {Jenkins},\ and\ \citenamefont
  {Gonzalez-Alonso}}]{Cirigliano:2009wk}%
  \BibitemOpen
  \bibfield  {author} {\bibinfo {author} {\bibfnamefont {V.}~\bibnamefont
  {Cirigliano}}, \bibinfo {author} {\bibfnamefont {J.}~\bibnamefont {Jenkins}},
  \ and\ \bibinfo {author} {\bibfnamefont {M.}~\bibnamefont
  {Gonzalez-Alonso}},\ }\href {\doibase 10.1016/j.nuclphysb.2009.12.020}
  {\bibfield  {journal} {\bibinfo  {journal} {Nucl. Phys.}\ }\textbf {\bibinfo
  {volume} {B830}},\ \bibinfo {pages} {95} (\bibinfo {year} {2010})},\ \Eprint
  {http://arxiv.org/abs/0908.1754} {arXiv:0908.1754 [hep-ph]} \BibitemShut
  {NoStop}%
\bibitem [{\citenamefont {Gonzalez-Alonso}\ and\ \citenamefont
  {Martin~Camalich}(2016)}]{Gonzalez-Alonso:2016etj}%
  \BibitemOpen
  \bibfield  {author} {\bibinfo {author} {\bibfnamefont {M.}~\bibnamefont
  {Gonzalez-Alonso}}\ and\ \bibinfo {author} {\bibfnamefont {J.}~\bibnamefont
  {Martin~Camalich}},\ }\href {\doibase 10.1007/JHEP12(2016)052} {\bibfield
  {journal} {\bibinfo  {journal} {JHEP}\ }\textbf {\bibinfo {volume} {12}},\
  \bibinfo {pages} {052} (\bibinfo {year} {2016})},\ \Eprint
  {http://arxiv.org/abs/1605.07114} {arXiv:1605.07114 [hep-ph]} \BibitemShut
  {NoStop}%
\bibitem [{\citenamefont {Bhattacharya}\ \emph {et~al.}(2012)\citenamefont
  {Bhattacharya}, \citenamefont {Cirigliano}, \citenamefont {Cohen},
  \citenamefont {Filipuzzi}, \citenamefont {Gonzalez-Alonso}, \citenamefont
  {Graesser}, \citenamefont {Gupta},\ and\ \citenamefont
  {Lin}}]{Bhattacharya:2011qm}%
  \BibitemOpen
  \bibfield  {author} {\bibinfo {author} {\bibfnamefont {T.}~\bibnamefont
  {Bhattacharya}}, \bibinfo {author} {\bibfnamefont {V.}~\bibnamefont
  {Cirigliano}}, \bibinfo {author} {\bibfnamefont {S.~D.}\ \bibnamefont
  {Cohen}}, \bibinfo {author} {\bibfnamefont {A.}~\bibnamefont {Filipuzzi}},
  \bibinfo {author} {\bibfnamefont {M.}~\bibnamefont {Gonzalez-Alonso}},
  \bibinfo {author} {\bibfnamefont {M.~L.}\ \bibnamefont {Graesser}}, \bibinfo
  {author} {\bibfnamefont {R.}~\bibnamefont {Gupta}}, \ and\ \bibinfo {author}
  {\bibfnamefont {H.-W.}\ \bibnamefont {Lin}},\ }\href {\doibase
  10.1103/PhysRevD.85.054512} {\bibfield  {journal} {\bibinfo  {journal} {Phys.
  Rev.}\ }\textbf {\bibinfo {volume} {D85}},\ \bibinfo {pages} {054512}
  (\bibinfo {year} {2012})},\ \Eprint {http://arxiv.org/abs/1110.6448}
  {arXiv:1110.6448 [hep-ph]} \BibitemShut {NoStop}%
\bibitem [{\citenamefont {Walker-Loud}\ \emph {et~al.}(2020)\citenamefont
  {Walker-Loud} \emph {et~al.}}]{Walker-Loud:2019cif}%
  \BibitemOpen
  \bibfield  {author} {\bibinfo {author} {\bibfnamefont {A.}~\bibnamefont
  {Walker-Loud}} \emph {et~al.},\ }\href {\doibase 10.22323/1.317.0020}
  {\bibfield  {journal} {\bibinfo  {journal} {PoS}\ }\textbf {\bibinfo {volume}
  {CD2018}},\ \bibinfo {pages} {020} (\bibinfo {year} {2020})},\ \Eprint
  {http://arxiv.org/abs/1912.08321} {arXiv:1912.08321 [hep-lat]} \BibitemShut
  {NoStop}%
\bibitem [{\citenamefont {Cirigliano}\ \emph
  {et~al.}(2013{\natexlab{b}})\citenamefont {Cirigliano}, \citenamefont
  {Gonzalez-Alonso},\ and\ \citenamefont {Graesser}}]{Cirigliano:2012ab}%
  \BibitemOpen
  \bibfield  {author} {\bibinfo {author} {\bibfnamefont {V.}~\bibnamefont
  {Cirigliano}}, \bibinfo {author} {\bibfnamefont {M.}~\bibnamefont
  {Gonzalez-Alonso}}, \ and\ \bibinfo {author} {\bibfnamefont {M.~L.}\
  \bibnamefont {Graesser}},\ }\href {\doibase 10.1007/JHEP02(2013)046}
  {\bibfield  {journal} {\bibinfo  {journal} {JHEP}\ }\textbf {\bibinfo
  {volume} {02}},\ \bibinfo {pages} {046} (\bibinfo {year}
  {2013}{\natexlab{b}})},\ \Eprint {http://arxiv.org/abs/1210.4553}
  {arXiv:1210.4553 [hep-ph]} \BibitemShut {NoStop}%
\bibitem [{\citenamefont {Hassan}\ \emph {et~al.}(2021)\citenamefont {Hassan}
  \emph {et~al.}}]{Hassan:2020hrj}%
  \BibitemOpen
  \bibfield  {author} {\bibinfo {author} {\bibfnamefont {M.~T.}\ \bibnamefont
  {Hassan}} \emph {et~al.},\ }\href {\doibase 10.1103/PhysRevC.103.045502}
  {\bibfield  {journal} {\bibinfo  {journal} {Phys. Rev. C}\ }\textbf {\bibinfo
  {volume} {103}},\ \bibinfo {pages} {045502} (\bibinfo {year} {2021})},\
  \Eprint {http://arxiv.org/abs/2012.14379} {arXiv:2012.14379 [nucl-ex]}
  \BibitemShut {NoStop}%
\bibitem [{\citenamefont {Cirigliano}\ \emph {et~al.}(2019)\citenamefont
  {Cirigliano}, \citenamefont {Garcia}, \citenamefont {Gazit}, \citenamefont
  {Naviliat-Cuncic}, \citenamefont {Savard},\ and\ \citenamefont
  {Young}}]{Cirgiliano:2019nyn}%
  \BibitemOpen
  \bibfield  {author} {\bibinfo {author} {\bibfnamefont {V.}~\bibnamefont
  {Cirigliano}}, \bibinfo {author} {\bibfnamefont {A.}~\bibnamefont {Garcia}},
  \bibinfo {author} {\bibfnamefont {D.}~\bibnamefont {Gazit}}, \bibinfo
  {author} {\bibfnamefont {O.}~\bibnamefont {Naviliat-Cuncic}}, \bibinfo
  {author} {\bibfnamefont {G.}~\bibnamefont {Savard}}, \ and\ \bibinfo {author}
  {\bibfnamefont {A.}~\bibnamefont {Young}},\ }\href@noop {} {\  (\bibinfo
  {year} {2019})},\ \Eprint {http://arxiv.org/abs/1907.02164} {arXiv:1907.02164
  [nucl-ex]} \BibitemShut {NoStop}%
\bibitem [{\citenamefont {Mampe}\ \emph {et~al.}(1993)\citenamefont {Mampe},
  \citenamefont {Bondarenko}, \citenamefont {Morozov}, \citenamefont {Panin},\
  and\ \citenamefont {Fomin}}]{Mampe:1993an}%
  \BibitemOpen
  \bibfield  {author} {\bibinfo {author} {\bibfnamefont {W.}~\bibnamefont
  {Mampe}}, \bibinfo {author} {\bibfnamefont {L.}~\bibnamefont {Bondarenko}},
  \bibinfo {author} {\bibfnamefont {V.}~\bibnamefont {Morozov}}, \bibinfo
  {author} {\bibfnamefont {Y.}~\bibnamefont {Panin}}, \ and\ \bibinfo {author}
  {\bibfnamefont {A.}~\bibnamefont {Fomin}},\ }\href@noop {} {\bibfield
  {journal} {\bibinfo  {journal} {JETP Lett.}\ }\textbf {\bibinfo {volume}
  {57}},\ \bibinfo {pages} {82} (\bibinfo {year} {1993})}\BibitemShut {NoStop}%
\bibitem [{\citenamefont {Byrne}\ and\ \citenamefont
  {Dawber}(1996)}]{Byrne:1996zz}%
  \BibitemOpen
  \bibfield  {author} {\bibinfo {author} {\bibfnamefont {J.}~\bibnamefont
  {Byrne}}\ and\ \bibinfo {author} {\bibfnamefont {P.}~\bibnamefont {Dawber}},\
  }\href {\doibase 10.1209/epl/i1996-00319-x} {\bibfield  {journal} {\bibinfo
  {journal} {Europhys. Lett.}\ }\textbf {\bibinfo {volume} {33}},\ \bibinfo
  {pages} {187} (\bibinfo {year} {1996})}\BibitemShut {NoStop}%
\bibitem [{\citenamefont {Serebrov}\ \emph {et~al.}(2005)\citenamefont
  {Serebrov} \emph {et~al.}}]{Serebrov:2004zf}%
  \BibitemOpen
  \bibfield  {author} {\bibinfo {author} {\bibfnamefont {A.}~\bibnamefont
  {Serebrov}} \emph {et~al.},\ }\href {\doibase 10.1016/j.physletb.2004.11.013}
  {\bibfield  {journal} {\bibinfo  {journal} {Phys. Lett. B}\ }\textbf
  {\bibinfo {volume} {605}},\ \bibinfo {pages} {72} (\bibinfo {year} {2005})},\
  \Eprint {http://arxiv.org/abs/nucl-ex/0408009} {arXiv:nucl-ex/0408009}
  \BibitemShut {NoStop}%
\bibitem [{\citenamefont {Pichlmaier}\ \emph {et~al.}(2010)\citenamefont
  {Pichlmaier}, \citenamefont {Varlamov}, \citenamefont {Schreckenbach},\ and\
  \citenamefont {Geltenbort}}]{Pichlmaier:2010zz}%
  \BibitemOpen
  \bibfield  {author} {\bibinfo {author} {\bibfnamefont {A.}~\bibnamefont
  {Pichlmaier}}, \bibinfo {author} {\bibfnamefont {V.}~\bibnamefont
  {Varlamov}}, \bibinfo {author} {\bibfnamefont {K.}~\bibnamefont
  {Schreckenbach}}, \ and\ \bibinfo {author} {\bibfnamefont {P.}~\bibnamefont
  {Geltenbort}},\ }\href {\doibase 10.1016/j.physletb.2010.08.032} {\bibfield
  {journal} {\bibinfo  {journal} {Phys. Lett. B}\ }\textbf {\bibinfo {volume}
  {693}},\ \bibinfo {pages} {221} (\bibinfo {year} {2010})}\BibitemShut
  {NoStop}%
\bibitem [{\citenamefont {Steyerl}\ \emph {et~al.}(2012)\citenamefont
  {Steyerl}, \citenamefont {Pendlebury}, \citenamefont {Kaufman}, \citenamefont
  {Malik},\ and\ \citenamefont {Desai}}]{Steyerl:2012zz}%
  \BibitemOpen
  \bibfield  {author} {\bibinfo {author} {\bibfnamefont {A.}~\bibnamefont
  {Steyerl}}, \bibinfo {author} {\bibfnamefont {J.}~\bibnamefont {Pendlebury}},
  \bibinfo {author} {\bibfnamefont {C.}~\bibnamefont {Kaufman}}, \bibinfo
  {author} {\bibfnamefont {S.}~\bibnamefont {Malik}}, \ and\ \bibinfo {author}
  {\bibfnamefont {A.}~\bibnamefont {Desai}},\ }\href {\doibase
  10.1103/PhysRevC.85.065503} {\bibfield  {journal} {\bibinfo  {journal} {Phys.
  Rev. C}\ }\textbf {\bibinfo {volume} {85}},\ \bibinfo {pages} {065503}
  (\bibinfo {year} {2012})}\BibitemShut {NoStop}%
\bibitem [{\citenamefont {Yue}\ \emph {et~al.}(2013)\citenamefont {Yue},
  \citenamefont {Dewey}, \citenamefont {Gilliam}, \citenamefont {Greene},
  \citenamefont {Laptev}, \citenamefont {Nico}, \citenamefont {Snow},\ and\
  \citenamefont {Wietfeldt}}]{Yue:2013qrc}%
  \BibitemOpen
  \bibfield  {author} {\bibinfo {author} {\bibfnamefont {A.}~\bibnamefont
  {Yue}}, \bibinfo {author} {\bibfnamefont {M.}~\bibnamefont {Dewey}}, \bibinfo
  {author} {\bibfnamefont {D.}~\bibnamefont {Gilliam}}, \bibinfo {author}
  {\bibfnamefont {G.}~\bibnamefont {Greene}}, \bibinfo {author} {\bibfnamefont
  {A.}~\bibnamefont {Laptev}}, \bibinfo {author} {\bibfnamefont
  {J.}~\bibnamefont {Nico}}, \bibinfo {author} {\bibfnamefont {W.}~\bibnamefont
  {Snow}}, \ and\ \bibinfo {author} {\bibfnamefont {F.}~\bibnamefont
  {Wietfeldt}},\ }\href {\doibase 10.1103/PhysRevLett.111.222501} {\bibfield
  {journal} {\bibinfo  {journal} {Phys. Rev. Lett.}\ }\textbf {\bibinfo
  {volume} {111}},\ \bibinfo {pages} {222501} (\bibinfo {year} {2013})},\
  \Eprint {http://arxiv.org/abs/1309.2623} {arXiv:1309.2623 [nucl-ex]}
  \BibitemShut {NoStop}%
\bibitem [{\citenamefont {Ezhov}\ \emph {et~al.}(2018)\citenamefont {Ezhov}
  \emph {et~al.}}]{Ezhov:2014tna}%
  \BibitemOpen
  \bibfield  {author} {\bibinfo {author} {\bibfnamefont {V.}~\bibnamefont
  {Ezhov}} \emph {et~al.},\ }\href {\doibase 10.1134/S0021364018110024}
  {\bibfield  {journal} {\bibinfo  {journal} {JETP Lett.}\ }\textbf {\bibinfo
  {volume} {107}},\ \bibinfo {pages} {671} (\bibinfo {year} {2018})},\ \Eprint
  {http://arxiv.org/abs/1412.7434} {arXiv:1412.7434 [nucl-ex]} \BibitemShut
  {NoStop}%
\bibitem [{\citenamefont {Arzumanov}\ \emph {et~al.}(2015)\citenamefont
  {Arzumanov}, \citenamefont {Bondarenko}, \citenamefont {Chernyavsky},
  \citenamefont {Geltenbort}, \citenamefont {Morozov}, \citenamefont
  {Nesvizhevsky}, \citenamefont {Panin},\ and\ \citenamefont
  {Strepetov}}]{Arzumanov:2015tea}%
  \BibitemOpen
  \bibfield  {author} {\bibinfo {author} {\bibfnamefont {S.}~\bibnamefont
  {Arzumanov}}, \bibinfo {author} {\bibfnamefont {L.}~\bibnamefont
  {Bondarenko}}, \bibinfo {author} {\bibfnamefont {S.}~\bibnamefont
  {Chernyavsky}}, \bibinfo {author} {\bibfnamefont {P.}~\bibnamefont
  {Geltenbort}}, \bibinfo {author} {\bibfnamefont {V.}~\bibnamefont {Morozov}},
  \bibinfo {author} {\bibfnamefont {V.}~\bibnamefont {Nesvizhevsky}}, \bibinfo
  {author} {\bibfnamefont {Y.}~\bibnamefont {Panin}}, \ and\ \bibinfo {author}
  {\bibfnamefont {A.}~\bibnamefont {Strepetov}},\ }\href {\doibase
  10.1016/j.physletb.2015.04.021} {\bibfield  {journal} {\bibinfo  {journal}
  {Phys. Lett. B}\ }\textbf {\bibinfo {volume} {745}},\ \bibinfo {pages} {79}
  (\bibinfo {year} {2015})}\BibitemShut {NoStop}%
\bibitem [{\citenamefont {Pattie}\ \emph {et~al.}(2018)\citenamefont {Pattie}
  \emph {et~al.}}]{Pattie:2017vsj}%
  \BibitemOpen
  \bibfield  {author} {\bibinfo {author} {\bibfnamefont {J.}~\bibnamefont
  {Pattie}, \bibfnamefont {R.W.}} \emph {et~al.},\ }\href {\doibase
  10.1126/science.aan8895} {\bibfield  {journal} {\bibinfo  {journal}
  {Science}\ }\textbf {\bibinfo {volume} {360}},\ \bibinfo {pages} {627}
  (\bibinfo {year} {2018})},\ \Eprint {http://arxiv.org/abs/1707.01817}
  {arXiv:1707.01817 [nucl-ex]} \BibitemShut {NoStop}%
\bibitem [{\citenamefont {Serebrov}\ \emph {et~al.}(2018)\citenamefont
  {Serebrov} \emph {et~al.}}]{Serebrov:2017bzo}%
  \BibitemOpen
  \bibfield  {author} {\bibinfo {author} {\bibfnamefont {A.}~\bibnamefont
  {Serebrov}} \emph {et~al.},\ }\href {\doibase 10.1103/PhysRevC.97.055503}
  {\bibfield  {journal} {\bibinfo  {journal} {Phys. Rev. C}\ }\textbf {\bibinfo
  {volume} {97}},\ \bibinfo {pages} {055503} (\bibinfo {year} {2018})},\
  \Eprint {http://arxiv.org/abs/1712.05663} {arXiv:1712.05663 [nucl-ex]}
  \BibitemShut {NoStop}%
\bibitem [{\citenamefont {Bopp}\ \emph {et~al.}(1986)\citenamefont {Bopp},
  \citenamefont {Dubbers}, \citenamefont {Hornig}, \citenamefont {Klemt},
  \citenamefont {Last}, \citenamefont {Schutze}, \citenamefont {Freedman},\
  and\ \citenamefont {Scharpf}}]{Bopp:1986rt}%
  \BibitemOpen
  \bibfield  {author} {\bibinfo {author} {\bibfnamefont {P.}~\bibnamefont
  {Bopp}}, \bibinfo {author} {\bibfnamefont {D.}~\bibnamefont {Dubbers}},
  \bibinfo {author} {\bibfnamefont {L.}~\bibnamefont {Hornig}}, \bibinfo
  {author} {\bibfnamefont {E.}~\bibnamefont {Klemt}}, \bibinfo {author}
  {\bibfnamefont {J.}~\bibnamefont {Last}}, \bibinfo {author} {\bibfnamefont
  {H.}~\bibnamefont {Schutze}}, \bibinfo {author} {\bibfnamefont
  {S.}~\bibnamefont {Freedman}}, \ and\ \bibinfo {author} {\bibfnamefont
  {O.}~\bibnamefont {Scharpf}},\ }\href {\doibase 10.1103/PhysRevLett.56.919}
  {\bibfield  {journal} {\bibinfo  {journal} {Phys. Rev. Lett.}\ }\textbf
  {\bibinfo {volume} {56}},\ \bibinfo {pages} {919} (\bibinfo {year} {1986})},\
  \bibinfo {note} {[Erratum: Phys.Rev.Lett. 57, 1192 (1986)]}\BibitemShut
  {NoStop}%
\bibitem [{\citenamefont {Liaud}\ \emph {et~al.}(1997)\citenamefont {Liaud},
  \citenamefont {Schreckenbach}, \citenamefont {Kossakowski}, \citenamefont
  {Nastoll}, \citenamefont {Bussiere}, \citenamefont {Guillaud},\ and\
  \citenamefont {Beck}}]{Liaud:1997vu}%
  \BibitemOpen
  \bibfield  {author} {\bibinfo {author} {\bibfnamefont {P.}~\bibnamefont
  {Liaud}}, \bibinfo {author} {\bibfnamefont {K.}~\bibnamefont
  {Schreckenbach}}, \bibinfo {author} {\bibfnamefont {R.}~\bibnamefont
  {Kossakowski}}, \bibinfo {author} {\bibfnamefont {H.}~\bibnamefont
  {Nastoll}}, \bibinfo {author} {\bibfnamefont {A.}~\bibnamefont {Bussiere}},
  \bibinfo {author} {\bibfnamefont {J.}~\bibnamefont {Guillaud}}, \ and\
  \bibinfo {author} {\bibfnamefont {L.}~\bibnamefont {Beck}},\ }\href {\doibase
  10.1016/S0375-9474(96)00325-9} {\bibfield  {journal} {\bibinfo  {journal}
  {Nucl. Phys. A}\ }\textbf {\bibinfo {volume} {612}},\ \bibinfo {pages} {53}
  (\bibinfo {year} {1997})}\BibitemShut {NoStop}%
\bibitem [{\citenamefont {Yerozolimsky}\ \emph {et~al.}(1997)\citenamefont
  {Yerozolimsky}, \citenamefont {Kuznetsov}, \citenamefont {Mostovoy},\ and\
  \citenamefont {Stepanenko}}]{Erozolimsky:1997wi}%
  \BibitemOpen
  \bibfield  {author} {\bibinfo {author} {\bibfnamefont {B.}~\bibnamefont
  {Yerozolimsky}}, \bibinfo {author} {\bibfnamefont {I.}~\bibnamefont
  {Kuznetsov}}, \bibinfo {author} {\bibfnamefont {Y.}~\bibnamefont {Mostovoy}},
  \ and\ \bibinfo {author} {\bibfnamefont {I.}~\bibnamefont {Stepanenko}},\
  }\href {\doibase https://doi.org/10.1016/S0370-2693(97)01004-6} {\bibfield
  {journal} {\bibinfo  {journal} {Phys. Lett. B}\ }\textbf {\bibinfo {volume}
  {412}},\ \bibinfo {pages} {240 } (\bibinfo {year} {1997})}\BibitemShut
  {NoStop}%
\bibitem [{\citenamefont {Mund}\ \emph {et~al.}(2013)\citenamefont {Mund},
  \citenamefont {Maerkisch}, \citenamefont {Deissenroth}, \citenamefont
  {Krempel}, \citenamefont {Schumann}, \citenamefont {Abele}, \citenamefont
  {Petoukhov},\ and\ \citenamefont {Soldner}}]{Mund:2012fq}%
  \BibitemOpen
  \bibfield  {author} {\bibinfo {author} {\bibfnamefont {D.}~\bibnamefont
  {Mund}}, \bibinfo {author} {\bibfnamefont {B.}~\bibnamefont {Maerkisch}},
  \bibinfo {author} {\bibfnamefont {M.}~\bibnamefont {Deissenroth}}, \bibinfo
  {author} {\bibfnamefont {J.}~\bibnamefont {Krempel}}, \bibinfo {author}
  {\bibfnamefont {M.}~\bibnamefont {Schumann}}, \bibinfo {author}
  {\bibfnamefont {H.}~\bibnamefont {Abele}}, \bibinfo {author} {\bibfnamefont
  {A.}~\bibnamefont {Petoukhov}}, \ and\ \bibinfo {author} {\bibfnamefont
  {T.}~\bibnamefont {Soldner}},\ }\href {\doibase
  10.1103/PhysRevLett.110.172502} {\bibfield  {journal} {\bibinfo  {journal}
  {Phys. Rev. Lett.}\ }\textbf {\bibinfo {volume} {110}},\ \bibinfo {pages}
  {172502} (\bibinfo {year} {2013})},\ \Eprint {http://arxiv.org/abs/1204.0013}
  {arXiv:1204.0013 [hep-ex]} \BibitemShut {NoStop}%
\bibitem [{\citenamefont {Brown}\ \emph {et~al.}(2018)\citenamefont {Brown}
  \emph {et~al.}}]{Brown:2017mhw}%
  \BibitemOpen
  \bibfield  {author} {\bibinfo {author} {\bibfnamefont {M.-P.}\ \bibnamefont
  {Brown}} \emph {et~al.} (\bibinfo {collaboration} {UCNA}),\ }\href {\doibase
  10.1103/PhysRevC.97.035505} {\bibfield  {journal} {\bibinfo  {journal} {Phys.
  Rev. C}\ }\textbf {\bibinfo {volume} {97}},\ \bibinfo {pages} {035505}
  (\bibinfo {year} {2018})},\ \Eprint {http://arxiv.org/abs/1712.00884}
  {arXiv:1712.00884 [nucl-ex]} \BibitemShut {NoStop}%
\bibitem [{\citenamefont {Kuznetsov}\ \emph {et~al.}(1995)\citenamefont
  {Kuznetsov}, \citenamefont {Serebrov}, \citenamefont {Stepanenko},
  \citenamefont {Aldushchenkov}, \citenamefont {Lasakov}, \citenamefont
  {Kokin}, \citenamefont {Mostovoi}, \citenamefont {Erozolimsky},\ and\
  \citenamefont {Dewey}}]{Kuznetsov:1995sk}%
  \BibitemOpen
  \bibfield  {author} {\bibinfo {author} {\bibfnamefont {I.}~\bibnamefont
  {Kuznetsov}}, \bibinfo {author} {\bibfnamefont {A.}~\bibnamefont {Serebrov}},
  \bibinfo {author} {\bibfnamefont {I.}~\bibnamefont {Stepanenko}}, \bibinfo
  {author} {\bibfnamefont {A.}~\bibnamefont {Aldushchenkov}}, \bibinfo {author}
  {\bibfnamefont {M.}~\bibnamefont {Lasakov}}, \bibinfo {author} {\bibfnamefont
  {A.}~\bibnamefont {Kokin}}, \bibinfo {author} {\bibfnamefont
  {Y.}~\bibnamefont {Mostovoi}}, \bibinfo {author} {\bibfnamefont
  {B.}~\bibnamefont {Erozolimsky}}, \ and\ \bibinfo {author} {\bibfnamefont
  {M.}~\bibnamefont {Dewey}},\ }\href {\doibase 10.1103/PhysRevLett.75.794}
  {\bibfield  {journal} {\bibinfo  {journal} {Phys. Rev. Lett.}\ }\textbf
  {\bibinfo {volume} {75}},\ \bibinfo {pages} {794} (\bibinfo {year}
  {1995})}\BibitemShut {NoStop}%
\bibitem [{\citenamefont {Serebrov}\ \emph {et~al.}(1998)\citenamefont
  {Serebrov} \emph {et~al.}}]{Serebrov:1998aj}%
  \BibitemOpen
  \bibfield  {author} {\bibinfo {author} {\bibfnamefont {A.}~\bibnamefont
  {Serebrov}} \emph {et~al.},\ }\href {\doibase 10.1134/1.558574} {\bibfield
  {journal} {\bibinfo  {journal} {J. Exp. Theor. Phys.}\ }\textbf {\bibinfo
  {volume} {86}},\ \bibinfo {pages} {1074} (\bibinfo {year}
  {1998})}\BibitemShut {NoStop}%
\bibitem [{\citenamefont {Kreuz}\ \emph {et~al.}(2005)\citenamefont {Kreuz}
  \emph {et~al.}}]{Kreuz:2005jz}%
  \BibitemOpen
  \bibfield  {author} {\bibinfo {author} {\bibfnamefont {M.}~\bibnamefont
  {Kreuz}} \emph {et~al.},\ }\href {\doibase 10.1016/j.physletb.2005.05.074}
  {\bibfield  {journal} {\bibinfo  {journal} {Phys. Lett. B}\ }\textbf
  {\bibinfo {volume} {619}},\ \bibinfo {pages} {263} (\bibinfo {year}
  {2005})}\BibitemShut {NoStop}%
\bibitem [{\citenamefont {Schumann}\ \emph {et~al.}(2007)\citenamefont
  {Schumann}, \citenamefont {Soldner}, \citenamefont {Deissenroth},
  \citenamefont {Gluck}, \citenamefont {Krempel}, \citenamefont {Kreuz},
  \citenamefont {Markisch}, \citenamefont {Mund}, \citenamefont {Petoukhov},\
  and\ \citenamefont {Abele}}]{Schumann:2007qe}%
  \BibitemOpen
  \bibfield  {author} {\bibinfo {author} {\bibfnamefont {M.}~\bibnamefont
  {Schumann}}, \bibinfo {author} {\bibfnamefont {T.}~\bibnamefont {Soldner}},
  \bibinfo {author} {\bibfnamefont {M.}~\bibnamefont {Deissenroth}}, \bibinfo
  {author} {\bibfnamefont {F.}~\bibnamefont {Gluck}}, \bibinfo {author}
  {\bibfnamefont {J.}~\bibnamefont {Krempel}}, \bibinfo {author} {\bibfnamefont
  {M.}~\bibnamefont {Kreuz}}, \bibinfo {author} {\bibfnamefont
  {B.}~\bibnamefont {Markisch}}, \bibinfo {author} {\bibfnamefont
  {D.}~\bibnamefont {Mund}}, \bibinfo {author} {\bibfnamefont {A.}~\bibnamefont
  {Petoukhov}}, \ and\ \bibinfo {author} {\bibfnamefont {H.}~\bibnamefont
  {Abele}},\ }\href {\doibase 10.1103/PhysRevLett.99.191803} {\bibfield
  {journal} {\bibinfo  {journal} {Phys. Rev. Lett.}\ }\textbf {\bibinfo
  {volume} {99}},\ \bibinfo {pages} {191803} (\bibinfo {year} {2007})},\
  \Eprint {http://arxiv.org/abs/0706.3788} {arXiv:0706.3788 [hep-ph]}
  \BibitemShut {NoStop}%
\bibitem [{\citenamefont {Mostovoi}\ \emph {et~al.}(2001)\citenamefont
  {Mostovoi} \emph {et~al.}}]{Mostovoi:2001ye}%
  \BibitemOpen
  \bibfield  {author} {\bibinfo {author} {\bibfnamefont {A.}~\bibnamefont
  {Mostovoi}} \emph {et~al.},\ }\href {\doibase 10.1134/1.1423745} {\bibfield
  {journal} {\bibinfo  {journal} {Phys. Atom. Nucl.}\ }\textbf {\bibinfo
  {volume} {64}},\ \bibinfo {pages} {1955} (\bibinfo {year} {2001})},\ \bibinfo
  {note} {[Yad. Fiz.64,2040(2001)]}\BibitemShut {NoStop}%
\bibitem [{\citenamefont {Darius}\ \emph {et~al.}(2017)\citenamefont {Darius}
  \emph {et~al.}}]{Darius:2017arh}%
  \BibitemOpen
  \bibfield  {author} {\bibinfo {author} {\bibfnamefont {G.}~\bibnamefont
  {Darius}} \emph {et~al.},\ }\href {\doibase 10.1103/PhysRevLett.119.042502}
  {\bibfield  {journal} {\bibinfo  {journal} {Phys. Rev. Lett.}\ }\textbf
  {\bibinfo {volume} {119}},\ \bibinfo {pages} {042502} (\bibinfo {year}
  {2017})}\BibitemShut {NoStop}%
\bibitem [{\citenamefont {Johnson}\ \emph {et~al.}(1963)\citenamefont
  {Johnson}, \citenamefont {Pleasonton},\ and\ \citenamefont
  {Carlson}}]{Johnson:1963zza}%
  \BibitemOpen
  \bibfield  {author} {\bibinfo {author} {\bibfnamefont {C.}~\bibnamefont
  {Johnson}}, \bibinfo {author} {\bibfnamefont {F.}~\bibnamefont {Pleasonton}},
  \ and\ \bibinfo {author} {\bibfnamefont {T.}~\bibnamefont {Carlson}},\ }\href
  {\doibase 10.1103/PhysRev.132.1149} {\bibfield  {journal} {\bibinfo
  {journal} {Phys. Rev.}\ }\textbf {\bibinfo {volume} {132}},\ \bibinfo {pages}
  {1149} (\bibinfo {year} {1963})}\BibitemShut {NoStop}%
\bibitem [{\citenamefont {Adelberger}\ \emph {et~al.}(1999)\citenamefont
  {Adelberger}, \citenamefont {Ortiz}, \citenamefont {Garcia}, \citenamefont
  {Swanson}, \citenamefont {Beck}, \citenamefont {Tengblad}, \citenamefont
  {Borge}, \citenamefont {Martel},\ and\ \citenamefont
  {Bichsel}}]{Adelberger:1999ud}%
  \BibitemOpen
  \bibfield  {author} {\bibinfo {author} {\bibfnamefont {E.}~\bibnamefont
  {Adelberger}}, \bibinfo {author} {\bibfnamefont {C.}~\bibnamefont {Ortiz}},
  \bibinfo {author} {\bibfnamefont {A.}~\bibnamefont {Garcia}}, \bibinfo
  {author} {\bibfnamefont {H.}~\bibnamefont {Swanson}}, \bibinfo {author}
  {\bibfnamefont {M.}~\bibnamefont {Beck}}, \bibinfo {author} {\bibfnamefont
  {O.}~\bibnamefont {Tengblad}}, \bibinfo {author} {\bibfnamefont
  {M.}~\bibnamefont {Borge}}, \bibinfo {author} {\bibfnamefont
  {I.}~\bibnamefont {Martel}}, \ and\ \bibinfo {author} {\bibfnamefont
  {H.}~\bibnamefont {Bichsel}} (\bibinfo {collaboration} {ISOLDE}),\ }\href
  {\doibase 10.1103/PhysRevLett.83.1299} {\bibfield  {journal} {\bibinfo
  {journal} {Phys. Rev. Lett.}\ }\textbf {\bibinfo {volume} {83}},\ \bibinfo
  {pages} {1299} (\bibinfo {year} {1999})},\ \bibinfo {note} {[Erratum:
  Phys.Rev.Lett. 83, 3101 (1999)]},\ \Eprint
  {http://arxiv.org/abs/nucl-ex/9903002} {arXiv:nucl-ex/9903002} \BibitemShut
  {NoStop}%
\bibitem [{\citenamefont {Gorelov}\ \emph {et~al.}(2005)\citenamefont {Gorelov}
  \emph {et~al.}}]{Gorelov:2004hv}%
  \BibitemOpen
  \bibfield  {author} {\bibinfo {author} {\bibfnamefont {A.}~\bibnamefont
  {Gorelov}} \emph {et~al.},\ }\href {\doibase 10.1103/PhysRevLett.94.142501}
  {\bibfield  {journal} {\bibinfo  {journal} {Phys. Rev. Lett.}\ }\textbf
  {\bibinfo {volume} {94}},\ \bibinfo {pages} {142501} (\bibinfo {year}
  {2005})},\ \Eprint {http://arxiv.org/abs/nucl-ex/0412032}
  {arXiv:nucl-ex/0412032} \BibitemShut {NoStop}%
\bibitem [{\citenamefont {Wauters}\ \emph {et~al.}(2010)\citenamefont {Wauters}
  \emph {et~al.}}]{Wauters:2010gh}%
  \BibitemOpen
  \bibfield  {author} {\bibinfo {author} {\bibfnamefont {F.}~\bibnamefont
  {Wauters}} \emph {et~al.},\ }\href {\doibase 10.1103/PhysRevC.82.055502}
  {\bibfield  {journal} {\bibinfo  {journal} {Phys. Rev. C}\ }\textbf {\bibinfo
  {volume} {82}},\ \bibinfo {pages} {055502} (\bibinfo {year} {2010})},\
  \Eprint {http://arxiv.org/abs/1005.5034} {arXiv:1005.5034 [nucl-ex]}
  \BibitemShut {NoStop}%
\bibitem [{\citenamefont {Soti}\ \emph {et~al.}(2014)\citenamefont {Soti} \emph
  {et~al.}}]{Soti:2014xua}%
  \BibitemOpen
  \bibfield  {author} {\bibinfo {author} {\bibfnamefont {G.}~\bibnamefont
  {Soti}} \emph {et~al.},\ }\href {\doibase 10.1103/PhysRevC.90.035502}
  {\bibfield  {journal} {\bibinfo  {journal} {Phys. Rev. C}\ }\textbf {\bibinfo
  {volume} {90}},\ \bibinfo {pages} {035502} (\bibinfo {year} {2014})},\
  \Eprint {http://arxiv.org/abs/1409.1824} {arXiv:1409.1824 [nucl-ex]}
  \BibitemShut {NoStop}%
\bibitem [{\citenamefont {Wauters}\ \emph {et~al.}(2009)\citenamefont
  {Wauters}, \citenamefont {Kraev}, \citenamefont {Tandecki}, \citenamefont
  {Traykov}, \citenamefont {Van~Gorp},\ and\ \citenamefont
  {Severijns}}]{Wauters:2009jw}%
  \BibitemOpen
  \bibfield  {author} {\bibinfo {author} {\bibfnamefont {F.}~\bibnamefont
  {Wauters}}, \bibinfo {author} {\bibfnamefont {I.}~\bibnamefont {Kraev}},
  \bibinfo {author} {\bibfnamefont {M.}~\bibnamefont {Tandecki}}, \bibinfo
  {author} {\bibfnamefont {E.}~\bibnamefont {Traykov}}, \bibinfo {author}
  {\bibfnamefont {S.}~\bibnamefont {Van~Gorp}}, \ and\ \bibinfo {author}
  {\bibfnamefont {N.}~\bibnamefont {Severijns}},\ }\href {\doibase
  10.1103/PhysRevC.80.062501} {\bibfield  {journal} {\bibinfo  {journal} {Phys.
  Rev. C}\ }\textbf {\bibinfo {volume} {80}},\ \bibinfo {pages} {062501}
  (\bibinfo {year} {2009})},\ \Eprint {http://arxiv.org/abs/0901.0081}
  {arXiv:0901.0081 [nucl-ex]} \BibitemShut {NoStop}%
\bibitem [{\citenamefont {Carnoy}\ \emph {et~al.}(1991)\citenamefont {Carnoy},
  \citenamefont {Deutsch}, \citenamefont {Girard},\ and\ \citenamefont
  {Prieels}}]{Carnoy:1991jd}%
  \BibitemOpen
  \bibfield  {author} {\bibinfo {author} {\bibfnamefont {A.}~\bibnamefont
  {Carnoy}}, \bibinfo {author} {\bibfnamefont {J.}~\bibnamefont {Deutsch}},
  \bibinfo {author} {\bibfnamefont {T.}~\bibnamefont {Girard}}, \ and\ \bibinfo
  {author} {\bibfnamefont {R.}~\bibnamefont {Prieels}},\ }\href {\doibase
  10.1103/PhysRevC.43.2825} {\bibfield  {journal} {\bibinfo  {journal} {Phys.
  Rev. C}\ }\textbf {\bibinfo {volume} {43}},\ \bibinfo {pages} {2825}
  (\bibinfo {year} {1991})}\BibitemShut {NoStop}%
\bibitem [{\citenamefont {Wichers}\ \emph {et~al.}(1987)\citenamefont
  {Wichers}, \citenamefont {Hageman}, \citenamefont {Van~Klinken},
  \citenamefont {Wilschut},\ and\ \citenamefont {Atkinson}}]{Wichers:1986es}%
  \BibitemOpen
  \bibfield  {author} {\bibinfo {author} {\bibfnamefont {V.}~\bibnamefont
  {Wichers}}, \bibinfo {author} {\bibfnamefont {T.}~\bibnamefont {Hageman}},
  \bibinfo {author} {\bibfnamefont {J.}~\bibnamefont {Van~Klinken}}, \bibinfo
  {author} {\bibfnamefont {H.}~\bibnamefont {Wilschut}}, \ and\ \bibinfo
  {author} {\bibfnamefont {D.}~\bibnamefont {Atkinson}},\ }\href {\doibase
  10.1103/PhysRevLett.58.1821} {\bibfield  {journal} {\bibinfo  {journal}
  {Phys. Rev. Lett.}\ }\textbf {\bibinfo {volume} {58}},\ \bibinfo {pages}
  {1821} (\bibinfo {year} {1987})}\BibitemShut {NoStop}%
\end{thebibliography}%

\end{document}